\documentclass[11pt]{article}
\pdfoutput=1

\usepackage{latexsym, graphicx}
\usepackage{amsmath, amssymb, bm}
\usepackage{hyperref}
\usepackage{longtable}
\usepackage{color}
\usepackage{rotating}
\usepackage{multicol}
\usepackage{authblk}

\begin{document}
\title{Off-diagonal coupling of supersymmetric SYK model}

\author[a]{Chenhao Zhang}
\author[a]{Wenhe Cai\thanks{whcai@shu.edu.cn}}
\date{}
\affil[a]{Department of Physics, Shanghai University, Shanghai, 200444, China}
\maketitle

\begin{abstract}
In this work, we consider the off-diagonal coupling between two supersymmetric SYK models, which preserves both supersymmetry and solvability. We found that the interaction terms of the N=2 supersymmetric SYK are holographically interpreted as a possible supersymmetric traversable wormhole. First, we introduce the coupling in the Homologic N=1 SYK model as a simplified example. Similar couplings can also be applied to the N=2 chiral SYK model with BPS states. We propose a special form of N=4 SYK by introducing supermultiplets, which also naturally include the coupling terms with additional symmetries. The N=4 holographic properties are investigated through the analysis of N=2 SYK theory. Furthermore, the effective actions are studied in both the thermal limit and the low-energy limit. We also investigate the SYK-like thermal field double states of the supersymmetric SYK model and the transmission amplitude between single-sided N=2 models in Lorentz time. Additionally, we study the multi-sided N=2,4 OTOCs.

\end{abstract}

\newpage
\tableofcontents
\newpage
\section{Introduction}
Recently, supersymmetric holographic wormholes have gained attention. However, these holographic wormholes still lack a simple and solvable model. In previous works\cite{1,2,3,4}, the SYK model was introduced as a Gaussian random Majorana fermionic model with good holographic properties. It is dual to a NAdS spacetime. The emergent IR conformal symmetry is also demonstrated in the description of NAdS2/NCFT1 duality\cite{5,6}. Additionally, the supersymmetric SYK model originates from this research on the cold horizon\cite{7,8}. Supersymmetric SYK (SSYK) models are supersymmetric extensions of the N=0 SYK model with dynamical bosons\cite{9,10,11}. The $ T\bar{T} $ deformations also exhibit an emergent supersymmetric conformal symmetry, which is dual to NAdS spacetime with additional N=1 and N=2 supersymmetry and Grassmann variables\cite{12,13,14,BRSZ}. These models have also been extensively studied\cite{15,16,17,18} and are characterized by valuable BPS states\cite{19}. For instance, in the double scaling limit\cite{20,21,22}, both non-supersymmetric and supersymmetric wormholes featuring statistical chords have been introduced\cite{23,24,25}. However, the intricate structures of these entangled systems remain unexamined within the context of an adequate holographic model.

In this paper, we introduce a first-order interaction into the supersymmetric SYK model, maintaining both supersymmetry and solvability. This model is similar to the approach used in the N=0 coupled SYK model (MQ)\cite{26,27}. We also explore its low-energy effective action\cite{28,29}, thermal phase structure, and causality. Previous work has proposed a simple supersymmetrization of JT gravity, which exhibits promising holographic properties \cite{30,31}. Notably, supersymmetric JT gravity closely resembles the N=0 model. Moreover, we extend this to supergravity with N=4 supersymmetry, which also exhibits duality due to the unique form of our N=4 SYK model.

The main goal of this paper is to examine the supersymmetric interaction term in both the N=2 and a possible N=4 SYK models. These models are formed using a special form of chiral supermultiplet that combines two N=2 SYK models and naturally represents spinors. Currently, supersymmetric SYK models are under extensive investigation, with a focus on aspects such as thermal effective actions, phase structures, and out-of-time correlators. However, they do not involve a detailed study of thermal phase interactions involving wormholes. Within this context, we aim to delve deeper into the causal properties involving excitation transmission and build upon prior research \cite{32,33,34,35,36,37}, thereby further enriching the understanding of supersymmetric extensions.

The second goal is to study the N=4 model. Unlike higher-dimensional N=1 and N=2 models \cite{38,39}, various N=4 SYK models have been proposed \cite{40,41,42}. The N=4 model includes chiral, vector, and tensor supermultiplets. In this paper, we utilize the 1D supermultiplet. The superfields in this context differ significantly from the bosonic case. Due to its unique structure, the holographic properties of this N=4 SYK model can be inferred from the N=2 model in superspace, although direct study poses challenges. We introduce interaction terms similar to those in the N=1 and N=2 models and examine black hole-wormhole structures.

This paper is organized as follows. In Section II, we briefly review the N=1 and N=2 supersymmetric SYK models and propose a simple form for the N=4 SYK model. We then introduce first-order interactions into the models. In Section III, we begin by examining the holographic duality between NAdS spacetime and the SYK model, and further investigate the supersymmetric NAdS model and the supersymmetric SYK model with interaction terms. The holographic properties of the N=4 model are also studied. In Section IV, we propose the thermal field double states in the supersymmetric SYK model. In Section V, we verify excitation transmission and out-of-time correlations (OTOCs). Section VI presents the conclusion and discussion.
\section{Definition of the model}
\subsection{A brief review of N=1 and N=2 SYK}
The N=1 SYK model has been the most widely studied attempt at a supersymmetric SYK model. Subsequent research has revealed that it has fewer physical interactions. Nevertheless, the N=1 SYK model remains a valuable case study for exploring supersymmetry, despite its limited significance in the field of supersymmetric physics. Its thermal properties exhibit similarities to those of the N=2 models. Furthermore, N=1 SYK theory can be formulated using independent supercharges, in a manner similar to that of the N=0 model
\begin{equation}
Q=i^{\frac{q-1}{2}}\sum_{i_1i_2...i_q}{C_{i_1i_2...i_q}\psi ^{i_1}\psi ^{i_2}}...\psi ^{i_q}.
\end{equation}
$\psi ^i$ represents Majorana fermions on sites ranging from $1...N$. In the initial definition, the interaction fermion $q$ is assigned a value of 3. The corresponding coupling $C_{ijk}$ is an $N\times N\times N$ Gaussian random antisymmetric tensor, specified by a constant $J$ with units of energy. The random average of the coupling $C$ exhibits a form similar to that in the N=0 SYK model
\begin{equation}
\overline{C_{i_1i_2...i_q}}=0,
$$
$$
\overline{C_{i_1i_2...i_q}^{2}}=\frac{2J}{N^2}.
\end{equation}
The supercharge of the N=1 model closely resembles that of the N=0 SYK model. The N=1 Hamiltonian is proportional to the square of the supercharge
\begin{equation}
\mathcal{H} =Q^2=E_0+\sum_{1\leqslant i<j<k\leqslant N}{J_{ijkl}\psi ^i}\psi ^j\psi ^k\psi ^l.
\end{equation}
It could be considered a special form of the N=0 SYK model that addresses
\begin{equation}
E_0=\sum_{1\leqslant i<j<k\leqslant N}{C_{ijk}^{2}},
$$
$$
J_{ijkl}=-\frac{1}{8}\sum_a{C_{aij}C_{kla}}.
\end{equation}
Here, $E_0$ is a disorder-averaged constant, while $J_{ijkl}$ constitutes a random matrix but is not Gaussian independent. The supercharge is constructed from Majorana fermions and should adhere to fermionic anticommutation relations. For simplicity, an auxiliary bosonic field can be introduced.
\begin{equation}
\left\{ Q,\psi ^i \right\} =Q\psi ^i=i\sum_{1\leqslant j<k\leqslant N}{C_{ijk}}\psi ^j\psi ^k\equiv b_i
$$
$$
Qb^i=H\psi ^i=\partial _{\tau}\psi ^i.
\end{equation}
The Lagrangian consists of kinetic fermionic and auxiliary bosonic field operators, as well as potential terms
\begin{equation}
\mathcal{L} =\sum_i{\left[ \frac{1}{2}\psi ^i\partial _{\tau}\psi ^i-\frac{1}{2}b^ib^i+i\sum_{1\leqslant j<k\leqslant N}{C_{ijk}}b^i\psi ^j\psi ^k \right]}.
\end{equation}

For simplicity, we can define a superfield within the corresponding superspace. This allows us to constrain both the bosons and fermions while describing the supersymmetric theory without sacrificing generality.
\begin{equation}
\varPsi \left( \tau ,\theta \right) =\psi \left( \tau \right) +\theta b\left( \tau \right),
\end{equation}
and introduce the super covariant derivative
\begin{equation}
D_{\theta}\equiv \partial _{\theta}+\theta \partial _{\tau},
$$
$$
D_{\theta}^{2}=\partial _{\tau}.
\end{equation}
We can define a bilinear superfield correlation function, which contains the components of f-f, f-b, b-f and b-b
\begin{align}
\mathcal{G} \left( \tau ,\theta ;\tau \prime,\theta \prime \right) &=\left< \varPsi \left( \tau ,\theta \right) \varPsi \left( \tau \prime,\theta \prime \right) \right> =\left< \left( \psi \left( \tau \right) +\theta b\left( \tau \right) \right) \left( \psi \left( \tau ' \right) +\theta 'b\left( \tau ' \right) \right) \right>
\\
&\equiv G_{\psi \psi}\left( \tau ,\tau ' \right) +\sqrt{2}\theta G_{b\psi}\left( \tau ,\tau ' \right) -\sqrt{2}\theta 'G_{\psi b}\left( \tau ,\tau ' \right) +2\theta \theta 'G_{bb}\left( \tau ,\tau ' \right) .
\end{align}
Then, we can return to the original SSYK model and rewrite the effective action into superspace
\begin{equation}
S_{EFF}=\int{d\theta d\tau \left( -\frac{1}{2}\varPsi ^iD_{\theta}\varPsi ^i \right)}+\frac{J}{3N^2}\int{d\theta _1d\tau _1d\theta _2d\tau _2\left( \varPsi ^i\varPsi ^i \right) ^3}.
\end{equation}
This action can be diagonalized into the odd N=0 SYK form using superfields and covariant derivatives. Additionally, the saddle point provides descriptions for the equations of motion, incorporating disorder averaging
\begin{equation}
D_{\theta}\mathcal{G} \left( \tau ,\theta ;\tau '',\theta '' \right) +\int{d\tau'd\theta'}\mathcal{G} \left( \tau ',\theta ';\tau '',\theta '' \right) \left( J\mathcal{G} \left( \tau ',\theta ';\tau '',\theta '' \right) ^2 \right) =\left( \theta -\theta '' \right) \delta \left( \tau -\tau '' \right) .
\end{equation}

N=2 supersymmetric SYK models can be generated from the symmetry breaking of N=1 theory. It reorganizes the square of supercharges into a pair of identical, conjugate supercharges with SU(2) symmetry
\begin{equation}
Q=i\sum_{1\leqslant i<j<k\leqslant N}{C_{ijk}\psi ^i}\psi ^j\psi ^k,
$$
$$
\bar{Q}=i\sum_{1\leqslant i<j<k\leqslant N}{\bar{C}^{ijk}\bar{\psi}_i}\bar{\psi}_j\bar{\psi}_k.
\end{equation}
$C$ and $\bar{C}$ represent independent Gaussian random coefficients, and the Majorana fermionic fields are replaced by a pair of chiral conjugated complex fields. Notice that fermions here only anticommute with fields in conjugated components, where the N=1 supersymmetry is broken
\begin{equation}
\left\{ \psi ^i,\bar{\psi}_j \right\} =\delta _{j}^{i},
$$
$$
\left\{ \psi ^i,\psi ^j \right\} =0,
$$
$$
\left\{ \bar{\psi}_i,\bar{\psi}_j \right\} =0.
\end{equation}
The N=2 SYK Hamiltonian is determined by the anticommutator of the conjugated supercharges
\begin{equation}
Q^2=0,
$$
$$
\bar{Q}^2=0,
$$
$$
\mathcal{H} =\left\{ Q,\bar{Q} \right\} =\left| C \right|^2+\sum_{ijkl}{J_{ij}^{kl}}\psi ^i\psi ^j\bar{\psi}_k\bar{\psi}_l.
\end{equation}
We can also average the square of the Gaussian random coefficient $C$ with an energy-dependent parameter
\begin{equation}
\overline{C_{ijk}\bar{C}^{ijk}}=\frac{2J}{N^2},
\end{equation}
Here, $J$ is also a random parameter, independently distributed according to a Gaussian distribution. Analogous to the N=1 model, the auxiliary bosonic fields are introduced for both chiral fermions through the anti-commutators of conjugate supercharges
\begin{equation}
\left\{ Q,\psi ^i \right\} =0,
$$
$$
\left\{ \bar{Q},\bar{\psi}_i \right\} =0,
$$
$$
b_i\equiv i\left\{ \bar{Q},\psi ^i \right\} =\sum_{1\leqslant j<k\leqslant N}{\bar{C}^{ijk}}\bar{\psi}_j\bar{\psi}_k,
$$
$$
\bar{b}^i\equiv i\left\{ Q,\bar{\psi}_i \right\} =\sum_{1\leqslant j<k\leqslant N}{C_{ijk}}\psi ^j\psi ^k.
\end{equation}

The covariant derivative operators are expressed in terms of conjugate supersymmetric components.
\begin{equation}
D\equiv \partial _{\theta}+\bar{\theta}\partial _{\tau},
$$
$$
\bar{D}\equiv \partial _{\bar{\theta}}+\theta \partial _{\tau}.
\end{equation}
In N=2 algebra, various approaches are permitted for structuring the superfield. Additionally, it is convenient to opt for a method known as the chiral superfield, which satisfies
\begin{equation}
\bar{D}\varPsi ^i=0,
$$
$$
D\bar{\varPsi}_i=0.
\end{equation}
And we obtain the effective actions with superfields
\begin{equation}
S_{eff}=\frac{1}{2}\int{d\theta d\bar{\theta}d\tau}\bar{\varPsi}_i\varPsi ^i+\int{d\theta d\bar{\theta}d\tau}\bar{\varPsi}_{i1}\bar{\varPsi}_{i2}\bar{\varPsi}_{i3}\left< \bar{C}^{i1i2i3}C_{j1j2j3} \right> \varPsi ^{j1}\varPsi ^{j2}\varPsi ^{j3},
$$
$$
\bar{S}_{eff}=\frac{1}{2}\int{d\bar{\theta}d\theta d\tau}\varPsi ^i\bar{\varPsi}_i+\int{d\bar{\theta}d\theta d\tau}\varPsi ^{j1}\varPsi ^{j2}\varPsi ^{j3}\left< C_{j1j2j3}\bar{C}^{i1i2i3} \right> \bar{\varPsi}_{i1}\bar{\varPsi}_{i2}\bar{\varPsi}_{i3}.
\end{equation}

Then, we can define the creation and annihilation operators using covariant derivatives, and subsequently construct the bilinear Fock space from the superfield
\begin{equation}
\varPsi ^i\left( \tau ,\theta ,\bar{\theta} \right) =\psi ^i\left( \tau +\theta \bar{\theta} \right) +\theta b^i,
$$
$$
\bar{\varPsi}_i\left( \tau ,\theta ,\bar{\theta} \right) =\bar{\psi}_i\left( \tau +\bar{\theta}\theta \right) +\bar{\theta}\bar{b}_i.
\end{equation}
Notice that the superfields adhere to anti-commutation relations and possess chiral conjugated properties. Additionally, we can define the bi-chiral correlation function using superfields
\begin{equation}
\begin{split}
&\mathcal{G} \left( \tau ,\theta ,\bar{\theta};\tau ',\theta ',\bar{\theta}' \right) =\left< \bar{\varPsi}\left( \tau ,\theta ,\bar{\theta} \right) \varPsi \left( \tau ',\theta ',\bar{\theta}' \right) \right>
\\
&\equiv G_{\psi \psi}\left( \tau -\theta \bar{\theta},\tau '+\theta '\bar{\theta}' \right) +\sqrt{2}\bar{\theta}G_{b\psi}\left( \tau ,\tau '+\theta '\bar{\theta}' \right) -\sqrt{2}\theta 'G_{\psi b}\left( \tau -\theta \bar{\theta},\tau ' \right) +2\bar{\theta}\theta 'G_{bb}\left( \tau ,\tau ' \right) .
\\
&\bar{\mathcal{G}}\left( \tau ,\theta ,\bar{\theta};\tau ',\theta ',\bar{\theta}' \right) =\left< \varPsi \left( \tau ,\theta ,\bar{\theta} \right) \bar{\varPsi}\left( \tau ',\theta ',\bar{\theta}' \right) \right>
\\
&\equiv \bar{G}_{\psi \psi}\left( \tau +\theta \bar{\theta},\tau '-\theta '\bar{\theta}' \right) +\sqrt{2}\theta \bar{G}_{b\psi}\left( \tau ,\tau '-\theta '\bar{\theta}' \right) -\sqrt{2}\bar{\theta}'\bar{\mathrm{G}}_{\psi b}\left( \tau +\theta \bar{\theta},\tau ' \right) +2\theta \bar{\theta}\bar{\mathrm{G}}_{bb}\left( \tau ,\tau \right) .
\end{split}
\end{equation}
We can further formulate the coupled Schwinger-Dyson equation for chiral-anti-chiral, and describe the N=2 SYK system with the conjugate components
\begin{equation}
\begin{split}
&D_{\theta}\mathcal{G} \left( \tau ,\theta ,\bar{\theta};\tau '',\theta '',\bar{\theta}'' \right) +\int{d\tau 'd\theta '}d\bar{\theta}'\mathcal{G}\left( \tau ,\theta ,\bar{\theta};\tau '',\theta '',\bar{\theta}'' \right) \left( J\mathcal{G} \left( \tau ,\theta ,\bar{\theta};\tau '',\theta '',\bar{\theta}'' \right) ^q \right)
\\
&=\left( \bar{\theta}-\bar{\theta}'' \right) \delta \left( \tau -\tau ''-\theta \bar{\theta}+\theta ''\bar{\theta}'' \right) ,
 \end{split}
\end{equation}
\begin{equation}
\begin{split}
&D_{\bar{\theta}}\mathcal{G} \left( \tau ,\theta ,\bar{\theta};\tau '',\theta '',\bar{\theta}'' \right) +\int{d\tau 'd\theta 'd\bar{\theta}'\bar{\mathcal{G}} \left( \tau ,\theta ,\bar{\theta};\tau ',\theta ',\bar{\theta}' \right) \left( J\bar{\mathcal{G}}\left( \tau ',\theta ',\bar{\theta}';\tau '',\theta '',\bar{\theta}'' \right) ^q \right)}
\\
&=\left( \theta -\theta '' \right) \delta \left( \tau -\tau ''-\theta \bar{\theta}+\theta ''\bar{\theta}'' \right).
 \end{split}
\end{equation}
In order to understand these supersymmetric SYK models more precisely, we can also consider the supercharges with a chord diagram. As shown in Figure 1(a), we use the blue line on the left-hand side (LHS) to represent a single SYK system with coupling $C_{ijk}$ (which is known as a supercharge). The black line on the right-hand side (RHS) represents the same SYK system that is to be interacted with. We assume the interaction, represented by chords that start on the LHS and end on the RHS. The pink area includes the interactions between a single fermion $\psi^{i}$ on the LHS and the entire SYK system on the RHS, as well as the classical SYK coupling $C_{ijk}$ inside the RHS. We define this pink region as the auxiliary boson $b^{i}$. It is easy to prove that the dimension of fermions is equal to the dimension of those auxiliary bosons. As depicted in Figure 1(b), N=2 supersymmetry prohibits interactions within a single chiral component. When we define the blue line as representing the chiral supercharge with $\psi^{i}$, the red line represents the anti-chiral supercharge with $\bar{\psi}^{i}$. The green line represents the anti-chiral supercharge to be interacted with, and the black line represents the anti-chiral supercharge. We also define the pink region as the chiral boson $b^{i}$ and the yellow region as the chiral boson $\bar{b}^{i}$.

\begin{figure}[!t]
\begin{minipage}{0.48\linewidth}
\centerline{\includegraphics[width=8cm]{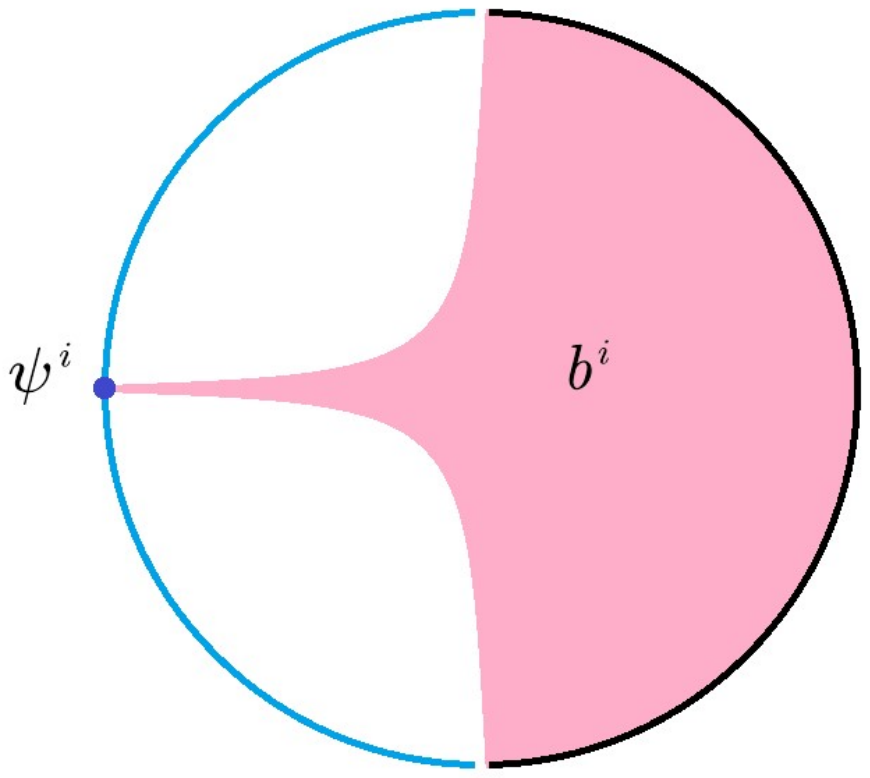}}
\centerline{(a)}
\end{minipage}
\hfill
\begin{minipage}{0.48\linewidth}
\centerline{\includegraphics[width=8cm]{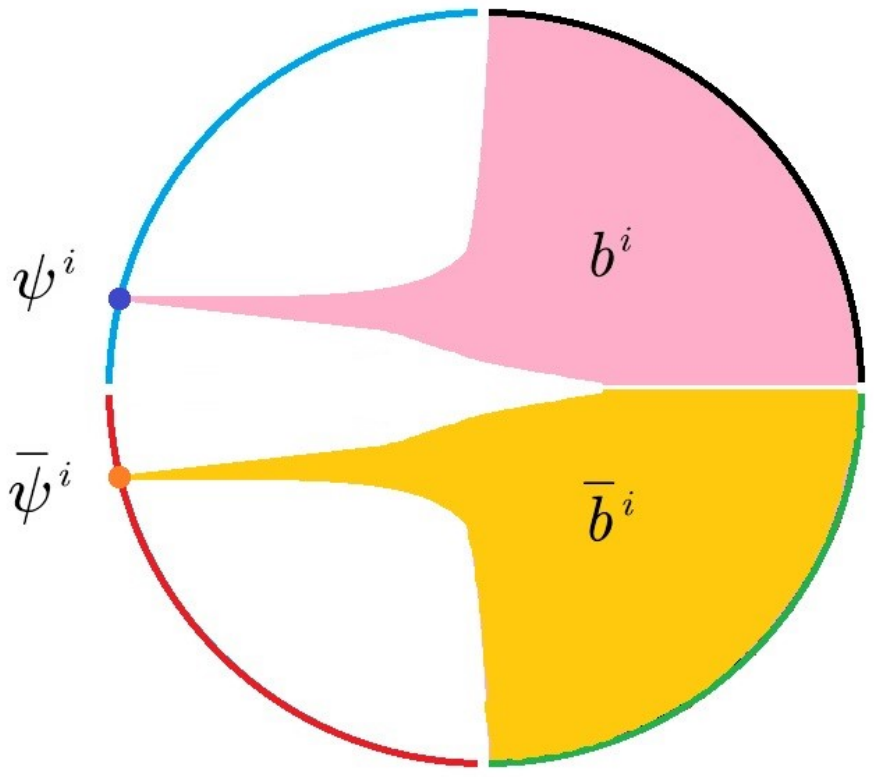}}
\centerline{(b)}
\end{minipage}

\caption{\label{fig:Figure 1}  Interpretation of the chord diagram as the supersymmetric SYK model (a) N=1 SYK model (b)N=2 SYK model with chirals}
\end{figure}
\subsection*{N=4 Bilinear SYK}

The N=4 SYK model was derived from considerations of a cold horizon. The operator $\varPhi$ represents a chiral multiplet that exhibits both bosonic and fermionic properties, and it is characterized by Gaussian randomness.
\begin{equation}
\mathcal{L}=\int{d^2\theta}\varOmega _{\alpha \beta \gamma}\varPhi ^{\alpha}\varPhi ^{\beta}\varPhi ^{\gamma}+h.c,
\end{equation}
the coefficients $\varOmega _{\alpha \beta \gamma}$ are Gaussian random variables.
\begin{equation}
\left< \varOmega _{\alpha \beta \gamma} \right> =0,
$$
$$
\left< \varOmega _{\alpha \beta \gamma}^{2} \right> =J.
\end{equation}
Numerous supersymmetric models exhibit the N=4 SYK-like Hamiltonian form (for example, see \cite{39,42}). We choose the bilinear scalar supermultiplet approach, which differs from the chiral multiplet approach described in \cite{40,41}. Here, we define
\begin{equation}
\mathcal{L} =\sum_{i,\alpha}{\left[ \partial _{\tau}\bar{\phi}_i\partial _{\tau}\phi ^i+\bar{\psi}_{i}^{\alpha}\partial _{\tau}\psi _{\alpha}^{i}+\bar{F}_iF^i \right]}+3C_{ijk}\left( F^i\phi ^j\phi ^k+\epsilon ^{\alpha \beta}\psi _{\alpha}^{i}\psi _{\beta}^{j}\phi ^k \right) +h.c,
\end{equation}
with $\alpha,\beta=1,2$, which correspond to supercharge indices 1 and 2. We define $\epsilon ^{1 1}=\epsilon ^{2 2}=0$, $\epsilon ^{1 2}=\epsilon ^{2 1}=1$.

The construction is commenced within the framework of identical N=2 supercharges
\begin{equation}
\left\{ \psi _{\alpha}^{i},\bar{\psi}_{j}^{\beta} \right\} =\delta _{\alpha}^{\beta}\delta _{j}^{i}.
\end{equation}

We consider two N=2 SYK models, as shown in Figure 1(b), denoted by indices 1 and 2. An additional U(1) symmetry can also emerge between models 1 and 2 (which is non-reparameterizing), whereas the N=2 supercharges are associated with SU(2)
\begin{equation}
Q_{\alpha}=i\sum_{1\leqslant i<j<k\leqslant N}{C_{ijk}^{\beta \gamma}\psi _{\alpha}^{\prime i}\psi _{ \beta}^{\prime j}\psi _{\gamma}^{\prime k}},
$$
$$
\bar{Q}^{\beta}=i\sum_{1\leqslant i<j<k\leqslant N}{\bar{C}_{\alpha \gamma}^{ijk}\bar{\psi}_{i}^{\prime \alpha}\bar{\psi}_{j}^{\prime \beta}\bar{\psi}_{k}^{\prime \gamma}}.
\end{equation}
Supercharges are essentially Gaussian random, and the fermions $\psi'$ also involve multiple indices $\alpha$,$\beta$, and $\gamma$, which vanish with the random coefficient. These indices can disappear through transformations, and we will see that they have less physical meaning and would vanish in integrals. For simplicity, we choose $\alpha=\beta=\gamma$ in a single supercharge to make the supercharges identical. Thus, the indices $\alpha$ and $\beta$ associated with supercharges at positions 1 and 2 exhibit rotational symmetry. The auxiliary boson $F$ is pertinent to higher-order cross terms between the indices. And $C_{ijk}=C_{ijk}^{1}C_{ijk}^{2}$ represents the multiple of two independent indices, which indicates $\overline{C_{ijk}\bar{C}^{ijk}}=\frac{2J}{N^2}$. We can also generate the corresponding bosonic operator $b$ and $\bar{b}$
\begin{equation}
b^i=\left\{ Q_{\alpha},\bar{\psi}_{i}^{\alpha} \right\} ,
$$
$$
\bar{b_i}=\left\{ \bar{Q}^{\alpha},\psi _{\alpha}^{i} \right\} .
\end{equation}
Considering the interaction potential in Eq. (27), the N=4 Hamiltonian is given by
\begin{equation}
\begin{split}
H&=\psi ^{\alpha}\psi ^{\beta}\left\{ \bar{Q}_{\alpha},\psi ^{\alpha} \right\} ^2\psi ^{\alpha}\psi ^{\beta}+\psi ^{\alpha}\psi ^{\beta}\left\{ \bar{Q}_{\alpha},\psi ^{\alpha} \right\} \psi ^{\alpha}\left\{ \bar{Q}_{\alpha},\psi ^{\alpha} \right\} \psi ^{\beta}+h.c
\\
&=\left\{ \bar{Q}_{\alpha},Q^{\alpha} \right\} \left\{ \bar{Q}_{\beta},Q^{\beta} \right\} +T\left\{ \bar{Q}_{\alpha},Q^{\alpha} \right\} \left\{ \bar{Q}_{\beta},Q^{\beta} \right\} +h.c
\\
&=\left\{ \bar{Q}_{\alpha},Q^{\alpha} \right\} \left\{ \bar{Q}_{\beta},Q^{\beta} \right\} .
\end{split}
\end{equation}

Here we use the operator $T$ to exchange the sorting order of the chiral fermionic pair, it vanishes when we ignore the $\alpha-\beta$ coupling in supercharges. To establish the N=4 SYK theory (as shown in Eq. (25)), the superfields are defined as follows
\begin{equation}
\varPhi ^i=\phi ^i+\theta ^{\alpha}\psi _{\alpha}^{i}+\theta^2  F^i=\left( \psi\prime_{\alpha}^{i}+\theta _{\alpha} b^i \right) \left( \psi\prime_{\beta}^{i}+\theta _{\beta} b^i \right)
=
\psi \prime_{\alpha}^{i}\psi\prime_{\beta}^{i}+\theta _{\alpha}b^i\psi\prime_{\beta}^{i}-\theta _{\beta} b^i\psi\prime_{\alpha}^{i}+\theta _{\alpha}\theta _{\beta} b^i b^i
$$
$$
\bar{\varPhi}_i=\phi _i+\bar{\theta}_{\alpha}\bar{\psi}_{i}^{\alpha}+\bar{\theta}^2\bar{F}_i=\left( \bar{\psi}_{i}^{\prime\alpha}+\bar{\theta}^{\alpha}\bar{b}_i \right) \left( \bar{\psi}_{i}^{\prime\beta}+\bar{\theta}^{\beta}\bar{b}_i \right) =\bar{\psi}_{i}^{\prime\alpha}\bar{\psi}_{i}^{\prime\beta}+\bar{\theta}^{\alpha}\bar{b}_i\bar{\psi}_{i}^{\prime\beta}-\bar{\theta}^{\beta}\bar{b}\bar{\psi}_{i}^{\prime\alpha}+\bar{\theta}^{\alpha}\bar{\theta}^{\beta}\bar{b}_i\bar{b}_i,
\end{equation}
We can then define the bosonic operators $\phi$ and $F$, as well as the fermionic operators $\psi _{\alpha}$ and $\psi _{\beta}$, all of which exhibit multiple chiral aspects within the Lagrangian.
\begin{equation}
\phi ^i=\epsilon ^{\alpha \beta}\psi \prime_{\alpha}^{i}\psi \prime_{\beta}^{i},
$$
$$
\psi _{\alpha}^{i}=\epsilon _{\alpha}^{\beta}b^i\psi \prime_{\beta}^{i},
$$
$$
\psi _{\beta}^{i}=\epsilon _{\beta}^{\alpha}b^i\psi \prime_{\alpha}^{i},
$$
$$
F^i=b^ib^i,
$$
$$
\bar{\phi}_i=\epsilon _{\alpha \beta}\bar{\psi}_{i}^{\prime\alpha}\bar{\psi}_{i}^{\prime\beta},
$$
$$
\bar{\psi}_{i}^{\alpha}=\epsilon _{\beta}^{\alpha}\bar{b}_i\bar{\psi}_{i}^{\prime\beta},
$$
$$
\bar{\psi}_{i}^{\beta}=\epsilon _{\alpha}^{\beta}\bar{b}_i\bar{\psi}_{i}^{\prime\alpha},
$$
$$
\bar{F}_i=\bar{b}_i\bar{b}_i.
\end{equation}

Here we use $\psi^{i}_{\alpha}$ and $\bar{\psi}^{\alpha}_{i}$ to represent fermions in two identical N=2 SYK models. We can also decouple the N=4 theory back to N=2. Given that the superfield theory can also be incorporated into the N=4 theory, the covariant derivative is defined as

\begin{equation}
D_{\alpha}=\frac{\partial}{\partial \theta ^{\alpha}}+\bar{\theta}_{\alpha}\frac{\partial}{\partial \tau},
$$
$$
\bar{D}^{\beta}=\frac{\partial}{\partial \bar{\theta}_{\beta}}+\theta ^{\beta}\frac{\partial}{\partial \tau},
$$
$$
D_{\theta}^{2}\equiv \frac{1}{4}\epsilon ^{\alpha \beta}D_{\alpha}D_{\beta}=D_1D_2,
$$
$$
D_{\bar{\theta}}^{2}\equiv \frac{1}{4}\epsilon _{\alpha \beta}\bar{D}^{\alpha}\bar{D}^{\beta}=\bar{D}^1\bar{D}^2.
\end{equation}

The action of the N=4 SYK model can be explicitly expressed using auxiliary parameters and superfields
\begin{equation}
\begin{split}
S&=\int{d\tau d\theta ^1d\theta ^2d\bar{\theta}_1d\bar{\theta}_2}\left( \bar{\varPhi}_i\varPhi ^i \right) +\int{d\tau d\theta ^1d\theta ^2}C_{ijk}\varPhi ^i\varPhi ^j\varPhi ^k+\int{d\tau d\bar{\theta}_1d\bar{\theta}_2}\bar{C}_{ijk}\bar{\varPhi}_i\bar{\varPhi}_j\bar{\varPhi}_k
\\
&=\int{d\tau}\left[ \dot{\bar{\phi}}_i\dot{\phi}^i+\bar{\psi}_{i}^{\alpha}\dot{\psi}_{\alpha}^{i}+\bar{F}_iF^i+3C_{ijk}\left( F_i\phi _j\phi _k+\epsilon ^{\alpha \beta}\psi _{\alpha}^{i}\psi _{\beta}^{j}\phi _k \right) +h.c \right] .
\end{split}
\end{equation}

This action can also be constrained by the auxiliary equation of motion
\begin{equation}
\begin{split}
&D_{\theta}^{2}\mathcal{G} \left( \tau ,\theta ^1,\theta ^2,\bar{\theta}_1,\bar{\theta}_2;\tau '',\theta ''^1,\theta ''^2,\bar{\theta}''_1,\bar{\theta}''_2 \right) +\int{d}\tau ' d^2\theta '\mathcal{G} _{AB}\left( \tau ,\theta ^1,\theta ^2,\bar{\theta}_1,\bar{\theta}_2;\tau '',\theta ''^1,\theta ''^2,\bar{\theta}''_1,\bar{\theta}''_2 \right)
\\
& \left( J\mathcal{G} _{AB}\left( \tau ,\theta ^1,\theta ^2,\bar{\theta}_1,\bar{\theta}_2;\tau '',\theta ''^1,\theta ''^2,\bar{\theta}''_1,\bar{\theta}''_2 \right) ^2 \right) =\left( \bar{\theta}^{1}-\bar{\theta}^{1\prime\prime} \right) \left( \bar{\theta}^{2}-\bar{\theta}^{2\prime\prime} \right) \delta \left( \tau -\tau '' \right) .
\end{split}
\end{equation}
\begin{equation}
\begin{split}
&\bar{D}_{\theta}^{2}\bar{\mathcal{G}}\left( \tau ,\theta ^1,\theta ^2,\bar{\theta}_1,\bar{\theta}_2;\tau '',\theta ''^1,\theta ''^2,\bar{\theta}''_1,\bar{\theta}''_2 \right) +\int{d}\tau ' d^2\bar{\theta}'\bar{\mathcal{G}}_{AB}\left( \tau ,\theta ^1,\theta ^2,\bar{\theta}_1,\bar{\theta}_2;\tau '',\theta ''^1,\theta ''^2,\bar{\theta}''_1,\bar{\theta}''_2 \right)
\\
& \left( J\bar{\mathcal{G}}_{AB}\left( \tau ,\theta ^1,\theta ^2,\bar{\theta}_1,\bar{\theta}_2;\tau '',\theta ''^1,\theta ''^2,\bar{\theta}''_1,\bar{\theta}''_2 \right) ^2 \right) =\left( \theta_{1} -\theta _{1}'' \right) \left( \theta_{2} -\theta_{2} '' \right) \delta \left( \tau -\tau '' \right).
\end{split}
\end{equation}
Where the N=4 correlation function is represented by
\begin{equation}
\begin{split}
&\mathcal{G} \left( \tau ,\theta _1,\bar{\theta}_1,\theta _2,\bar{\theta}_2;\tau ',\theta ',\bar{\theta}',\theta _2',\bar{\theta}_2' \right) =\left< \bar{\varPhi}\left( \tau ,\theta _1,\bar{\theta}_1,\theta _2,\bar{\theta}_2;\tau ',\theta ',\bar{\theta}',\theta _2',\bar{\theta}_2' \right) \varPhi \left( \tau ,\theta _1,\bar{\theta}_1,\theta _2,\bar{\theta}_2;\tau ',\theta ',\bar{\theta}',\theta _2',\bar{\theta}_2' \right) \right>
\\
&=G_{\phi \phi}\left( \tau -\theta _1\bar{\theta}_1-\theta _2\bar{\theta}_2,\tau '+\theta _1'\bar{\theta}_1'+\theta _2'\bar{\theta}_2' \right) +\sqrt{2}\theta _{\alpha}G_{\phi \psi \alpha}\left( \tau -\theta _1\bar{\theta}_1-\theta _2\bar{\theta}_2,\tau '+\theta _1'\bar{\theta}_1'+\theta _2'\bar{\theta}_2'-\theta _{\alpha}'\bar{\theta}_{\alpha}' \right)
\\
&-\sqrt{2}\bar{\theta}_{\alpha}'G_{\psi \alpha \phi}\left( \tau -\theta _1\bar{\theta}_1-\theta _2\bar{\theta}_2+\theta _{\alpha}\bar{\theta}_{\alpha},\tau '+\theta _1'\bar{\theta}_1'+\theta _2'\bar{\theta}_2' \right) +\bar{\theta}_1\bar{\theta}_2\theta _1\theta _2G_{FF}\left( \tau ,\tau ' \right)
\\
&+2\bar{\theta}_{\alpha}\theta _{\beta}'G_{\psi \alpha \psi \beta}\left( \tau ,\tau ' \right) +\left( \tau -\theta _1\bar{\theta}_1-\theta _2\bar{\theta}_2+\theta _{\alpha}\bar{\theta}_{\alpha},\tau '+\theta _1'\bar{\theta}_1'+\theta _2'\bar{\theta}_2'-\theta _{\beta}'\bar{\theta}_{\beta}' \right)
\\
&+\theta _1\theta _2G_{\phi F}\left( \tau -\theta _1\bar{\theta}_1-\theta _2\bar{\theta}_2,\tau ' \right) +\bar{\theta}_1\bar{\theta}_2G_{F\phi}\left( \tau ,\tau '+\theta _1'\bar{\theta}_1'+\theta _2'\bar{\theta}_2' \right)
\\
&+\sqrt{2}\bar{\theta}_{\alpha}\theta _1\theta _2G_{\psi \alpha F}\left( \tau -\theta _1\bar{\theta}_1-\theta _2\bar{\theta}_2+\theta _{\alpha}\bar{\theta}_{\alpha},\tau ' \right) +\sqrt{2}\bar{\theta}_1\bar{\theta}_2\theta _{\alpha}G_{F\psi \alpha}\left( \tau ,\tau '+\theta _1'\bar{\theta}_1'+\theta _2'\bar{\theta}_2'-\theta _{\beta}'\bar{\theta}_{\beta}' \right) ,
\end{split}
\end{equation}
\begin{equation}
\begin{split}
&\bar{\mathcal{G}}\left( \tau ,\theta _1,\bar{\theta}_1,\theta _2,\bar{\theta}_2;\tau ',\theta ',\bar{\theta}',\theta _2',\bar{\theta}_2' \right) =\left< \varPhi \left( \tau ,\theta _1,\bar{\theta}_1,\theta _2,\bar{\theta}_2;\tau ',\theta ',\bar{\theta}',\theta _2',\bar{\theta}_2' \right) \bar{\varPhi}\left( \tau ,\theta _1,\bar{\theta}_1,\theta _2,\bar{\theta}_2;\tau ',\theta ',\bar{\theta}',\theta _2',\bar{\theta}_2' \right) \right>
\\
&=\bar{G}_{\phi \phi}\left( \tau +\theta _1\bar{\theta}_1+\theta _2\bar{\theta}_2,\tau '-\theta _1'\bar{\theta}_1'-\theta _2'\bar{\theta}_2' \right) +\sqrt{2}\bar{\theta}_{\alpha}\bar{G}_{\phi \psi \alpha}\left( \tau +\theta _1\bar{\theta}_1+\theta _2\bar{\theta}_2,\tau '-\theta _1'\bar{\theta}_1'-\theta _2'\bar{\theta}_2'+\theta _{\alpha}'\bar{\theta}_{\alpha}' \right)
\\
&-\sqrt{2}\theta _{\alpha}'\bar{G}_{\psi \alpha \phi}\left( \tau +\theta _1\bar{\theta}_1+\theta _2\bar{\theta}_2-\theta _{\alpha}\bar{\theta}_{\alpha},\tau '-\theta _1'\bar{\theta}_1'-\theta _2'\bar{\theta}_2' \right) +\bar{\theta}_1\bar{\theta}_2\theta _1\theta _2\bar{G}_{FF}\left( \tau ,\tau ' \right)
\\
&+2\theta _{\alpha}\bar{\theta}_{\beta}'\bar{G}_{\psi \alpha \psi \beta}\left( \tau ,\tau ' \right) +\left( \tau +\theta _1\bar{\theta}_1+\theta _2\bar{\theta}_2-\theta _{\alpha}\bar{\theta}_{\alpha},\tau '-\theta _1'\bar{\theta}_1'-\theta _2'\bar{\theta}_2'+\theta _{\beta}'\bar{\theta}_{\beta}' \right)
\\
&+\bar{\theta}_1\bar{\theta}_2\bar{G}_{\phi F}\left( \tau +\theta _1\bar{\theta}_1+\theta _2\bar{\theta}_2,\tau ' \right) +\bar{\theta}_1\bar{\theta}_2\bar{G}_{F\phi}\left( \tau ,\tau '-\theta _1'\bar{\theta}_1'-\theta _2'\bar{\theta}_2' \right)
\\
&+\sqrt{2}\theta _{\alpha}\bar{\theta}_1\bar{\theta}_2\bar{G}_{\psi \alpha F}\left( \tau +\theta _1\bar{\theta}_1+\theta _2\bar{\theta}_2-\theta _{\alpha}\bar{\theta}_{\alpha},\tau ' \right) +\sqrt{2}\theta _1\theta _2\bar{\theta}_{\alpha}\bar{G}_{F\psi \alpha}\left( \tau ,\tau '-\theta _1'\bar{\theta}_1'-\theta _2'\bar{\theta}_2'+\theta _{\beta}'\bar{\theta}_{\beta}' \right) .
\end{split}
\end{equation}
Some of these components would vanish, and we will discuss this problem later.

\subsection{Introduce the coupling}
In order to describe a holographic wormhole, we can propose a similar first-order interaction term to the one used in MQ theory within superspace for consistency. For N=1 SYK, the coupled theory encompasses two separate N=1 SYK models, along with the interaction term
\begin{equation}
H_{int}=i\mu \partial _{\theta}\sum_j{\left( \varPsi _{L}^{j}\varPsi _{R}^{j}-\varPsi _{R}^{j}\varPsi _{L}^{j} \right)}.
\end{equation}

Here, we have introduced the derivative $\partial _{\theta}$ to constrain the theory with a supersymmetric delta function in a single-sided superposition. Subsequently, we obtain the supersymmetric action that includes the interaction term
\begin{equation}
\begin{split}
\frac{S_{eff}}{N}&=-\log Pf\left( D_{half\,\,integer} \right) +\log\det \left( \frac{D_{integer}}{\left( iw_n \right) ^2} \right) +\sum_{A,B=L,R}{\frac{1}{2}}\int{d\tau d\tau \prime}\left[ \varSigma _{\psi \psi ,AB}\left( \tau ,\tau \prime \right) G_{\psi \psi ,AB}\left( \tau ,\tau \prime \right)\right.
\\
& +\varSigma _{\psi b,AB}\left( \tau ,\tau \prime \right) G_{\psi b,AB}\left( \tau ,\tau \prime \right) +\varSigma _{b\psi ,AB}\left( \tau ,\tau \prime \right) G_{b\psi ,AB}\left( \tau ,\tau \prime \right) +\varSigma _{bb,AB}\left( \tau ,\tau \prime \right) G_{bb,AB}\left( \tau ,\tau \prime \right)
\\
& -\left( \delta _{AB}+\left( 1-\delta _{AB} \right) \left( -1 \right) ^{\left( q-1 \right) /2} \right) J\left( G_{\psi \psi ,AB}^{q-1}\left( \tau ,\tau \prime \right) G_{bb,AB}\left( \tau ,\tau \prime \right) -\frac{\left( q-1 \right)}{2}G_{\psi \psi ,AB}^{q-2}\left( \tau ,\tau \prime \right) G_{\psi b,AB}^{A}\left( \tau ,\tau \prime \right) \right.
\\
&\left. \left. G_{b\psi ,AB}^{S}\left( \tau ,\tau \prime \right) -\frac{\left( q-1 \right)}{2}G_{\psi \psi ,AB}^{q-2}\left( \tau ,\tau \prime \right) G_{\psi b,AB}^{S}\left( \tau ,\tau \prime \right) G_{b\psi ,AB}^{A}\left( \tau ,\tau \prime \right) \right) \right] .
\end{split}
\end{equation}
We can further utilize the variation of Green's function $G$ and self-energy $\varSigma$ in the effective action, and derive the equations of motion with the coupling term

\begin{equation}
D_{\theta}\mathcal{G} _{AB}\left( \tau ,\tau ' \right) -\sum_C{\left( -i\mu \epsilon _{AC}\partial _{\theta}\mathcal{G} _{AB}\left( \tau ,\tau ' \right) -\int{d\tau ''d\theta ''}\varSigma _{AC}\left( \tau ,\tau '' \right) \mathcal{G} _{BC}\left( \tau ,\tau '' \right) \right)}=\delta _{AB}\left( \theta -\theta ' \right) \delta \left( \tau -\tau ' \right) .
\end{equation}

In the generalization of the N=2 supersymmetric SYK model, the interaction Hamiltonian contains the correlation between chiral and anti-chiral supercharges, while the interactions between single-sided terms should vanish. Additionally, we can consider the solvable first-order interaction term
\begin{equation}
H_{int}=i\mu \partial _{\theta}\sum_j{\left( \varPsi _{L}^{j}\bar{\varPsi}_{R}^{j}-\varPsi _{R}^{j}\bar{\varPsi}_{L}^{j} \right)},
$$
$$
\bar{H}_{int}=i\mu \partial _{\bar{\theta}}\sum_j{\left( \bar{\varPsi}_{L}^{j}\varPsi _{R}^{j}-\bar{\varPsi}_{R}^{j}\varPsi _{L}^{j} \right)}.
\end{equation}

The corresponding N=2 chiral and anti-chiral actions, when combined with the interaction term, lead to the equations of motion
\begin{equation}
D_{\theta}\mathcal{G} _{AB}\left( Z,Z' \right) -\sum_C{\left( -i\mu \epsilon _{AC}\partial _{\theta}\mathcal{G} _{AB}\left( Z,Z' \right) -\int{d\tau ''d\theta ''}\varSigma _{AC}\left( Z,Z'' \right) \mathcal{G} _{BC}\left( Z,Z'' \right) \right)}=\delta _{AB}\delta \left( Z-Z' \right) ,
$$
$$
D_{\bar{\theta}}\bar{\mathcal{G}}_{AB}\left( Z,Z' \right) -\sum_C{\left( -i\mu \epsilon _{AC}\partial _{\bar{\theta}}\bar{\mathcal{G}}_{AB}\left( Z,Z' \right) -\int{d\tau ''d\bar{\theta}''}\bar{\varSigma}_{AC}\left( Z,Z'' \right) \bar{\mathcal{G}}_{BC}\left( Z,Z'' \right) \right)}=\delta _{AB}\delta \left( Z-Z' \right) .
\end{equation}

We assume that there is symmetry between the chiral and anti-chiral Green's functions
\begin{equation}
\mathcal{G} \left( \tau ,\theta ,\bar{\theta};\tau '',\theta '',\bar{\theta}'' \right) =\bar{\mathcal{G}}\left( \tau ,\theta ,\bar{\theta};\tau '',\theta '',\bar{\theta}'' \right) .
\end{equation}
We can further express the Schwinger-Dyson equations. Notably, the forms of the N=2 equations are similar to those of the N=1 equations in their components
\begin{equation}
\begin{split}
\varSigma _{\psi \psi ,AB}\left( \tau \right) &=\left( q-1 \right) \left( -1 \right) ^{\left( q-1 \right) /2}JG_{\psi \psi ,AB}^{q-2}\left( \tau \right) G_{bb,AB}\left( \tau \right) -\frac{\left( q-1 \right)}{2}\left( q-2 \right) \left( -1 \right) ^{\left( q-1 \right) /2}JG_{\psi \psi ,AB}^{q-3}\left( \tau \right) ,
\\
&G_{\psi b,AB}^{A}\left( \tau \right) G_{b\psi ,AB}^{S}\left( \tau \right) -\frac{\left( q-1 \right)}{2}\left( q-2 \right) \left( -1 \right) ^{\left( q-1 \right) /2}JG_{\psi \psi ,AB}^{q-2}\left( \tau \right) G_{\psi b,AB}^{S}\left( \tau \right) G_{b\psi ,AB}^{A}\left( \tau \right) ,
\\
\varSigma _{bb,AB}\left( \tau \right) &=JG_{\psi \psi ,AB}^{q}\left( \tau \right) ,
\\
\varSigma _{b\psi ,AB}^{S}\left( \tau \right) &=-\frac{\left( q-1 \right)}{2}JG_{\psi b,AB}^{A}\left( \tau \right) G_{\psi \psi ,AB}^{q-2}\left( \tau \right) ,
\\
\varSigma _{b\psi ,AB}^{A}\left( \tau \right) &=-\left( -1 \right) ^{\left( q-1 \right) /2}\frac{\left( q-1 \right)}{2}JG_{\psi b,AB}^{S}\left( \tau \right) G_{\psi \psi ,AB}^{q-2}\left( \tau \right) ,
\\
\varSigma _{\psi b ,AB}^{S}\left( \tau \right) &=\frac{\left( q-1 \right)}{2}JG_{b\psi,AB}^{A}\left( \tau \right) G_{\psi \psi ,AB}\left( \tau \right) ,
\\
\varSigma _{\psi b,AB}^{A}\left( \tau \right) &=\left( -1 \right) ^{\left( q-1 \right) /2}\frac{\left( q-1 \right)}{2}J G_{b\psi ,AB}^{S}\left( \tau \right) G_{\psi \psi ,AB}\left( \tau \right) .
\end{split}
\end{equation}

Factor $\left( -1 \right) ^{\left( q-1 \right) /2}$ is generated from the fermionic sign. In N=2 theory, we need to adjust the evolution time of $\psi$ variable as follows: $\tau \rightarrow \tau -\theta \bar{\theta}$ and $\tau '\rightarrow \tau '+\theta '\bar{\theta}'$.

Additionally, it is noteworthy that the components of Green's functions $G_{\psi b,AB}$ and $G_{b\psi ,AB}$ are neither symmetric nor anti-symmetric. Nevertheless, these components can be regularized through the exchange of Grassmann variables $\theta $ and $\bar{\theta} $
\begin{equation}
G_{\psi b,AB}\theta \rightarrow \theta \bar{\theta}G_{\psi b,AB},
$$
$$
\bar{\theta}G_{b\psi ,AB}\rightarrow G_{b\psi ,AB}.
\end{equation}
It is easy to check
\begin{equation}
-\theta G_{\psi b,AB}\bar{\theta}G_{b\psi ,AB}\rightarrow -\bar{\theta}\theta G_{\psi b,AB}G_{b\psi ,AB},
\end{equation}
where we have defined the component $-\theta \bar{\theta}G_{\psi b,AB}$ as anti-symmetric, whereas the component $G_{b\psi ,AB}$ is symmetric. Alternatively, we can also define the component $G_{\psi b,AB}$ as symmetric and $G_{b\psi ,AB}$ as anti-symmetric without affecting both the physical and numerical results. This is due to the symmetry between these two components.

Similarly, we can apply the same process to make the component $G_{b\psi ,AB}$ anti-symmetric and make $G_{\psi b ,AB}$ symmetric
$$
G_{b\psi,AB}\theta \rightarrow \theta \bar{\theta}G_{b\psi,AB},
$$
$$
\bar{\theta}G_{\psi b ,AB}\rightarrow G_{\psi b ,AB}.
$$
Another important observation is that these components must satisfy the periodic boundary condition due to the diagonal term, and their low-energy ansatz is precisely zero. This allows us to separate them into symmetric and anti-symmetric components and apply the Matsubara method without considering additional subdivisions. According to Eq. (20), the multiple combination of the components $\theta \bar{\theta}G_{\psi b ,AB} G_{b\psi ,AB}$should be defined as antisymmetric, similar to the combination $\theta \bar{\theta}G_{\psi \psi ,AB} G_{bb ,AB}$. We can still split the Green function into anti-symmetric parts $G_{\psi b ,AB}^{A}$ and symmetric parts $G_{\psi b ,AB}^{S}$, with a similar division for the anti-symmetric $G_{b\psi ,AB}^{A}$ and symmetric $G_{b\psi ,AB}^{S}$.

\begin{figure}[!t]
\begin{minipage}{0.48\linewidth}
\centerline{\includegraphics[width=8cm]{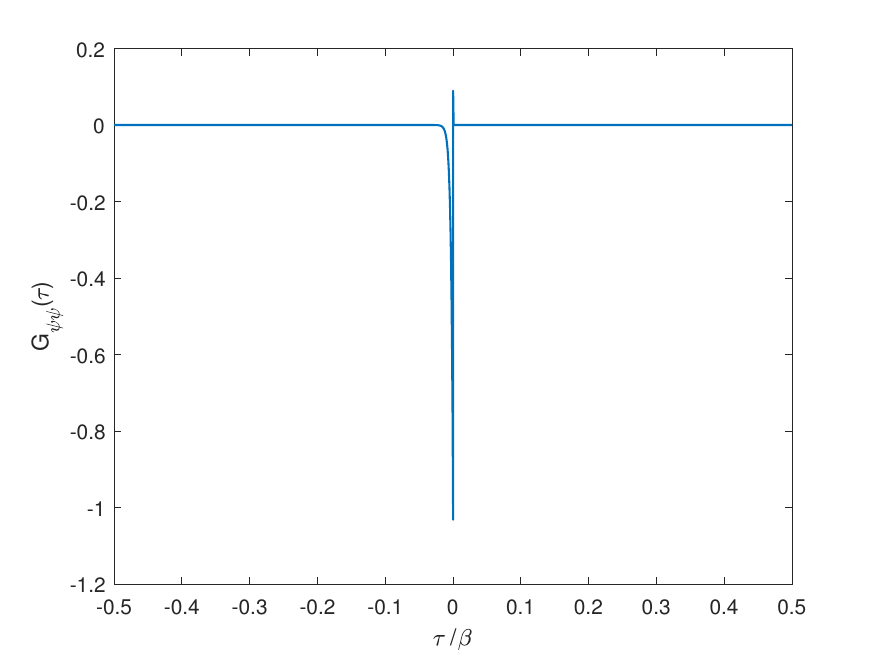}}
\centerline{(a)}
\end{minipage}
\hfill
\begin{minipage}{0.48\linewidth}
\centerline{\includegraphics[width=8cm]{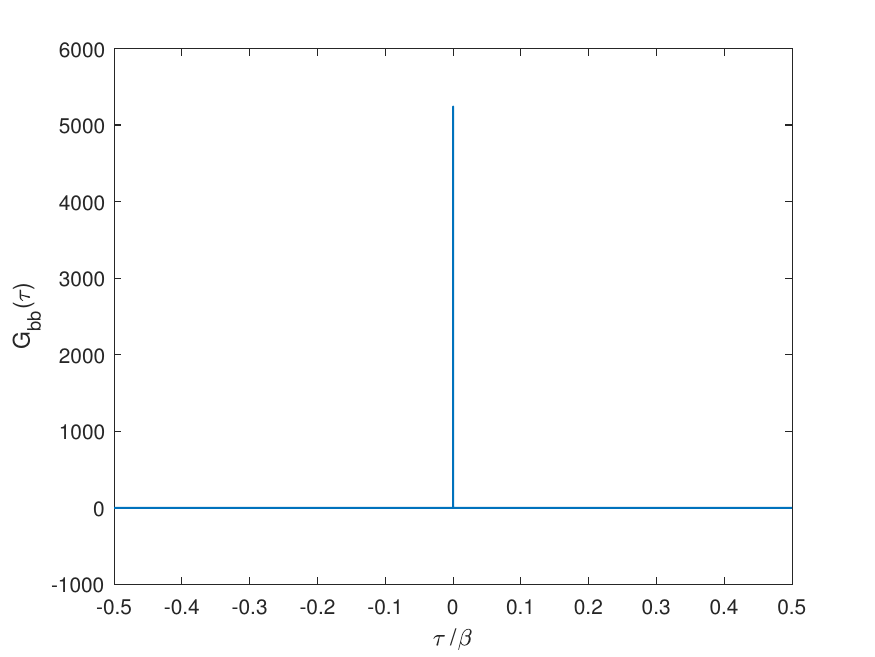}}
\centerline{(b)}
\end{minipage}
\vfill
\begin{minipage}{0.48\linewidth}
\centerline{\includegraphics[width=8cm]{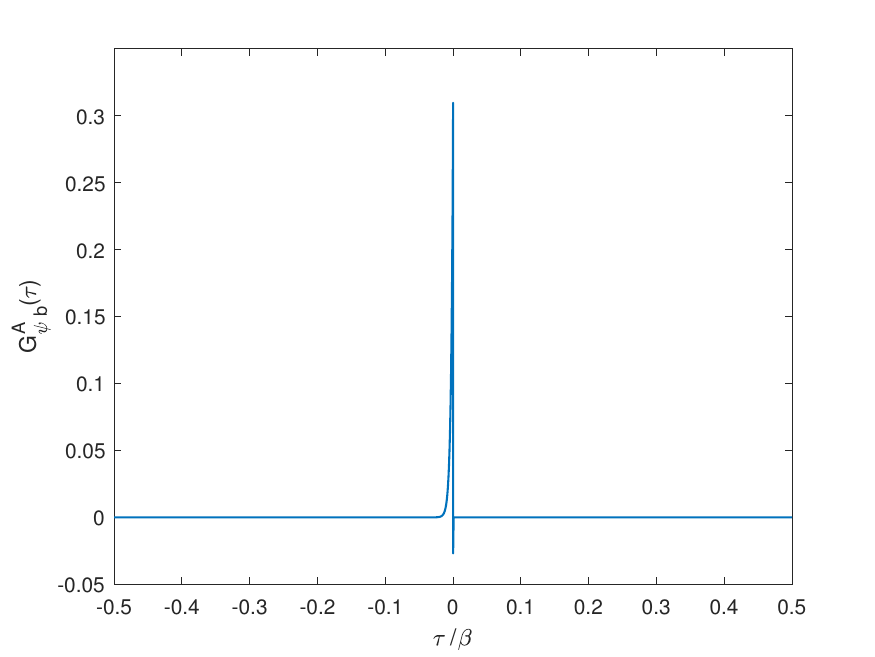}}
\centerline{(c)}
\end{minipage}
\hfill
\begin{minipage}{0.48\linewidth}
\centerline{\includegraphics[width=8cm]{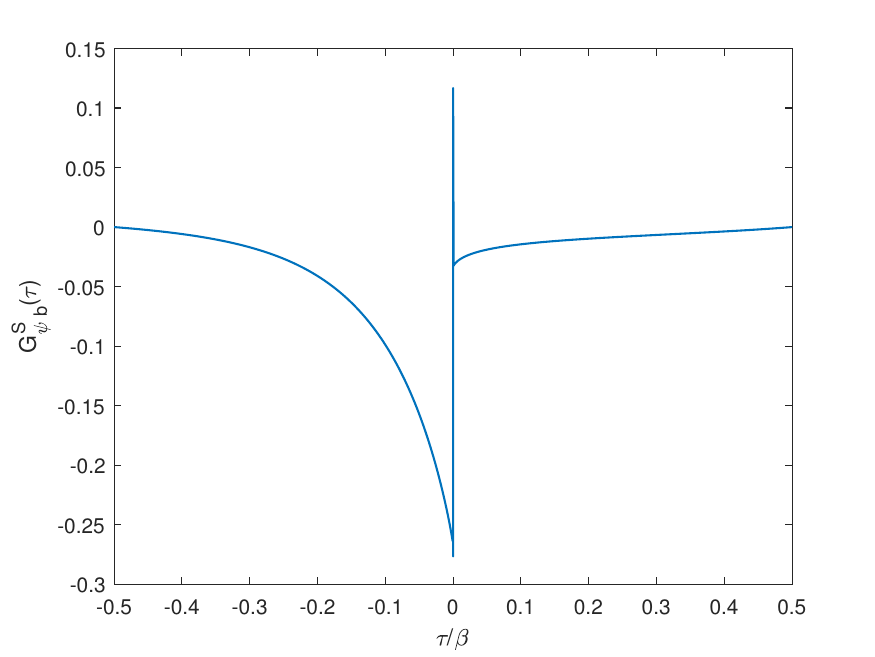}}
\centerline{(d)}
\end{minipage}
\caption{\label{fig:Figure 2} N=2 supersymmetric Green's function in components fixed $J=3,\mu=0.3,T=0.01,q=5$ (a) Single-side fermionic Green's function $G_{\psi \psi,AA}$, (b) Single-side bosonic Green's function $G_{bb,AA}$ (c) multi-side Anti-symmetric fermionic-bosonic Correlation function $G_{\psi b,LR}^{A}$. (d) multi-side symmetric fermionic-bosonic Correlation function $G_{\psi b,LR}^{S}$}
\end{figure}

The numerical results in Figure (2) show that supersymmetric Green's function in components fixed $J=3,\mu=0.1,T=0.01$. In figure 2(a), the fermionic Green's function behaves like N=0 SYK, but no longer symmetry of Euclidean time $\tau $. Its Retarded side ($\tau>0 $) decays faster than its advanced side ($\tau<0 $). This is very close to the supersymmetric SYK with fermionic chemical potential results in \cite{18}. However, this is different from cSYK\cite{43} and Maldacena-Qi\cite{26}. The model, characterized by supersymmetric correlations, implies that only non-vanishing fermionic-bosonic and bosonic-fermionic correlations exist. And it cannot undergo renormalization similar to that described in \cite{18}. In Figure 2(b), the bosonic component approximates a delta function, which is an inherent consequence of the original supersymmetric SYK model. We have also plotted both the anti-symmetric and symmetric fermionic-bosonic components in Figure 2(c), Figure 2(d). Upon including the coupling term, these off-diagonal components are no longer regarded as negligible, as suggested by previous studies. In Figure 2(c), the anti-symmetric fermionic-bosonic correlation function closely resembles the fermionic correlator $G_{\psi \psi}$. Meanwhile, the symmetric fermionic-bosonic correlation function in Figure 2(d) decreases gradually and approaches the start time of 0, influenced by the delta function present in the bosonic Green's function. One can verify that setting the components $G_{\psi \psi,LR},G_{bb,LR},G_{\psi b,AA}$ equal to zero yields a stable solution. Additionally, other Green's functions, such as the boson-fermion conjugate and the left-right conjugate, can be derived from the components discussed in Figure 2 using anti-symmetric Eq. (22).

And the equations of motion in frequency space are
\begin{equation}
G_{\psi \psi ,AB}\left( \omega \right) =\frac{Det\left( A_{\psi \psi ,AB}\left( D_{half\,\,integer} \right) \right)}{Det\left( D_{half\,\,integer} \right)},
$$
$$
G_{bb,AB}\left( \omega \right) =-\frac{Det\left( A_{bb,AB}\left( D_{integer} \right) \right)}{Det\left( D_{integer} \right)},
$$
$$
G_{\psi b,AB}^{A}\left( \omega \right) =\frac{Det\left( A_{\psi b,AB}^{A}\left( D_{half\,\,integer} \right) \right)}{Det\left( D_{half\,\,integer} \right)},
$$
$$
G_{b\psi ,AB}^{S}\left( \omega \right) =-\frac{Det\left( A_{b\psi ,AB}^{S}\left( D_{integer} \right) \right)}{Det\left( D_{integer} \right)},
$$
$$
G_{b\psi,AB}^{A}\left( \omega \right) =\frac{Det\left( A_{b\psi,AB}^{A}\left( D_{half\,\,integer} \right) \right)}{Det\left( D_{half\,\,integer} \right)},
$$
$$
G_{\psi b ,AB}^{S}\left( \omega \right) =-\frac{Det\left( A_{\psi b ,AB}^{S}\left( D_{integer} \right) \right)}{Det\left( D_{integer} \right)}.
\end{equation}
In this work, we introduce auxiliary fermions with bosonic integral frequency $\varSigma _{\psi \psi}\left( \omega _{intergral} \right) $ and bosons with fermionic half-integral frequency $\varSigma _{bb}\left( \omega _{half\,\,intergral} \right) $, which should statistically equal zero.
$$
D_{half\,\,integer}=\left( \begin{matrix}
	-iw_n-\varSigma _{LL,\psi \psi}&		-\varSigma _{LR,\psi \psi}&		-\varSigma _{LL,\psi b}^{A}&		-i\mu -\varSigma _{LR,\psi b}^{A}\\
	-\varSigma _{RL,\psi \psi}&		-iw_n-\varSigma _{RR,\psi \psi}&		i\mu -\varSigma _{RL,\psi b}^{A}&		-\varSigma _{RR,\psi b}^{A}\\
	-\varSigma _{LL,b\psi}^{A}&		i\mu -\varSigma _{LR,b\psi}^{A}&		-1-0&		-0\\
	-i\mu -\varSigma _{RL,b\psi}^{A}&		-\varSigma _{RR,b\psi}^{A}&		-0&		-1-0\\
\end{matrix} \right) ,
$$
$$
D_{integer}=\left( \begin{matrix}
	-iw_n-0&		-0&		-\varSigma _{LL,\psi b}^{S}&		i\mu -\varSigma _{LR,\psi b}^{S}\\
	-0&		-iw_n-0&		-i\mu -\varSigma _{RL,\psi b}^{S}&		-\varSigma _{RR,\psi b}^{S}\\
	-\varSigma _{LL,b\psi}^{S}&		-i\mu -\varSigma _{LR,b\psi}^{S}&		-1-\varSigma _{LL,bb}&		-\varSigma _{LR,bb}\\
	i\mu -\varSigma _{RL,b\psi}^{S}&		-\varSigma _{RR,b\psi}^{S}&		-\varSigma _{RL,bb}&		-1-\varSigma _{RR,bb}\\
\end{matrix} \right) .
$$
The matrices $D_{half\,\,integer}$ and $D_{integer}$ are introduced to represent the variational Jacobian matrix with symmetric and anti-symmetric properties respectively, and the coefficient -1 serves as a bosonic factor in this context. We can also absorb this factor and invert the $\varsigma$ as in previous works. As demonstrated in the Eq. (41), these algebraic cofactors of the matrix originate from the Jacobi method applied to Schwinger-Dyson equations. Zero elements in a matrix represent the possible values of $\varSigma$. We assume that the bosonic $\varSigma$s are zero, indicate that they also exhibit zero values at half-integer frequencies. In contrast, the fermionic $\varSigma$ also exhibit zero values at integer frequencies.

The energy can be evaluated in the thermal limit
\begin{equation}
\frac{E}{N}=\int{d\bar{\theta}}\left[ \frac{2}{q-1}D\mathcal{G} _{LL}+i\mu \left( 1-\frac{2}{q-1} \right) \mathcal{G} _{LR} \right] _{\tau \rightarrow 0^+}+h.c.
\end{equation}
We can precisely calculate these values within the frequency branches
\begin{equation}
\begin{split}
\frac{E}{N}&=\frac{2}{q-1}\left[ T\sum_{half\,\,integral}{\left( \varSigma _{LL,\psi \psi}\left( iw_n \right) G_{LL,\psi \psi}\left( iw_n \right) +\mu \left( \frac{q-3}{2} \right) \mathrm{Im}\left[ G_{LR,\psi b}^{A}\left( iw_n \right) +G_{LR,b\psi}^{A}\left( iw_n \right) \right] \right)} \right]
\\
&+\frac{2}{q-1}\left[ T\sum_{integral}{\left( \varSigma _{LL,bb}\left( iw_n \right) G_{LL,bb}\left( iw_n \right) +\mu \left( \frac{q-3}{2} \right) \mathrm{Im}\left[ G_{LR,\psi b}^{S}\left( iw_n \right) +G_{LR,b\psi}^{S}\left( iw_n \right) \right] \right)} \right].
 \end{split}
\end{equation}
Additionally, the free energy can also be derived from the saddle point effective action
\begin{equation}
\begin{split}
\frac{F}{N}&=-T\frac{\log Z}{N}=T\frac{S_{eff}}{N}=
\\
&-T\left[ 2\log \left( 2 \right) +\sum_{half\,\,integral}{\left( \log \frac{D_{half\,\,integral}\left( iw_n \right)}{\left( iw_n \right) ^2}+\frac{2q-2}{q}\varSigma _{LL,\psi \psi}\left( iw_n \right) G_{LL,\psi \psi}\left( iw_n \right) \right)} \right.
\\
&\left. +\frac{2q-2}{q}\varSigma _{LR,\psi b}^{A}\left( iw_n \right) G_{LR,\psi b}^{A}\left( iw_n \right) +\varSigma _{LR,\psi b}^{A}\left( iw_n \right) G_{LR,b\psi}^{A}\left( iw_n \right) \right] -\sum_{integral}{\left( \log \frac{D_{integral}\left( iw_n \right)}{\left( iw_n \right) ^2} \right.}
\\
&\left. -\frac{2q-2}{q}\varSigma _{LL,bb}\left( iw_n \right) G_{LL,bb}\left( iw_n \right) -\frac{2q-2}{q}\varSigma _{LR,\psi b}^{S}\left( iw_n \right) G_{LR,\psi b}^{S}\left( iw_n \right) -\varSigma _{LR,\psi b}^{S}\left( iw_n \right) G_{LR,b\psi}^{S}\left( iw_n \right) \right) .
\end{split}
\end{equation}

In this section, we have renormalized the determinant terms
$$
\sum_{half\,\,integral}{\left( \log \left( iw_n \right) \right)}=\ln 2.
$$

We have introduced $iw_n$ to eliminate the physical influence of auxiliary fermions on $D_{integer}$. Since there are no natural fermions with integral frequencies, the integral part of $\left( \log \left( iw_n \right) \right) $ does not contribute any additional $\ln 2$.
\begin{figure}[!t]
\begin{minipage}{0.48\linewidth}
\centerline{\includegraphics[width=8cm]{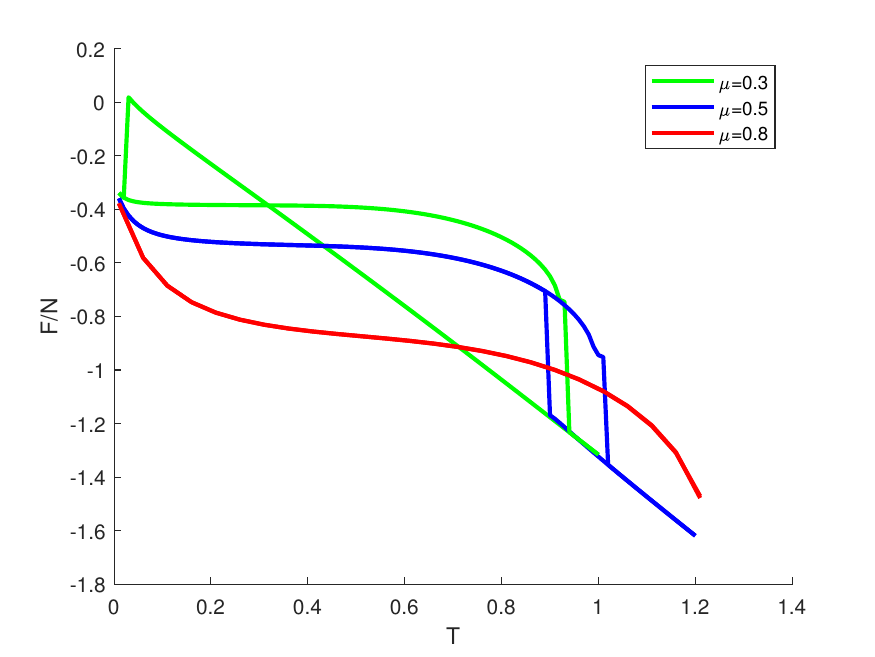}}
\centerline{(a)}
\end{minipage}
\hfill
\begin{minipage}{0.48\linewidth}
\centerline{\includegraphics[width=8cm]{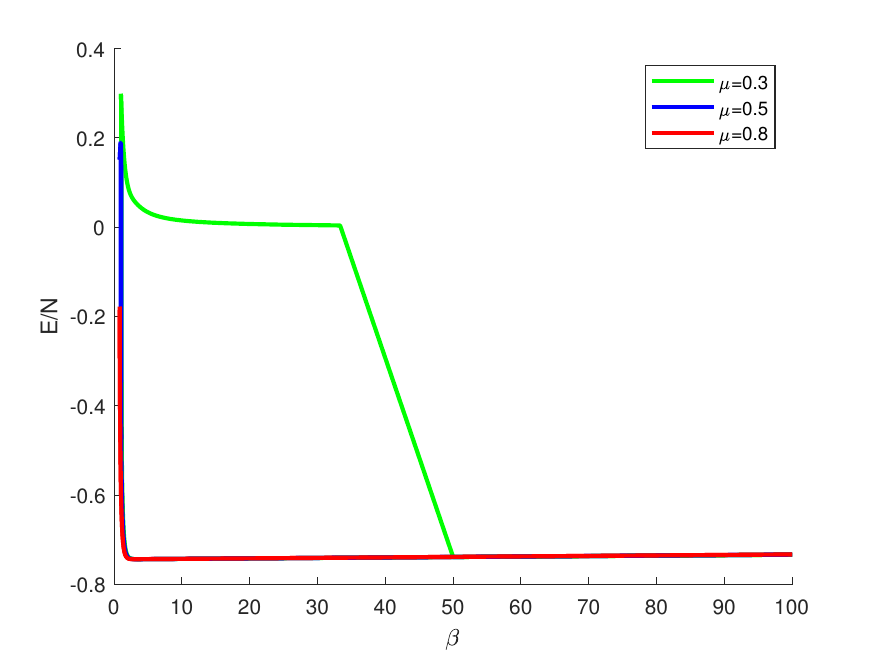}}
\centerline{(b)}
\end{minipage}

\caption{\label{fig:Figure 3} Free energy and energy of N=2 Supersymmetric SYK fixed $J=3,q=5$  (a)Free energy under condition $\mu=0.3,0.5,0.8$, (b) Energy under condition $\mu=0.3,0.5,0.8$, }
\end{figure}

The free energy as a function of temperature is plotted in Figure 3(a). We first cool the system from its highest temperature and then heat it back to its original temperature. We consider the number of couplings $q=5$ in Eq. (1), which means the theory should correspond to a non-supersymmetric q=4 coupled SYK\cite{48,51}. We observe a phase transition in Figure 3(a). The observed free energy is analogous to the Hawking-Page transition in standard MQ and cSYK models. We consider the low-temperature phase as wormholes, which would evolve into two gapless systems as the temperature increases. We will further discuss this interpretation in the next section.  However, after introducing the supersymmetric coupling, the bosonic and multistate contributions to the free energy term gradually transform the 'wormhole-phase' into a state different from the eternal wormhole in previous works. Furthermore, the phase transition is also influenced by this effect (as shown in Figure 3(a)). The energy as a function of inverse temperature $\beta$ is plotted in Figure 3(b). Similarly, we can observe a comparable ensemble result regarding the energy behaviors. For $T<T_c$, there are two distinct solutions, and their saddle points constitute the phase transition. We can also obtain a smooth curve outside the transition region.

The second-order transition ends at a critical point ($T_{c}=1.15,\mu_{c}=0.7$). We observe that the phase transition emerges upon introducing the super-coupling. The wormhole phase at low energy will also be influenced by additional bosonic contributions, and the free energy at low temperatures is no longer constant as we increase the coupling. 

For the N=4 model, we can include the first-order interaction term in the Hamiltonian, while maintaining supersymmetry and ensuring that the model remains solvable
\begin{equation}
H_{int}=\partial _{\theta}^{2}i\mu \left( \varPhi _{L}^{i}\bar{\varPhi}_{iR}+\varPhi _{R}^{i}\bar{\varPhi}_{iL} \right) =i\mu F_{L}^{i}\bar{\phi}_{iR}+i\mu F_{R}^{i}\bar{\phi}_{iL},
$$
$$
\bar{H}_{int}=\partial _{\bar{\theta}}^{2}i\mu \left( \bar{\varPhi}_{iL}\varPhi _{R}^{i}+\bar{\varPhi}_{iR}\varPhi _{L}^{i} \right) =i\mu \bar{\phi}_{iL}F_{R}^{i}+i\mu \bar{\phi}_{iR}F_{L}^{i}.
\end{equation}
\begin{equation}
\begin{split}
\mathcal{G} ^3&=\left( \varPhi \varPhi \right) ^3=\left( FF \right) \left( \phi \phi \right) \left( \phi \phi \right) -\left( \phi \phi \right) \left( \psi _{\alpha}\psi _{\alpha} \right) \left( \psi _{\beta}\psi _{\beta} \right) +\left( \phi \phi \right) \left( \psi _{\alpha}\psi _{\beta} \right) \left( \psi _{\beta}\psi _{\alpha} \right) +2\left( \phi \phi \right) \left( F\phi \right) \left( \phi F \right)
\\
&-\left( \psi _{\alpha}\phi \right) \left( \phi \psi _{\alpha} \right) \left( \psi _{\beta}\psi _{\beta} \right) -\left( \psi _{\beta}\phi \right) \left( \phi \psi _{\beta} \right) \left( \psi _{\alpha}\psi _{\alpha} \right) +\left( \psi _{\beta}\phi \right) \left( \phi \psi _{\alpha} \right) \left( \psi _{\alpha}\psi _{\beta} \right) +\left( \psi _{\alpha}\phi \right) \left( \phi \psi _{\beta} \right) \left( \psi _{\beta}\psi _{\alpha} \right) .
\end{split}
\end{equation}

Then we can write down the N=4 action, including the specific interaction term between the left and right fields
\begin{equation}
\begin{split}
\frac{S_{eff}}{N}&=-\frac{1}{2}\log Pf\left( D_{half\,\,integer} \right) +\log\det \left( D_{integer} \right) +\sum_{A,B=L,R}{\frac{1}{2}}\int{d\tau d\tau \prime}\left[ \varSigma _{\psi 1\psi 1,AB}\left( \tau ,\tau \prime \right) G_{\psi 1\psi 1,AB}\left( \tau ,\tau \prime \right) \right.
\\
& +\varSigma _{\psi 2\psi 2,AB}\left( \tau ,\tau \prime \right) G_{\psi 2\psi 2,AB}\left( \tau ,\tau \prime \right) +\varSigma _{FF,AB}\left( \tau ,\tau \prime \right) G_{FF,AB}\left( \tau ,\tau \prime \right) +\varSigma _{\phi \phi ,AB}\left( \tau ,\tau \prime \right) G_{\phi \phi ,AB}\left( \tau ,\tau \prime \right)
\\
& +\varSigma _{F\phi ,AB}\left( \tau ,\tau \prime \right) G_{F\phi ,AB}\left( \tau ,\tau \prime \right) +\varSigma _{\phi F,AB}\left( \tau ,\tau \prime \right) G_{\phi F,AB}\left( \tau ,\tau \prime \right) -J\left( \left( q-1 \right) G_{\phi \phi ,AB}^{q-2}\left( \tau ,\tau \prime \right) G_{\psi 1\psi 1,AB}\left( \tau ,\tau \prime \right) \right.
\\
&\left. \left. G_{\psi 2\psi 2,AB}\left( \tau ,\tau \prime \right) -G_{\phi \phi ,AB}^{q-1}\left( \tau ,\tau \prime \right) G_{FF,AB}\left( \tau ,\tau \prime \right) -\left( q-1 \right) G_{\phi \phi ,AB}^{q-2}\left( \tau ,\tau \prime \right) G_{F\phi ,AB}\left( \tau ,\tau \prime \right) G_{\phi F,AB}\left( \tau ,\tau \prime \right) \right) \right].
\end{split}
\end{equation}

and
\begin{equation}
D_{\theta}^{2}\mathcal{G} _{AB}\left( Z,Z' \right) -\sum_C{\left( -i\mu \epsilon _{AC}\partial _{\theta}^{2}\mathcal{G} _{AB}\left( Z,Z' \right) -\int{d\tau ''d\theta _{1}^{''}d\theta _{2}^{''}}\varSigma _{AC}\left( Z,Z'' \right) \mathcal{G} _{BC}\left( Z,Z'' \right) \right)}=\delta _{AB}\delta \left( Z-Z' \right) ,
$$
$$
D_{\bar{\theta}}^{2}\bar{\mathcal{G}}_{AB}\left( Z,Z' \right) -\sum_C{\left( -i\mu \epsilon _{AC}\partial _{\bar{\theta}}^{2}\bar{\mathcal{G}}_{AB}\left( Z,Z' \right) -\int{d\tau ''d\bar{\theta}_{1}^{''}\bar{\theta}_{2}^{''}}\bar{\varSigma}_{AC}\left( Z,Z'' \right) \bar{\mathcal{G}}_{BC}\left( Z,Z'' \right) \right)}=\delta _{AB}\delta \left( Z-Z' \right) .
\end{equation}

For each component, we have determined the solution through variations in the associated correlators
\begin{equation}
\begin{split}
\varSigma _{\psi 1\psi 1,AB}\left( \tau \right) &=J\left( q-1 \right) G_{\phi \phi ,AB}^{q-2}\left( \tau ,\tau ' \right) G_{\psi 2\psi 2,AB}\left( \tau ,\tau ' \right) ,
\\
\varSigma _{\psi 2\psi 2,AB}\left( \tau \right) &=J\left( q-1 \right) G_{\phi \phi ,AB}^{q-2}\left( \tau ,\tau ' \right) G_{\psi 1\psi 1,AB}\left( \tau ,\tau ' \right) ,
\\
\varSigma _{\phi \phi ,AB}\left( \tau \right) &=-\left( q-1 \right) \left( q-2 \right) JG_{\phi \phi ,AB}^{q-3}\left( \tau ,\tau ' \right) G_{\psi 1\psi 1,AB}\left( \tau ,\tau ' \right) G_{\psi 2\psi 2,AB}\left( \tau ,\tau ' \right) +\left( q-1 \right) JG_{\phi \phi ,AB}^{q-2}\left( \tau ,\tau ' \right)
\\
& G_{FF,AB}\left( \tau ,\tau ' \right) +\left( q-1 \right) \left( q-2 \right) JG_{\phi \phi ,AB}^{q-3}\left( \tau ,\tau ' \right) G_{F\phi ,AB}\left( \tau ,\tau ' \right) G_{\phi F,AB}\left( \tau ,\tau ' \right) ,
\\
\varSigma _{FF,AB}\left( \tau \right) &=-JG_{\phi \phi ,AB}^{q-1}\left( \tau ,\tau ' \right) ,
\\
\varSigma _{\phi F,AB}\left( \tau \right) &=-J\left( q-1 \right) G_{\phi \phi ,AB}^{q-2}\left( \tau ,\tau ' \right) G_{F\phi ,AB}\left( \tau ,\tau ' \right) ,
\\
\varSigma _{F\phi ,AB}\left( \tau \right) &=-J\left( q-1 \right) G_{\phi \phi ,AB}^{q-2}\left( \tau ,\tau ' \right) G_{\phi F,AB}\left( \tau ,\tau ' \right) .
\end{split}
\end{equation}
Similarly, we can express the saddle-point Green's functions in frequency space
\begin{equation}
G_{\psi 1\psi 1,AB}\left( \omega \right) =\frac{Det\left( A_{\psi 1\psi 1,AB}\left( D_{half\,\,integer} \right) \right)}{Det\left( D_{half\,\,integer} \right)},
$$
$$
G_{\psi 2\psi 2,AB}\left( \omega \right) =\frac{Det\left( A_{\psi 2\psi 2,AB}\left( D_{half\,\,integer} \right) \right)}{Det\left( D_{half\,\,integer} \right)},
$$
$$
G_{\phi \phi ,AB}\left( \omega \right) =-\frac{Det\left( A_{\phi \phi ,AB}\left( D_{integer} \right) \right)}{Det\left( D_{integer} \right)},
$$
$$
G_{FF,AB}\left( \omega \right) =-\frac{Det\left( A_{FF,AB}\left( D_{integer} \right) \right)}{Det\left( D_{integer} \right)},
$$
$$
G_{\phi F,AB}\left( \omega \right) =-\frac{Det\left( A_{\phi F,AB}\left( D_{integer} \right) \right)}{Det\left( D_{integer} \right)},
$$
$$
G_{F\phi ,AB}\left( \omega \right) =-\frac{Det\left( A_{F\phi ,AB}\left( D_{integer} \right) \right)}{Det\left( D_{integer} \right)}.
\end{equation}
The corresponding Jacobi matrix is

$$
D_{integer}=\left( \begin{matrix}
	\omega _{n}^{2}-\varSigma _{LL,\phi \phi}&		-\varSigma _{LR,\phi \phi}&		-\varSigma _{LL,\phi F}&		i\mu -\varSigma _{LR,\phi F}\\
	-\varSigma _{RL,\phi \phi}&		\omega _{n}^{2}-\varSigma _{RR,\phi \phi}&		i\mu -\varSigma _{RL,\phi F}&		-\varSigma _{RR,\psi F}\\
	-\varSigma _{LL,F\phi}&		i\mu -\varSigma _{LR,F\phi}&		1-\varSigma _{LL,FF}&		-\varSigma _{LR,FF}\\
	i\mu -\varSigma _{RL,F\phi}&		-\varSigma _{RR,F\phi}&		-\varSigma _{RL,FF}&		1-\varSigma _{RR,FF}\\
\end{matrix} \right) ,
$$
$$
D_{half\,\,integer}=\left( \begin{matrix}
	-iw_n-\varSigma _{LL,\psi 1\psi 1}&		-\varSigma _{LL,\psi 1\psi 2}&		-\varSigma _{LR,\psi 1\psi 1}&		-\varSigma _{LR,\psi 1\psi 2}\\
	-\varSigma _{LL,\psi 2\psi 1}&		-iw_n-\varSigma _{LL,\psi 2\psi 2}&		-\varSigma _{LR,\psi 2\psi 1}&		-\varSigma _{LR,\psi 2\psi 2}\\
	-\varSigma _{RL,\psi 1\psi 1}&		-\varSigma _{RL,\psi 1\psi 2}&		-iw_n-\varSigma _{RR,\psi 1\psi 1}&		-\varSigma _{RR,\psi 1\psi 2}\\
	-\varSigma _{RL,\psi 2\psi 1}&		-\varSigma _{RL,\psi 2\psi 2}&		-\varSigma _{RR,\psi 2\psi 1}&		-iw_n-\varSigma _{RR,\psi 2\psi 2}\\
\end{matrix} \right) .
$$

\begin{figure}[!t]
\begin{minipage}{0.48\linewidth}
\centerline{\includegraphics[width=8cm]{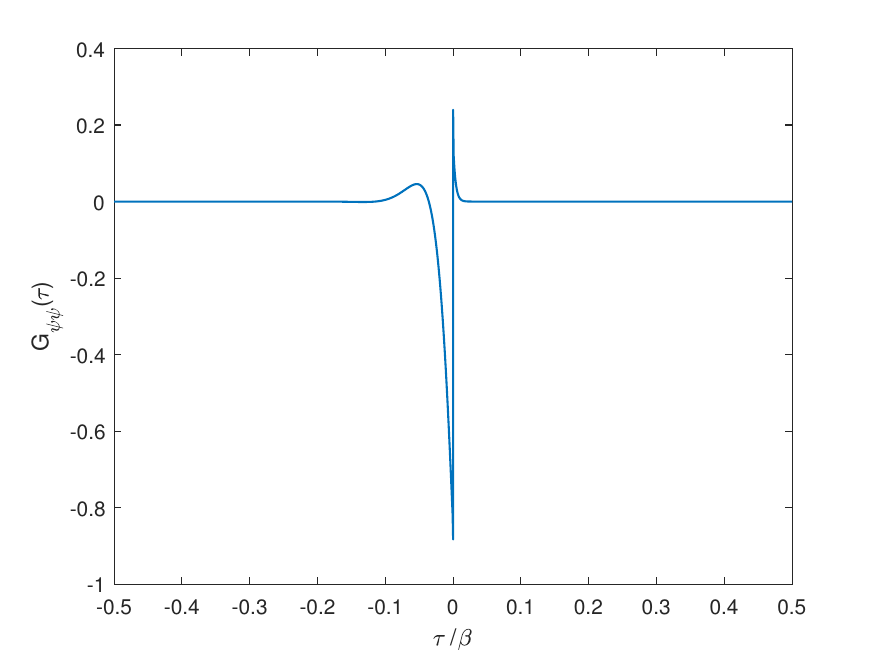}}
\centerline{(a)}
\end{minipage}
\hfill
\begin{minipage}{0.48\linewidth}
\centerline{\includegraphics[width=8cm]{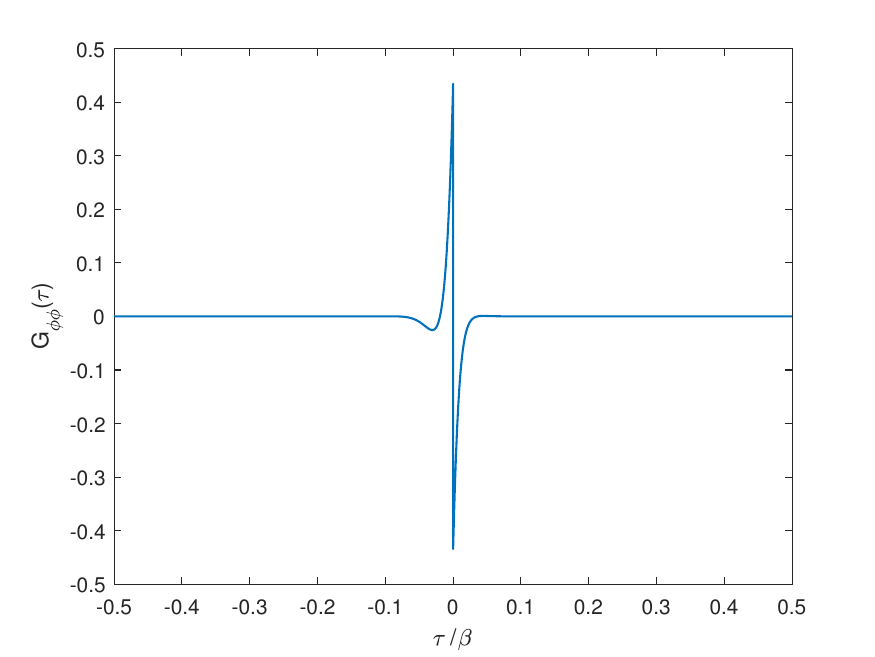}}
\centerline{(b)}
\end{minipage}
\vfill
\begin{minipage}{0.48\linewidth}
\centerline{\includegraphics[width=8cm]{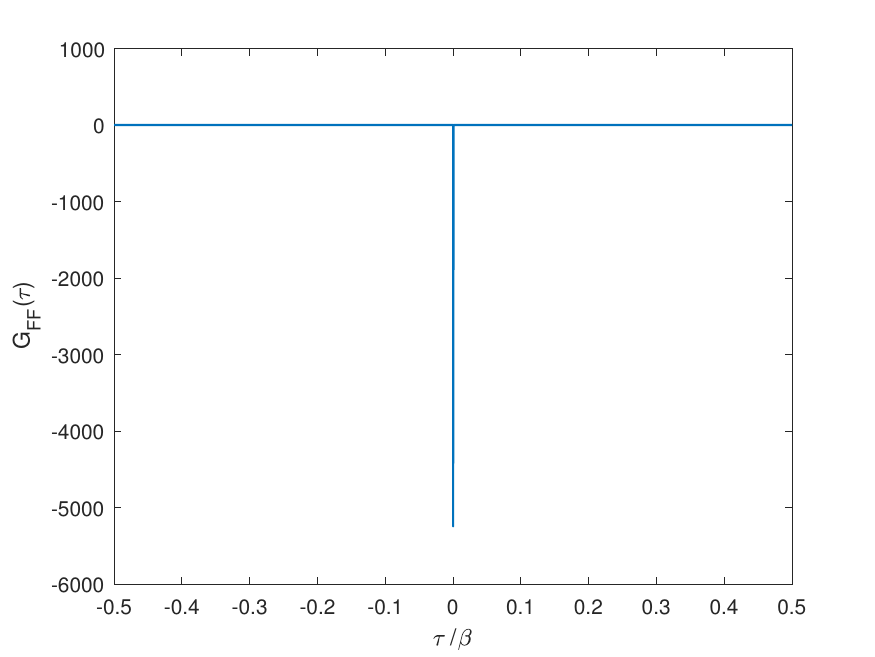}}
\centerline{(c)}
\end{minipage}
\hfill
\begin{minipage}{0.48\linewidth}
\centerline{\includegraphics[width=8cm]{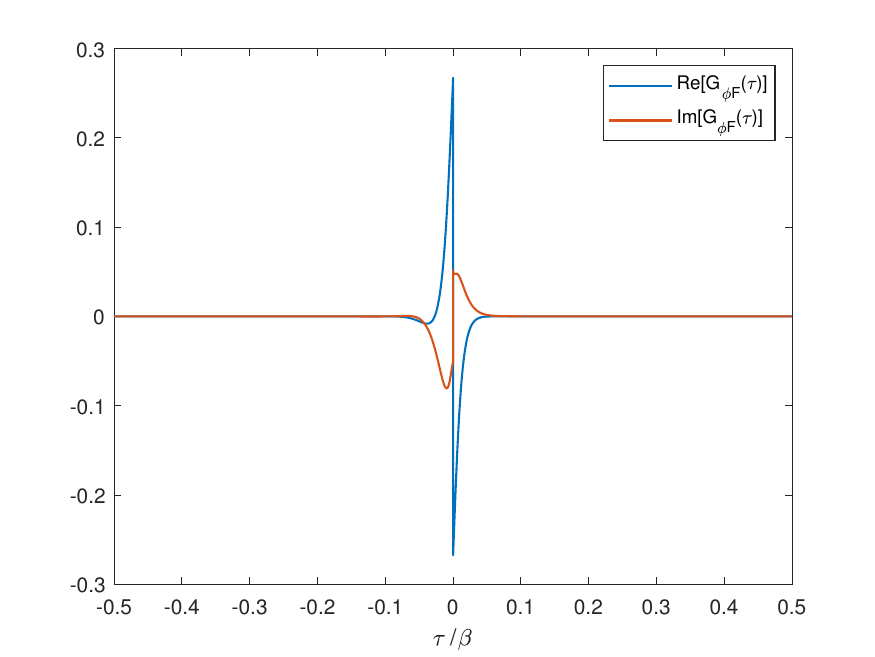}}
\centerline{(d)}
\end{minipage}
\caption{\label{fig:Figure 4}  Supersymmetric Green's function in components fixed $J=3,q=5$ (a) Single-side fermionic Green's function $G_{\psi \psi,AA}$ fixed $\mu=0.4,T=0.01$, (b) Single-side bosonic Green's function $G_{\phi \phi,AA}$ fixed $\mu=0.4i,T=0.01$ (c) Single-side auxiliary bosonic Green's function $G_{FF,AA}$ fixed $\mu=0.4i,T=0.01$ (d) multi-side Correlator between different bosons in real $Re[G_{\phi F,LR}]$ and imaginary $Im[G_{\phi F,LR}]$}
\end{figure}

In Figure 4, we have plotted the Green's functions for the N=4 SYK model. In order to compare with Eq. (34), we have chosen $\mu$ to be imaginary. In Figure 4(a), we consider the fermionic Green's function, which is very similar to the Green's function of the N=2 SYK model \cite{15} and it includes a fermionic chemical potential \cite{18}. Additionally, we consider the bosonic Green's function, which is related to the supersymmetric sector of the theory. In Figure 4(c), the Green's function of auxiliary bosons approaches the delta function, exhibiting behavior similar to that of bosonic N=2 Green's functions in Figure 2(b). Furthermore, we can generate it from the function in Eq. (34). Unlike the N=1 and N=2 models, the N=4 SYK model also contains original bosons that have no direct counterparts in lower-dimensional Grassmann theories. In Figure 4(b) these bosons $\phi$ decay over time. Additionally, we numerically obtain the multiside correlators, which are constrained by supersymmetry within the context of $G_{\phi F,LR}$ and its conjugates. In Figure 4(d), the multi-sided correlators have both real and imaginary components, which influence the phase constant of the transformations. The real component in Figure 4(c) behaves like the bosonic part in Figure 4(b), while the imaginary component is also influenced by the fermionic Green's function in Figure 4(a). One can also make the super-coupling constant $i\mu$ to be real, thus the real and imaginary parts would transform to each other under certain conditions. Additionally, the components $G_{\phi \phi,LR}$,$G_{FF,LR}$,$G_{\phi F,AA}$,$G_{F \phi,AA}$,$G_{\psi\alpha \psi\beta,AB}$,$G_{\psi\alpha \phi,AB}$,$G_{\phi,AB \psi\alpha}$ that equal to zero are stable solutions, while $G_{\psi\alpha F,AB}$,$G_{F \psi\alpha,AB}$ should be integrated out in the analysis.

We can further analyze this result from two N=2 theories with R symmetry. In N=2 formalism, the component $G_{FF,AA}$ enables only two N=2 $G_{bb,AA}$, and the Green's function approaches a delta function, as the $G_{bb,AA}$ in N=2 theory does. While the $G_{\phi \phi,AA}$ enables only two $G_{\psi \psi,AA}$ components, the combined parts also make it symmetric with dependence on $\tau$. However, the N=4 $G_{\psi \psi,AA}$ component with R symmetry is weakly restricted, containing the 0-order N=2 $G_{\psi \psi,AA}$,$G_{bb,AA}$, and all the higher-order corrections which satisfy supersymmetry. We can also derive the multiside components with real and imaginary components due to the factor $i\mu$. The real part mainly depends on the trivial correlators between both two N=2 correlators, while the imaginary part contains the extra correlations.

The energy can be calculated as
\begin{equation}
\begin{split}
\frac{E}{N}&=\int{d\bar{\theta}_1d\bar{\theta}_2}\left[ \frac{2}{q-1}D^2\mathcal{G} _{LL}+i\mu \left( 1-\frac{2}{q-1} \right) \mathcal{G} _{LR} \right] _{\tau \rightarrow 0^+}+h.c
\\
&=\frac{2}{q-1}\left[ T\sum_{half\,\,integral}{\left( \varSigma _{LL,\psi \psi}\left( iw_n \right) G_{LL,\psi \psi}\left( iw_n \right) \right)} \right]
\\
&+\frac{2}{q-1}\left[ T\sum_{integral}{\left( \varSigma _{LL,\phi \phi}\left( iw_n \right) G_{LL,\phi \phi}\left( iw_n \right) +\varSigma _{LL,FF}\left( iw_n \right) G_{LL,FF}\left( iw_n \right) \right.} \right.
\\
&\left. \left. +\mu Re\left( \frac{q-3}{2} \right) \left( G_{LL,\phi F}\left( iw_n \right) +G_{LL,F\phi}\left( iw_n \right) \right)\right) \right] +h.c.
\end{split}
\end{equation}
Here we use real or imaginary component of the correlation function, depending on how we define the interaction parameter $\mu$ and make the energy real.
We obtain the free energy from saddle point solutions
\begin{equation}
\begin{split}
\frac{F}{N}&=-T\frac{\log Z}{N}=T\frac{S_{eff}}{N}=
\\
&-T\left[ 2\log \left( 2 \right) +\sum_{half\,\,integral}{\left( \log \frac{D_{half\,\,integral}\left( iw_n \right)}{\left( iw_n \right) ^2}+\frac{4q-4}{q}\varSigma _{LL,\psi \psi}\left( iw_n \right) G_{LL,\psi \psi}\left( iw_n \right) \right)} \right.
\\
&-\sum_{integral}{\left( \log \frac{D_{integral}\left( iw_n \right)}{\left( iw_n \right) ^2}-\frac{2q-2}{q}\varSigma _{LL,\phi \phi}\left( iw_n \right) G_{LL,\phi \phi}\left( iw_n \right) -\frac{2q-2}{q}\varSigma _{LL,FF}\left( iw_n \right. G_{LL,FF}\left( iw_n \right) \right)}
\\
&\left.\left. -\frac{2q-2}{q}\varSigma _{LL,\phi F}\left( iw_n \right) G_{LL,\phi F}\left( iw_n \right) -\varSigma _{LL,F\phi}\left( iw_n \right) G_{LL,F\phi}\left( iw_n \right) \right) \right].
\end{split}
\end{equation}

In summary, we consider the cross term in N=4 theory, which is characterized by the component $G_{\phi\phi}$, the N=4 theory turns to bosonic model. We have studied the free energy and energy as a function of temperature, and there still exists two identical phases. However, the wormhole-black hole picture is altered due to the contribution of $G_{\phi\phi}$.  Because it is mainly a bosonic model, the free energy returns to zero when the temperature is high. Thus, the thermal structure is no longer a valid solution via the variation of inverse temperature $\beta$. We will provide further discussion in the context of the low-energy effective action.

\section{$NAdS_2$ sugra and expanded supersymmetic JT gravity}
\subsection{Low energy effective action}
The holographic picture of N=0 is relevant to the emergent conformal symmetry in the low-energy limit\cite{5,6}. In previous works\cite{12}, it has been shown that the N=1 supersymmetric algebra contains an extra supersymmetric reparametrization. We can define sets of reparametrization transformations using more general coordinates
\begin{equation}
\tau \rightarrow \tau '\left( \tau ,\theta \right) ,
$$
$$
\theta \rightarrow \theta'\left( \tau ,\theta \right) .
\end{equation}
Under a more general reparameterization, the correlation function with certain dimensions should be invariant under coordinate transformation rescaling
\begin{equation}
\mathcal{G} \left( \tau _1,\theta _1;\tau _2,\theta _2 \right) =Ber\left( \tau _{1}^{'},\theta _{1}^{'};\tau _1,\theta _1 \right) ^{\frac{1}{q}}Ber\left( \tau _{2}^{'},\theta _{2}^{'};\tau _2,\theta _2 \right) ^{\frac{1}{q}}\mathcal{G} \left( \tau _{1}^{'},\theta _{1}^{'};\tau _{2}^{'},\theta _{2}^{'} \right) ,
\end{equation}
where the derivative terms in superspace are with Berezinian
\begin{equation}
Ber\left( \tau ',\theta';\tau ,\theta \right) \equiv Ber\left( \begin{matrix}
	\partial _{\tau}\tau '&		\partial _{\tau}\theta '\\
	\partial _{\theta}\tau '&		\partial _{\theta}\theta '\\
\end{matrix} \right) .
\end{equation}
In low-energy limit, the corresponding Berezinian can be simplified to a super Jacobian derivative factor, and this simplification gives rise to a nearly super-conformal symmetry, with conformal dimension 1/q
\begin{equation}
Ber\left( \tau ',\theta ';\tau ,\theta \right) =D_{\theta}\theta '.
\end{equation}
The correlation functions exhibit a super conformal symmetry in the IR limit which is similar to single-side SSYK, and this symmetry is accompanied by the preservation of covariant derivative transformations
\begin{equation}
\mathcal{G} \left( \tau _1,\theta _1;\tau _2,\theta _2 \right) =\left( D_{\theta _1}\theta _{1}^{'} \right) ^{\frac{1}{q}}\left( D_{\theta _2}\theta _{2}^{'} \right) ^{\frac{1}{q}}\mathcal{G} \left( \tau _{1}^{'},\theta _{1}^{'};\tau _{2}^{'},\theta _{2}^{'} \right) .
\end{equation}
which is invariant under the super transformation
\begin{equation}
\tau'=\tau +\epsilon +\theta \eta ,
$$
$$
\theta'=\theta +\eta .
\end{equation}
Here $\epsilon $ represents an arbitrary bosonic translation, and $\eta$ is a Grassmann variable. Additionally, we perform a special reparameterization of the SDiff N=1 supersymmetry
\begin{equation}
\tau \rightarrow \tau '=f\left( \tau \right) ,
$$
$$
\theta \rightarrow \theta '=\sqrt{\partial _{\tau}f\left( \tau \right)}\theta .
\end{equation}
We briefly review the discussion in \cite{9}. Furthermore, a convenient generator of transformations with super-conformal symmetry is a reparameterization of inversion
\begin{equation}
\tau \rightarrow \tau '=-\frac{1}{\tau},
$$
$$
\theta \rightarrow \theta '=\frac{\theta}{\tau}.
\end{equation}

In the IR limit, superconformal symmetry in supersymmetric SYK allows us to construct a correlation operator between two decoupled models through super-conformal reparameterizations $h$ in single-side SSYK. The correlation function in the low-energy limit admits solutions with ansatz
\begin{equation}
\mathcal{G} \left( \tau _1,\theta _1;\tau _2,\theta _2 \right) =\frac{b}{\left| \tau _1-\tau _2-\theta _1\theta _2 \right|^{2\varDelta}},
\end{equation}
where the conformal dimension is restricted by $\varDelta=1/q$. Furthermore, we can incorporate the coupling action using hyperbolic reparameterizations
\begin{equation}
S_{int}=\frac{\mu}{2}\int{d\tau d\theta \left[ \frac{bD_{\theta}\left( \theta ' \right) _LD_{\theta}\left( \theta ' \right) _R}{\cosh ^2\frac{h_L\left( \tau +\theta \eta \left( \tau \right) \right) -h_R\left( \tau +\theta \eta \left( \tau \right) \right) -\left( \theta ' \right) _L\left( \theta ' \right) _R}{2}} \right]}^{\frac{1}{q}}.
\end{equation}
Thanks to the mathematical relations in the superspace and the terms we have included, the interaction terms in low energy preserve super-conformal symmetry and behave similarly to an N=0 theory, except that correlations and derivatives have been replaced by their supercovariant counterparts. Moreover, the correlation form reverts to a non-supersymmetric case when the superspace prolongation vanishes.

We can introduce an infinitesimal N=2 reparameterization involving two conjugate Grassmann variables $\bar{\eta}$,$\eta$, as well as one bosonic variable $h$
\begin{equation}
\tau '\rightarrow h\left( \tau +\theta \bar{\eta}\left( \tau \right) +\bar{\theta}\eta \left( \tau \right) \right) ,
$$
$$
\theta '\rightarrow \exp \left( ia\left( \tau \right) \right) \sqrt{\partial _{\tau}h\left( \tau \right)}\left[ \theta +\eta \left( \tau \right) \left( \tau +\theta \bar{\theta} \right) \right] ,
$$
$$
\bar{\theta}'\rightarrow \exp \left( -ia\left( \tau \right) \right) \sqrt{\partial _{\tau}f\left( \tau \right)}\left[ \bar{\theta}+\bar{\eta}\left( \tau \right) \left( \tau -\theta \bar{\theta} \right)  \right] .
\end{equation}
This relationship leads to an N=2 super-Schwarzian action with the low-energy symmetry breaking.
\begin{equation}
S_A=-N\alpha _S\int{d\tau d\theta d\bar{\theta}\,\,S\left[ \tau ',\theta ',\bar{\theta}';\tau ,\theta ,\bar{\theta} \right]}.
\end{equation}
$\alpha _S$ is also a constant that arises from four-point and higher-order modifications. The low-energy N=2 Schwarzian action in the thermal bosonic sector should exhibit an additional U(1) reparameterization symmetry, denoted by
\begin{equation}
S_b=N\alpha _S\int{d\tau \left( -Sch\left( \tanh \frac{h_A}{2},\tau \right) +\left( \partial _{\tau}a \right) ^2 \right)}.
\end{equation}

Compared to the N=2 IR superconformal limit, low-energy interaction terms can be classified into chiral and antichiral components, and we can similarly reparameterize the function using a thermal form akin to that of the N=1 case.
\begin{equation}
\mathcal{G} \left( \tau _1,\theta _1,\bar{\theta}_1;\tau _2,\theta _2,\bar{\theta}_2 \right) =\frac{b}{\left| \tau _1-\tau _2-\theta _1\bar{\theta}_1-\theta _2\bar{\theta}_2+2\theta _1\bar{\theta}_2 \right|^{2\varDelta}},
$$
$$
\bar{\mathcal{G}}\left( \tau _1,\theta _1,\bar{\theta}_1;\tau _2,\theta _2,\bar{\theta}_2 \right) =\frac{b}{\left| \tau _1-\tau _2+\theta _1\bar{\theta}_1+\theta _2\bar{\theta}_2-2\theta _1\bar{\theta}_2 \right|^{2\varDelta}},
$$
$$
S_{int}=\mu \int{d\tau d\theta d\bar{\theta}\left[ \frac{bD_{\bar{\theta}}\left( \bar{\theta}' \right) _LD_{\theta}\left( \theta ' \right) _R}{\cosh ^2\frac{h_L\left( \tau ,\theta ',\bar{\theta}' \right) -h_R\left( \tau ,\theta ',\bar{\theta}' \right) -\left( \theta ' \right) _L\left( \bar{\theta}' \right) _L-\left( \theta ' \right) _R\left( \bar{\theta}' \right) _R+2\left( \theta ' \right) _L\left( \bar{\theta}' \right) _R}{2}} \right]}^{\Delta},
$$
$$
\bar{S}_{int}=\mu \int{d\tau d\theta d\bar{\theta}\left[ \frac{bD_{\theta}\left( \theta ' \right) _LD_{\bar{\theta}}\left( \bar{\theta}' \right) _R}{\cosh ^2\frac{h_L\left( \tau ,\theta ',\bar{\theta}' \right) -h_R\left( \tau ,\theta ',\bar{\theta}' \right) +\left( \theta ' \right) _L\left( \bar{\theta}' \right) _L+\left( \theta ' \right) _R\left( \bar{\theta}' \right) _R-2\left( \theta ' \right) _L\left( \bar{\theta}' \right) _R}{2}} \right]}^{\Delta}.
\end{equation}
The total action is given by a chiral combination
\begin{equation}
S_{int,total}=S_{int}+\bar{S}_{int}.
\end{equation}

As demonstrated in \cite{9}, a similar N=2 supertransformation involves one bosonic variable $\epsilon$ and two fermionic Grassmann variables, which are $\eta$ and $\bar{\eta}$
\begin{equation}
\tau \rightarrow \tau '=\tau +\epsilon +\theta \bar{\eta}+\bar{\theta}\eta ,
$$
$$
\theta \rightarrow \theta '=\theta +\eta ,
$$
$$
\bar{\theta}\rightarrow \bar{\theta}'=\bar{\theta}+\bar{\eta}.
\end{equation}

In the IR limit, both chiral and anti-chiral correlation functions exhibit super-conformal symmetry. The N=2 coordinate transformations in the low-energy limit are also constrained by the supercovariant derivative and the chiral and anti-chiral Jacobians
\begin{equation}
\mathcal{G} \left( \tau _1,\theta _1,\bar{\theta}_1;\tau _2,\theta _2,\bar{\theta}_2 \right) =\left( D_{\bar{\theta}_1}\bar{\theta}_{1}^{'} \right) ^{\frac{1}{q}}\left( D_{\theta _2}\theta _{2}^{'} \right) ^{\frac{1}{q}}\mathcal{G} \left( \tau _{1}^{'},\theta _{1}^{'},\bar{\theta}_{1}^{'};\tau _{2}^{'},\theta _{2}^{'},\bar{\theta}_{2}^{'} \right) ,
$$
$$
\bar{\mathcal{G}}\left( \tau _1,\theta _1,\bar{\theta}_1;\tau _2,\theta _2,\bar{\theta}_2 \right) =\left( D_{\theta _1}\theta _{1}^{'} \right) ^{\frac{1}{q}}\left( D_{\bar{\theta}_2}\bar{\theta}_{2}^{'} \right) ^{\frac{1}{q}}\bar{\mathcal{G}}\left( \tau _{1}^{'},\theta _{1}^{'},\bar{\theta}_{1}^{'};\tau _{2}^{'},\theta _{2}^{'},\bar{\theta}_{2}^{'} \right) .
\end{equation}
For instance, we can also introduce chiral-anti-chiral coordinates similar to those in \cite{9}.
\begin{equation}
\tau _{\pm}=\tau \pm \theta \bar{\theta}.
\end{equation}
The reparameterization of inversion can also yield a global conformal symmetry
\begin{equation}
\tau \rightarrow \tau '=-\frac{1}{\tau},
$$
$$
\theta \rightarrow \theta '=\frac{\theta}{\tau},
$$
$$
\bar{\theta}\rightarrow \bar{\theta}'=\frac{\bar{\theta}}{\tau},
$$
$$
\tau _{\pm}\rightarrow \tau _{\pm}^{'}=-\frac{1}{\tau _{\pm}}.
\end{equation}

Compared to N=0 and N=1, explicit symmetry breaking at low energies also results in an N=2 super-Schwarzian action in superderivative form
\begin{equation}
\begin{split}
S\left[ \tau ',\theta ',\bar{\theta}';\tau ,\theta ,\bar{\theta} \right] &=\frac{\partial _{\tau}\bar{D}\bar{\theta}'}{\bar{D}\bar{\theta}'}-\frac{\partial _{\tau}D\theta '}{D\theta '}-2\frac{\partial _{\tau}\theta '\partial _{\tau}\bar{\theta}'}{\left( \bar{D}\bar{\theta}' \right) \left( D\theta ' \right)}
\\
&=S_f\left( \tau ',\theta ',\bar{\theta}';\tau ,\theta ,\bar{\theta} \right) +\theta S_{\theta}\left( \tau ',\theta ',\bar{\theta}';\tau ,\theta ,\bar{\theta} \right) +\bar{\theta}S_{\bar{\theta}}\left( \tau ',\theta ',\bar{\theta}';\tau ,\theta ,\bar{\theta} \right) +\theta \bar{\theta}S_b\left( \tau ',\theta ',\bar{\theta}';\tau ,\theta ,\bar{\theta} \right) .
\end{split}
\end{equation}

For the N=4 theory, we begin with the ladder diagram in the N=2 case. The propagator involving two identical superfields corresponds to a four-point contribution, known as solvable ladder diagrams in the SYK model. In the low-energy limit, we consider the 0th-order kernel
$$
\mathcal{F} _0\left( \tau _1,\tau _2;\tau _3,\tau _4 \right) =-G\left( \tau _1,\tau _3 \right) G\left( \tau _2,\tau _4 \right) +G\left( \tau _1,\tau _4 \right) G\left( \tau _2,\tau _3 \right) .
$$
It can return to two fermionic superfields, and we have obtained the correlation function for the N=4 SYK model.
\begin{equation}
\begin{split}
\mathcal{G} \left( \tau ,\theta _1,\bar{\theta}_1,\theta _2,\bar{\theta}_2;\tau ',\theta _1',\bar{\theta}_1',\theta _2',\bar{\theta}_2' \right) &=-\frac{b}{\left| \tau -\tau '-\theta _1\bar{\theta}_1-\theta _1'\bar{\theta}_1'+2\theta _1\bar{\theta}_1' \right|^{2\varDelta}\left| \tau -\tau '-\theta _2\bar{\theta}_2-\theta _2'\bar{\theta}_2'+2\theta _2\bar{\theta}_2' \right|^{2\varDelta}}
\\
&+\frac{b}{\left| \tau -\tau '-\theta _1\bar{\theta}_2-\theta _1'\bar{\theta}_2'+2\theta _1\bar{\theta}_2' \right|^{2\varDelta}\left| \tau -\tau '-\theta _2\bar{\theta}_1-\theta _2'\bar{\theta}_1'+2\theta _2\bar{\theta}_1' \right|^{2\varDelta}}.
\end{split}
\end{equation}
The conformal limit still holds for a single propagator but not for kernels. The second term arises from the possible off-diagonal dependence under general reparametrization. The formal result reduces to a copy of N=2 theories when we perform a trivial reparametrization of the theory (the Grassmann variable of which should also be integrated and leads to different physics). We can derive a simple form for the N=4 reparametrization.
\begin{equation}
\theta _1'\rightarrow \exp \left( ia_1\left( \tau \right) \right) \left( \partial _{\tau}\tanh \left( \frac{\tau +\theta \bar{\theta}+\theta \bar{\eta}+\bar{\theta}\eta}{2} \right) \right) ^{\frac{1}{2}}\left[ \theta _1+\eta _1\left( \tau +\theta _1\bar{\theta}_1 \right) \right] ,
$$
$$
\bar{\theta}_1'\rightarrow \exp \left( -ia_1\left( \tau \right) \right) \left( \partial _{\tau}\tanh \left( \frac{\tau -\theta \bar{\theta}+\theta \bar{\eta}+\bar{\theta}\eta}{2} \right) \right) ^{\frac{1}{2}}\left[ \bar{\theta}_1+\bar{\eta}_1\left( \tau -\theta _1\bar{\theta}_1 \right) \right] ,
$$
$$
\theta _2'\rightarrow \exp \left( ia_2\left( \tau \right) \right) \left( \partial _{\tau}\tanh \left( \frac{\tau +\theta \bar{\theta}+\theta \bar{\eta}+\bar{\theta}\eta}{2} \right) \right) ^{\frac{1}{2}}\left[ \theta _2+\eta _2\left( \tau +\theta _2\bar{\theta}_2 \right) \right] ,
$$
$$
\bar{\theta}_2'\rightarrow \exp \left( -ia_2\left( \tau \right) \right) \left( \partial _{\tau}\tanh \left( \frac{\tau -\theta \bar{\theta}+\theta \bar{\eta}+\bar{\theta}\eta}{2} \right) \right) ^{\frac{1}{2}}\left[ \bar{\theta}_2+\bar{\eta}_2\left( \tau -\theta _2\bar{\theta}_2 \right) \right] ,
$$
$$
h_1=\tanh \frac{\tau +\theta _1\bar{\theta}_1+\theta _1\bar{\eta}_1+\bar{\theta}_1\eta _1+\theta _2\bar{\theta}_2+\theta _2\bar{\eta}_2+\bar{\theta}_2\eta _2}{2},
$$
$$
h_2=\tanh \frac{\tau -\theta _1\bar{\theta}_1+\theta _1\bar{\eta}_1+\bar{\theta}_1\eta _1+\theta _2\bar{\theta}_2+\theta _2\bar{\eta}_2+\bar{\theta}_2\eta _2}{2}.
\end{equation}
Here, we use $h_1$ and $h_2$ to denote the reparameterized initial time and final time respectively, and the U(1) variables $a1$ and $a2$ in N=2 theory are also allowed.

Furthermore, we consider the effective reparameterized interaction action under a special constraint
\begin{equation}
\begin{split}
S_{int}&=\mu \int{d\tau d\theta d\bar{\theta}\left[ \frac{{bD_{\bar{\theta}}}^2\left( \bar{\theta}_1'\bar{\theta}_2' \right) _L{D_{\theta}}^2\left( \theta _1'\theta _2' \right) _R}{\cosh ^2\frac{\left( h_L-h_R-\left( \theta _1' \right) _L\left( \bar{\theta}_1' \right) _L-\left( \theta _1' \right) _R\left( \bar{\theta}_1' \right) _R+2\left( \theta _1' \right) _L\left( \bar{\theta}_1' \right) _R \right) \left( h_L-h_R-\left( \theta _2' \right) _L\left( \bar{\theta}_2' \right) _L-\left( \theta _2' \right) _R\left( \bar{\theta}_2' \right) _R+2\left( \theta _2' \right) _L\left( \bar{\theta}_2' \right) _R \right)}{2}} \right.}
\\
&-\left. \frac{{bD_{\bar{\theta}}}^2\left( \bar{\theta}_1'\bar{\theta}_2' \right) _L{D_{\theta}}^2\left( \theta _1'\theta _2' \right) _R}{\cosh ^2\frac{\left( h_L-h_R-\left( \theta _1' \right) _L\left( \bar{\theta}_2' \right) _L-\left( \theta _1' \right) _R\left( \bar{\theta}_2' \right) _R+2\left( \theta _1' \right) _L\left( \bar{\theta}_2' \right) _R \right) \left( h_L-h_R-\left( \theta _2' \right) _L\left( \bar{\theta}_1' \right) _L-\left( \theta _2' \right) _R\left( \bar{\theta}_1' \right) _R+2\left( \theta _2' \right) _L\left( \bar{\theta}_1' \right) _R \right)}{2}} \right] .
\end{split}
\end{equation}

The N=4 model allows four independent Grassmann variables. We consider an infinitesimal transformation. For simplicity, we assume that there is no dependence between $\theta_1 '$ and $\theta_2$, it is easy to check
\begin{equation}
D_{\theta _1}\theta _{1}^{'}D_{\theta _2}\theta _{2}^{'}=D_{\theta}^{2}\left( \theta _{1}^{'}\theta _{2}^{'} \right) ,
$$
$$
\theta ^{'}\rightarrow \theta _{1}^{'}\theta _{2}^{'}.
\end{equation}
We consider an arbitrary physical function $\mathcal{F} $, which should be invariant under a coordinate transformation
\begin{equation}
\begin{split}
&D_{\theta}^{2}F\left( \tau ,\theta ^1,\theta ^2,\bar{\theta}_1,\bar{\theta}_2 \right) =D_{\theta}^{2}\tau '\partial _{\tau '}\mathcal{F} +D_{\theta}^{2}\theta ^{'2}\partial^{'2} _{\theta }\mathcal{F} +D_{\bar{\theta}}^{2}\bar{\theta}^{'2}\partial^{'2} _{\bar{\theta}}\mathcal{F}
\\
&=D_{\theta}^{2}\theta ^2D'^{2}_{\theta }\mathcal{F} +D_{\bar{\theta}}^{2}\bar{\theta}^2D'^2_{\bar{\theta}}\mathcal{F} +\left( D_{\theta}^{2}\tau '-\bar{\theta}_{1}^{'}\partial _{\theta ^{'2}}+\bar{\theta}_{2}^{'}\partial _{\theta ^{'1}}-\bar{\theta}_{1}^{'}\bar{\theta}_{2}^{'}\partial _{\tau ^{'}} \right) \partial _{\tau '}\mathcal{F} .
\end{split}
\end{equation}
One interesting special condition in mathematical manipulations is the reduction of higher-order terms
\begin{equation}
D_{\theta}^{2}\tau '=\left( \bar{\theta}_{1}^{'}\partial _{\theta ^{'2}}-\bar{\theta}_{2}^{'}\partial _{\theta ^{'1}}+\bar{\theta}_{1}^{'}\bar{\theta}_{2}^{'}\partial _{\tau ^{'}} \right) D_{\theta}^{2}\theta ^{'2},
$$
$$
D\bar{\theta}_{\alpha}=D_{\theta}^{2}\bar{\theta}_{\alpha}=D_{\theta}^{2}\bar{\theta}^2=0.
\end{equation}
This restriction also leads to a transformation-invariant quantity, which exhibits similarities to a certain aspect of superconformal theories\cite{9}. Furthermore, we will see that this condition is highly beneficial in discovering a holographic theory within the context of gravitational physics. We can also consider the analogous approach utilizing Berezian integrals, which are equivalent in certain respects to the methods previously discussed. The Berezin integral is given by
\begin{equation}
\begin{split}
Ber\left( \begin{matrix}
	\partial _{\tau}\tau '&		\partial _{\tau}\theta ^{'2}\\
	\partial _{\theta ^2}\tau '&		\partial _{\theta ^2}\theta ^2\\
\end{matrix} \right) =Ber\left( \begin{matrix}
	\partial _{\tau}\tau '&		\partial _{\tau}\theta ^{'2}\\
	D_{\theta}^{2}\tau '-\left( \bar{\theta}_1\partial _{\theta ^2}-\bar{\theta}_2\partial _{\theta ^1}+\bar{\theta}_1\bar{\theta}_2\partial _{\tau} \right) \partial _{\tau}\tau '&		 D_{\theta}^{2}\theta ^{'2}-\left( \bar{\theta}_1\partial _{\theta ^2}-\bar{\theta}_2\partial _{\theta ^1}+\bar{\theta}_1\bar{\theta}_2\partial _{\tau} \right) \partial _{\tau}\theta ^{'2}\\
\end{matrix} \right)
\\
=Ber\left( \begin{matrix}
	\partial _{\tau}\tau '&		\partial _{\tau}\theta ^{'2}\\
	D_{\theta}^{2}\tau '&		D_{\theta}^{2}\theta ^{'2}\\
\end{matrix} \right) =\left( D_{\theta}^{2}\theta ^{'2} \right) ^{-1}Ber\left( \begin{matrix}
	\partial _{\tau}\tau '&		\partial _{\tau}\theta ^{'2}\\
	 \bar{\theta}_{1}^{'}\partial _{\theta ^{'2}}-\bar{\theta}_{2}^{'}\partial _{\theta ^{'1}}+\bar{\theta}_{1}^{'}\bar{\theta}_{2}^{'}\partial _{\tau ^{'}} &		1\\
\end{matrix} \right) =\bar{D}_{\theta}^{2}\bar{\theta}^{'2},
\end{split}
\end{equation}
where we have assumed that the infinitesimal translation exhibits a symmetry between the two Grassmann variables $\theta_1$ and $\theta_2$

\begin{equation}
\mathcal{G} \left( \tau _1,\theta _1,\bar{\theta}_1;\tau _2,\theta _2,\bar{\theta}_2 \right) =\left( D_{\bar{\theta}_1}^{2}\bar{\theta}_{1}^{2'} \right) ^{\frac{1}{q}}\left( D_{\theta _2}^{2}\theta _{2}^{2'} \right) ^{\frac{1}{q}}\mathcal{G} \left( \tau _{1}^{'},\theta _{1}^{'},\bar{\theta}_{1}^{'};\tau _{2}^{'},\theta _{2}^{'},\bar{\theta}_{2}^{'} \right) ,
$$
$$
\bar{\mathcal{G}}\left( \tau _1,\theta _1,\bar{\theta}_1;\tau _2,\theta _2,\bar{\theta}_2 \right) =\left( D_{\theta _2}^{2}\theta _{1}^{2'} \right) ^{\frac{1}{q}}\left( D_{\bar{\theta}_1}^{2}\bar{\theta}_{2}^{2'} \right) ^{\frac{1}{q}}\bar{\mathcal{G}}\left( \tau _{1}^{'},\theta _{1}^{'},\bar{\theta}_{1}^{'};\tau _{2}^{'},\theta _{2}^{'},\bar{\theta}_{2}^{'} \right) .
\end{equation}
Terms with two identical Grassmann variables and the 0th-order(or second-order) derivatives of $\tau$ will break the usual superconformal limit of reparametrization in N=4 theory. However, an important feature is that, when one of the Grassmann variables is ignored, the N=4 theory should reduce to the N=2 theory. Furthermore, we can construct the N=4 effective action based on the N=2 theory. The aforementioned approach of constructing the N=4 effective action from N=2 theory can also be applied to the gravitational sector. The behavior of the infinitesimal translation in the context of N=4 theory should exhibit some similarities to that in the N=2 theory, but with some notable differences, as follows
\begin{equation}
\theta _1\rightarrow \xi -i\epsilon \rho ,
$$
$$
\theta _2\rightarrow \xi -i\epsilon \rho ,
$$
$$
\bar{\theta}_1\rightarrow \bar{\xi}-i\epsilon \bar{\rho},
$$
$$
\bar{\theta}_2\rightarrow \bar{\xi}-i\epsilon \bar{\rho}.
\end{equation}
Since we have the first-order restrictions specified in the equation, we can proceed to express the infinitesimal generators $\rho$ and $\bar{\rho}$ in a manner similar to that in the N=2 theory
\begin{equation}
\rho =-\xi '
$$
$$
\bar{\rho}=-\bar{\xi}' .
\end{equation}
Then we can characterize the effective action by super Schwarzian
\begin{equation}
\begin{split}
Schw\left( x,\xi ,\bar{\xi};\tau ,\theta ,\bar{\theta} \right) &=\frac{\partial _{\tau}\left( D_{\bar{\theta}1}\bar{\xi} \right) \left( D_{\bar{\theta}2}\bar{\xi} \right)}{\left( D_{\bar{\theta}1}\bar{\xi} \right) \left( D_{\bar{\theta}2}\bar{\xi} \right)}+\frac{\left( D_{\bar{\theta}1}\bar{\xi} \right) \partial _{\tau}\left( D_{\bar{\theta}2}\bar{\xi} \right)}{\left( D_{\bar{\theta}1}\bar{\xi} \right) \left( D_{\bar{\theta}2}\bar{\xi} \right)}
\\
&+\frac{\partial _{\tau}\left( D_{\theta 1}\xi \right) \left( D_{\theta 2}\xi \right)}{\left( D_{\theta 1}\xi \right) \left( D_{\theta 2}\xi \right)}+\frac{\left( D_{\theta 1}\xi \right) \partial _{\tau}\left( D_{\theta 2}\xi \right)}{\left( D_{\theta 1}\xi \right) \left( D_{\theta 2}\xi \right)}-2\frac{\partial _{\tau}\xi ^2\partial _{\tau}\bar{\xi}^2}{\left( D_{\theta}^{2}\xi ^2 \right) \left( D_{\bar{\theta}}^{2}\bar{\xi}^2 \right)}
\\
&=\frac{\left( D_{\bar{\theta}}^{2}\partial _{\tau}\bar{\xi}^2 \right)}{D_{\bar{\theta}}^{2}\bar{\xi}^2}+\frac{\left( D_{\theta}^{2}\partial _{\tau}\xi ^2 \right)}{D_{\theta}^{2}\xi ^2}-2\frac{\partial _{\tau}\xi ^2\partial _{\tau}\bar{\xi}^2}{\left( D_{\theta}^{2}\xi ^2 \right) \left( D_{\bar{\theta}}^{2}\bar{\xi}^2 \right)}.
\end{split}
\end{equation}
This effective action arises from two identical N=2 SYK models, including a final term that arises from the first-order crossing term. It can be verified that eliminating one of the Grassmann variables reduces the theory to the N=2 theory. Moreover, this result indicates that the effective action is predominantly Schwarzian in the absence of superspace, and the additional cross term would break the emergent superconformal symmetry.
\begin{equation}
D_{\theta}^{2}\xi ^2=\left( \frac{\partial}{\partial \theta ^1}+\bar{\theta}_1\frac{\partial}{\partial \tau} \right) \xi \left( \frac{\partial}{\partial \theta ^2}+\bar{\theta}_2\frac{\partial}{\partial \tau} \right) \xi =\partial _{\theta}^{2}\xi ^2+D_{\theta ^1}\xi \frac{\partial}{\partial \theta ^2}\xi +\frac{\partial}{\partial \theta ^1}\xi D_{\theta ^2}\xi +\bar{\theta}_1\bar{\theta}_2\frac{\partial}{\partial \tau}\frac{\partial}{\partial \tau}\xi ^2.
\end{equation}

Compared to the mean field methods and SD equations we have studied in the previous section, we can also derive the IR Green's function for the N=2 SYK model. Firstly, when we neglect the higher-order terms and concentrate on the low-energy theory, the effective propagators in the low-energy limit can be written as
\begin{equation}
G_{\psi \psi}\left( \tau \right) \sim \frac{sgn\left( \tau \right)}{c_{\psi}\left( \tau \right)}\sim \frac{sgn\left( \tau \right)}{\left| \tau \right|^{2\varDelta}},
$$
$$
G_{bb}\left( \tau \right) \sim -\frac{\delta \left( \tau \right)}{c_b\left( \tau \right)}\sim \frac{1}{\left| \tau \right|^{2\varDelta+1}}.
\end{equation}
where the functions $c$ are the conformal factors in propagators and are determined by the conformal dimension, which is related to the equation $G_{\psi \psi}\left( \tau \right) =-\partial _{\tau}G_{bb}\left( \tau \right) $.

The N=4 ansatz can be defined as a specific theory
\begin{equation}
G_{\psi \psi}\left( \tau \right) \sim \frac{sgn\left( \tau \right)}{c_{\psi}\left( \tau \right)}\sim \frac{sgn\left( \tau \right)}{\left| \tau \right|^{2\varDelta +1}},
$$
$$
G_{FF}\left( \tau \right) \sim \frac{\delta \left( \tau \right)}{c_F\left( \tau \right)}\sim \frac{1}{\left| \tau \right|^{2\varDelta +2}},
$$
$$
G_{\phi \phi}\left( \tau \right) \sim \frac{\left| \tau \right|}{c_{\phi}\left( \tau \right)}\sim \frac{1}{\left| \tau \right|^{2\varDelta}}.
\end{equation}
with $G_{\phi \phi}\left( \tau \right) =\partial _{\tau}G_{\psi \psi}\left( \tau \right) =\partial _{\tau}^{2}G_{FF}\left( \tau \right) $.
In Eq. (33), the component $G_{\phi \phi}$ consists of two identical chiral fermions, and the fermionic conformal dimension remains, while the component $G_{F F}$ consists of two identical chiral bosons. Additionally, the component $G_{\psi \psi}$ consists of one chiral fermion and the conjugate boson, with the conformal dimension averaged over fermion and boson.

Then, we can further qualitatively analyze the thermal entropy with the mean field ansatz, especially when considering the zero-temperature limit
\begin{equation}
\exp \left( -S \right) \sim \frac{1}{Z^2}\int{D\phi D\psi DF}\exp \left( -S_{eff}\left( \Sigma _{F F}\left( \tau \right) \right) \right).
\end{equation}
Since the partition function consists of a first-order dependence on the logarithm, it will not converge to a certain constant due to the contribution of $\Sigma_{\phi \phi}$, and it has a dramatic change around the periodic short time $\tau =\beta /2$(where $\beta$ is large enough). According to  \cite{39}, the components with certain dimensions will dominate the bare propagator. Considering the Replica trick, it is also a periodic function.

\subsection{Gravitational action on boundary}
In this subsection, we investigate the holographic duality of the supersymmetric SYK model. First, we briefly review the low-energy SYK models with N=0,1,2 supersymmetry and their holography duality, which can be described as JT gravity
\begin{equation}
S=-\frac{1}{16\pi G}\left[ \int_M{d^2x}\sqrt{g}\phi \left( R+2 \right) +2\int_{\partial M}{du}\sqrt{h}K \right] ,
\end{equation}
which is embedded in a global $AdS_{2}$ metric
\begin{equation}
ds^2=\frac{dt^2+dz^2}{z^2}.
\end{equation}
We can also apply a finite curve reparameterization to the boundary
\begin{equation}
\frac{1}{\epsilon ^2}=\frac{t^{'2}+z^{'2}}{z^2}
\end{equation}
This form leads to a Schwarzian form with NAdS boundary condition as $\epsilon $ approaches zero
\begin{equation}
S\left[ t\left( u \right) \right] =\int{du}\phi _r\left( u \right) Sch\left( t,u \right) .
\end{equation}
We can also consider N=1 super JT gravity according to \cite{12}
\begin{equation}
S=-\frac{1}{16\pi G}\left[ i\int{d^2zd^2\theta}E\varPhi \left( R_{+-}-2 \right) +2\int_{\partial M}{dud\theta}\varPhi K \right] ,
\end{equation}
Here we use $\varphi$ to represent supersymmetric dilatons \cite{44,45}, and $E$ is the supervielbein that to harmonize the spinor and scalar fields. In \cite{12}, it also gives a special superconformal gauge condition
\begin{equation}
\frac{du^2+2\theta d\theta du}{4\epsilon ^2}=dz^{\xi}E_{\xi}^{1}dz^{\pi}E_{\pi}^{\bar{1}}.
\end{equation}
The superconformal gauge condition is the most important and widely researched among those relevant to quantum gravity and supersymmetric field theories with a supersymmetry condition. To obtain the effective action, the vielbeins on both dimensions satisfy
\begin{equation}
Dz=\theta D\theta ,
$$
$$
z=t+i\epsilon \left( D\xi \right) ^2,
$$
$$
Dt\left( u,\theta \right) =\xi \left( u,\theta \right) D\xi \left( u,\theta \right) .
\end{equation}

Supersymmetric gravity will contribute an additional spinor component. To make these spinors consistent with scalar superspace, the supervielbein should also be involved. We follow the notation $E$ to denote vielbein components relevant to inverse density, which leads to
\begin{equation}
K=\frac{T^AD_Tn_A}{T^AT_A},
\end{equation}
where $T$ is a tangent vector along the boundary, normalizing the orthogonal vector $T^An_A=0$.
\begin{equation}
K=4\epsilon ^2S\left[ t,\xi ;u,\theta \right] .
\end{equation}
The covariant derivative $D_T$ contains contributions from the superderivative and spinor conjugate, as expressed in the equation $D_Tn_A=Dn_A+D\varOmega $. Furthermore, the boundary action exhibits properties similar to those of the Schwarzian derivative
\begin{equation}
S_{bdy}=\int{dud\theta}\varPhi _r\left( u,\theta \right) S\left[ t,\xi ;u,\theta \right] .
\end{equation}
In order to describe a wormhole theory, we can introduce sets of interaction operators on both the left and right boundaries
\begin{equation}
S_{int}=g\sum_i{\int{dud\theta}O_{L}^{i}\left( u,\theta \right)}O_{R}^{i}\left( u,\theta \right) .
\end{equation}
O is a set of N operators with super conformal dimension $\varDelta $. And the dimension of $g$ is given by [energy]$^{2\varDelta-1}$\cite{46,47}. We will discuss the wormhole traversability in Appendix B.

Then we can construct the holographic duality of the interaction part
\begin{equation}
\left< O\left( t_{P}^{1} \right) O\left( t_{P}^{2} \right) \right> =\left| t_{P}^{1}-t_{P}^{2}-\theta _{P}^{1}\theta _{P}^{2} \right|^{-2\varDelta}.
\end{equation}
After applying the thermal reparametrization
\begin{equation}
h=\tanh \frac{\left( \tau +\theta \eta \right)}{2},
$$
$$
\theta ^{'}=\left( \partial _{\tau}\tanh \left( \frac{\left( \tau +\theta \eta \right)}{2} \right) \right) ^{\frac{1}{2}}\left( \theta +\eta +\frac{1}{2}\eta \partial _{\tau}\eta \right) .
\end{equation}
It is easy to check that the interactions have the same superconformal form with low-energy N=1 SYK.

We can also write the N=2 JT gravity from \cite{13}
\begin{equation}
S=-\frac{1}{16\pi G}\left[ \int{d^2zd^2\theta}E\varPhi \left( R+2 \right) +\int{d^2zd^2\bar{\theta}}E\bar{\varPhi}\left( \bar{R}+2 \right) +2\int_{\partial M}{dud\theta d\bar{\theta}}\left( \varPhi +\bar{\varPhi} \right) \mathcal{K} \right] ,
\end{equation}
with the boundary external supercurvature
\begin{equation}
\int_{\partial M}{dud\theta d\bar{\theta}}\mathcal{K} =\int_{\partial M}{dud\theta}K+\int_{\partial M}{dud\bar{\theta}}\bar{K},
$$
$$
K=\frac{T^AD_Tn_A}{T^AT_A},
$$
$$
\bar{K}=\frac{T^A\bar{D}_Tn_A}{T^AT_A}.
\end{equation}
Consider the N=2 interaction with chirals, the action should consist two part
\begin{equation}
S_{int}=g\sum_i{\int{dud\theta d\bar{\theta}}}\left( \bar{O}_{L}^{i}\left( u,\theta ,\bar{\theta} \right) O_{R}^{i}\left( u,\theta ,\bar{\theta} \right) +O_{L}^{i}\left( u,\theta ,\bar{\theta} \right) \bar{O}_{R}^{i}\left( u,\theta ,\bar{\theta} \right) \right) ,
\end{equation}
with supersymmetrization
\begin{equation}
\frac{du^2+2\bar{\theta}d\theta du+2\theta d\bar{\theta}du+2\theta \bar{\theta}d\theta d\bar{\theta}}{4\epsilon ^2}=dz^{\xi}E_{\xi}^{l}dz^{\pi}E_{\pi}^{\bar{l}}.
\end{equation}
Here, we use the supervielbeins as the chiral densities to rewrite the supergravity in superspace.

Then we have the reparameterization. This involves a super Schwarzian effective action and exhibits holographic duality with the N=2 SYK model
\begin{equation}
Dz=\bar{\theta}D\theta ,
$$
$$
\bar{D}z=\theta \bar{D}\bar{\theta},
$$
$$
z=t+i\epsilon \left( D\xi \right) \left( \bar{D}\bar{\xi} \right) ,
$$
$$
Dt\left( u,\theta ,\bar{\theta} \right) =\bar{\xi}\left( u,\theta ,\bar{\theta} \right) D\xi \left( u,\theta ,\bar{\theta} \right) ,
$$
$$
\bar{D}t\left( u,\theta ,\bar{\theta} \right) =\xi \left( u,\theta ,\bar{\theta} \right) \bar{D}\bar{\xi}\left( u,\theta ,\bar{\theta} \right) .
\end{equation}

The extrinsic supercurvature is very similar to the theories involving N=1 supersymmetry.
\begin{equation}
K=-4\epsilon ^2\left[ \frac{\xi ''}{D\xi}-\frac{\xi ^{'}\left( D\xi ^{'} \right)}{\left( D\xi \right) ^2}+\frac{\left( \bar{D}\bar{\xi}^{'} \right) \xi ^{'}}{\left( D\xi \right) \left( \bar{D}\bar{\xi} \right)} \right] ,
$$
$$
K=-4\epsilon ^2\left[ \frac{\bar{\xi}''}{\bar{D}\xi}-\frac{\bar{\xi}^{'}\left( \bar{D}\bar{\xi}^{'} \right)}{\left( \bar{D}\bar{\xi} \right) ^2}+\frac{\left( D\xi ^{'} \right) \bar{\xi}^{'}}{\left( \bar{D}\bar{\xi} \right) \left( D\xi \right)} \right] .
\end{equation}
We can also rewrite the boundary action in terms of the super-Schwarzian
\begin{equation}
S_{bdy}=\int{dud\theta d\bar{\theta}}\left( \varPhi _r+\bar{\varPhi}_r \right) S\left[ t,\xi ,\bar{\xi};u,\theta ,\bar{\theta} \right] .
\end{equation}
The N=2 correlation function between boundaries can be studied. Since the fermions with N=2 supersymmetry interact with conjugate chiral particles, we have restricted the sets of operators $O$ and $\bar{O}$ based on their chirality. Likewise, we can utilize the thermal reparametrization solutions
\begin{equation}
h_{1}=\tanh \frac{\tau +\theta \bar{\theta}+\theta \bar{\eta}+\bar{\theta}\eta}{2},
$$
$$
h_{2}=\tanh \frac{\tau -\theta \bar{\theta}+\theta \bar{\eta}+\bar{\theta}\eta}{2},
$$
$$
\theta '\rightarrow \exp \left( ia\left( \tau \right) \right) \left( \partial _{\tau}\tanh \left( \frac{\tau +\theta \bar{\theta}+\theta \bar{\eta}+\bar{\theta}\eta}{2} \right) \right) ^{\frac{1}{2}}\left[ \theta +\eta \left( \tau +\theta \bar{\theta} \right) \right] ,
$$
$$
\bar{\theta}'\rightarrow \exp \left( -ia\left( \tau \right) \right) \left( \partial _{\tau}\tanh \left( \frac{\tau -\theta \bar{\theta}+\theta \bar{\eta}+\bar{\theta}\eta}{2} \right) \right) ^{\frac{1}{2}}\left[ \bar{\theta}+\bar{\eta}\left( \tau -\theta \bar{\theta} \right) \right] ,
$$
$$
S_{int}=\frac{g}{2^{2\Delta}}\int{d\tau d\theta d\bar{\theta}\left[ \frac{D_{\theta}\left( \theta ' \right) _LD_{\bar{\theta}}\left( \bar{\theta}' \right) _R}{\cosh ^2\frac{h_L\left( \tau +\theta \bar{\eta}\left( \tau \right) +\bar{\theta}\eta \left( \tau \right) \right) -h_R\left( \tau +\theta \bar{\eta}\left( \tau \right) +\bar{\theta}\eta \left( \tau \right) \right) -\left( \theta ' \right) _L\left( \bar{\theta}' \right) _R+\left( \bar{\theta}' \right) _L\left( \theta ' \right) _R}{2}} \right]}^{\frac{1}{q}},
$$
$$
\bar{S}_{int}=\frac{g}{2^{2\Delta}}\int{d\tau d\theta d\bar{\theta}\left[ \frac{D_{\bar{\theta}}\left( \bar{\theta}' \right) _LD_{\theta}\left( \theta ' \right) _R}{\cosh ^2\frac{h_L\left( \tau +\theta \bar{\eta}\left( \tau \right) +\bar{\theta}\eta \left( \tau \right) \right) -h_R\left( \tau +\theta \bar{\eta}\left( \tau \right) +\bar{\theta}\eta \left( \tau \right) \right) -\left( \theta ' \right) _L\left( \bar{\theta}' \right) _R+\left( \bar{\theta}' \right) _L\left( \theta ' \right) _R}{2}} \right]}^{\frac{1}{q}}.
\end{equation}
Here $h_1$ and $h_2$ denote the reparameterized initial and final time.

We also want to derive the effective action of N=4 holographic theory in the low temperature limit. First, consider the special form of N=4 super Schwarzian which is discussed in the context of the N=4 low energy limit

\begin{equation}
Schw\left( x,\xi ,\bar{\xi};\tau ,\theta ,\bar{\theta} \right) =\frac{\left( D_{\bar{\theta}}^{2}\partial _{\tau}\bar{\xi}^2 \right)}{D_{\bar{\theta}}^{2}\bar{\xi}^2}+\frac{\left( D_{\theta}^{2}\partial _{\tau}\xi ^2 \right)}{D_{\theta}^{2}\xi ^2}-2\frac{\partial _{\tau}\xi ^2\partial _{\tau}\bar{\xi}^2}{\left( D_{\theta}^{2}\xi ^2 \right) \left( D_{\bar{\theta}}^{2}\bar{\xi}^2 \right)}.
\end{equation}

This action is not globally conformal nor superconformal, unlike the N=0 and N=1,2 SYK models, which exhibit special reparameterization symmetries. Different from the 2D analogues of the SYK model\cite{38}, this gravitational action is also restricted to 1+1 dimensions. Since there exists only one time scale, higher-order terms emerge. These terms include cross terms between the Grassmann variables and are not invariant under conformal transformations. A special case is when SL(2,R) emerges in the low-energy limit of N=0 SYK, whereas $SU(1,1\mid 1)$ symmetry does not emerge under N=2 conditions.

However, it can still be derived from the N=2 SYK model. Since the infinitesimal translation group is similar to those in the N=2 SYK model, we can propose a generalization based on the N=2 NAdS spacetime. Similarly, in the N=4 case, we expect this N=4 NAdS spacetime to exhibit properties analogous to those of the N=2 SYK model. To extend to N=4 spacetime, we introduce supersymmetry into the relevant components, based on the N=2 NAdS spacetime framework. When we eliminate one Grassmann variable $\alpha$ by imposing certain constraints and symmetries
$$
\partial _{\theta _{\alpha}}\partial _{\bar{\theta}_{\alpha}}\left( \theta _{\alpha}\bar{\theta}_{\alpha}\mathcal{L} \right) .
$$
and the theory will return to N=2.

We can easily adopt an ansatz for the specific form of the metric within the potential framework of N=4 JT gravity
\begin{equation}
S_{JT}=-\frac{1}{16\pi G_N}\left( \int_{\mathcal{M}}{d^2zd^4\theta}\mathcal{E} ^{-1}\varPhi \left( R-2 \right) +\int_{\mathcal{M}}{d^2zd^4\bar{\theta}}\bar{\mathcal{E}}^{-1}\bar{\varPhi}\left( \bar{R}-2 \right) +2\int_{\mathcal{M}}{dzd^2\theta d^2\bar{\theta}\left( \varPhi _b+\bar{\varPhi}_b \right)}\mathcal{K} \right) .
\end{equation}
The parameters in this equation should also be constrained with certain conditions as N=4 SYK, which means that we should restrict the N=4 theory back to N=2 if it involves the first-order of the other Grassmann variable. This 'external curvature' term can be written in terms of Grassmann components
\begin{equation}
\int_{\mathcal{M}}{dzd^2\theta d^2\bar{\theta}}\mathcal{K} =\int_{\mathcal{M}}{dzd^2\theta}K+\int_{\mathcal{M}}{dzd^2\bar{\theta}}\bar{K}.
\end{equation}
Since our aim is to acquire the formula for the effective action, it is quite natural to define the external curvature in terms of Grassmann components, which are associated with and derived from the N=2 theory
\begin{equation}
K=\frac{T^A\bar{D}_{T1}n_{A1}\bar{D}_{T2}n_{A2}}{T^2}=\frac{T^A\bar{D}_{T}^{2}n_A}{T^2},
$$
$$
\bar{K}=\frac{T^AD_{T1}n_{A1}D_{T2}n_{A2}}{T^2}=\frac{T^AD_{T}^{2}n_A}{T^2}.
\end{equation}
Here we use $\bar{D}_{T1}$, $\bar{D}_{T2}$, $D_{T1}$, $D_{T2}$ to express the scalar-spinor covariant derivatives with different Grassmann variables. These derivatives are calculated using a super connection term $\varOmega _{\xi}$
\begin{equation}
D_{T}^{2}n_A=D_{T}^{2}n_A+\left( D_{T}^{2}z^{\xi}\varOmega _{\xi} \right) n_A,
$$
$$
\bar{D}_{T}^{2}n_A=\bar{D}_{T}^{2}n_A+\left( \bar{D}_{T}^{2}z^{\xi}\varOmega _{\xi} \right) n_A.
\end{equation}

We have preliminarily proposed a formalism for the curvature. It is easy to check that this curvature includes the N=2 curvature with a single Grassmann component, either $\theta$ or $\bar{\theta}$.

Additionally, it can also be written as the boundary effective action up to second order
\begin{equation}
\frac{\left( du^2+2\bar{\theta}_1d\theta _1du+2\theta _1d\bar{\theta}_1du+2\theta _1\bar{\theta}_1d\theta _1d\bar{\theta}_1 \right) \left( du^2+2\bar{\theta}_2d\theta _2du+2\theta _2d\bar{\theta}_2du+2\theta _2\bar{\theta}_2d\theta _2d\bar{\theta}_2 \right)}{4\epsilon ^2}=dz^{\xi}E_{\xi}^{l}dz^{\pi}E_{\pi}^{\bar{l}}.
\end{equation}
Compared to the N=2 theory, this equation can be generated from the N=4 supercovariant derivative. Parameters $E$ and $\bar{E}$ as chiral densities also lead to the supervielbein, which unifies the spinor and scalar components with indices. This equation also contains elements of the N=2 theories.

To develop a holographic theory of fields, certain constraints must be satisfied by the boundary reparameterization
\begin{equation}
Dz=\left( \bar{\theta}_1\partial _{\theta 2}-\bar{\theta}_2\partial _{\theta 1}+\bar{\theta}_1\bar{\theta}_2\partial _z \right) D_{\theta}^{2}\theta ^2,
$$
$$
\bar{D}z=\left( \theta _1\partial _{\bar{\theta}2}-\theta _2\partial _{\bar{\theta}1}+\theta _1\theta _2\partial _z \right) D_{\bar{\theta}}^{2}\bar{\theta}^2,
$$
$$
z=t+i\epsilon \left( D_{\theta}^{2}\xi ^2 \right) \left( D_{\bar{\theta}}^{2}\bar{\xi}^2 \right) ,
$$
$$
Dt\left( u,\theta _1,\theta _2,\bar{\theta}_1,\bar{\theta}_2 \right) =\left( \bar{\xi}\left( u,\theta _1,\theta _2,\bar{\theta}_1,\bar{\theta}_2 \right) \partial _{\theta 2}-\bar{\xi}\left( u,\theta _1,\theta _2,\bar{\theta}_1,\bar{\theta}_2 \right) \partial _{\theta 1}+\bar{\xi}^2\left( u,\theta _1,\theta _2,\bar{\theta}_1,\bar{\theta}_2 \right) \partial _z \right) D_{\theta}^{2}\xi ^2,
$$
$$
\bar{D}t\left( u,\theta _1,\theta _2,,\bar{\theta}_1,\bar{\theta}_2 \right) =\left( \xi \left( u,\theta _1,\theta _2,\bar{\theta}_1,\bar{\theta}_2 \right) \partial _{\bar{\theta}2}-\xi \left( u,\theta _1,\theta _2,\bar{\theta}_1,\bar{\theta}_2 \right) \partial _{\bar{\theta}1}+\xi ^2\left( u,\theta _1,\theta _2,\bar{\theta}_1,\bar{\theta}_2 \right) \partial _z \right) D_{\bar{\theta}}^{2}\bar{\xi}^2 .
\end{equation}

An important signature here is that this action should contain the N=2 gravity, which is dual to the N=2 SYK model. If we eliminate terms involving the Grassmann variables $\theta_{1}$ and $\bar{\theta_{1}}$ or $\theta_{2}$ and $\bar{\theta_{2}}$
$$
\partial _{\theta _{\alpha}}\partial _{\bar{\theta}_{\alpha}}\left( \theta _{\alpha}\bar{\theta}_{\alpha} S \right) ,
$$
the action now simplifies to an N=2 action, which contains a bosonic part that relates to the non-supersymmetric Schwarzian action. And we can easily verify this equation in terms of the effective action. However, the method for verifying the N=4 construction is incomplete. We naturally ignore terms of first order. These terms do not appear in the N=2 model. We should proceed to additional restrictions.

Notice that a stronger condition is required. This condition is necessary for the full consistency of the theory.
\begin{equation}
\bar{D}_{\bar{\theta}}^{2}\bar{\xi}^2=\bar{D}_{\bar{\theta}}\bar{\xi}\bar{D}_{\bar{\theta}}\bar{\xi},
$$
$$
D_{\theta}^{2}\xi ^2=D_{\theta}\xi D_{\theta}\xi ,
\end{equation}

the reparametrization $\xi$ has only first-order dependence on $\tau$. The single reparametrization $\xi$ should involve a Grassmann variable. Assuming the relevant equation holds, we can then proceed to consider the curvature tensor. The curvature tensor can be expressed in its component form
\begin{equation}
K=-2\epsilon ^2D_{\theta}^{2}\left( \frac{\left( D_{\theta}^{2}\partial _{\tau}\xi ^2 \right)}{D_{\theta}^{2}\xi ^2}-\frac{\partial _{\tau}\xi ^2\partial _{\tau}\bar{\xi}^2}{\left( D_{\theta}^{2}\xi ^2 \right) \left( D_{\bar{\theta}}^{2}\bar{\xi}^2 \right)} \right) ,
$$
$$
\bar{K}=-2\epsilon ^2\bar{D}_{\bar{\theta}}^{2}\left( \frac{\left( D_{\bar{\theta}}^{2}\partial _{\tau}\bar{\xi}^2 \right)}{D_{\bar{\theta}}^{2}\bar{\xi}^2}-\frac{\partial _{\tau}\xi ^2\partial _{\tau}\bar{\xi}^2}{\left( D_{\theta}^{2}\xi ^2 \right) \left( D_{\bar{\theta}}^{2}\bar{\xi}^2 \right)} \right) .
\end{equation}
One can easily verify that the curvature exhibits features characteristic of N=2 curvature. Moreover, this N=4 gravitation theory also has $SU(2)\times U(1)$ R symmetry within the four Grassmann variables, and this additional symmetry will not lead to a divergent result. We can easily check that the theory is invariant under the the transformation of the components: $\theta _1\rightarrow \theta _2$, $\theta _2\rightarrow -\theta _1$. These features are consistent with what is expected for an N=2 theory
\begin{equation}
D_{\theta}^{2}\xi ^2=\left( \frac{\partial}{\partial \theta ^1}+\bar{\theta}_1\frac{\partial}{\partial \tau} \right) \xi \left( \frac{\partial}{\partial \theta ^2}+\bar{\theta}_2\frac{\partial}{\partial \tau} \right) \xi =\partial _{\theta}^{2}\xi ^2+D_{\theta ^1}\xi \frac{\partial}{\partial \theta ^2}\xi +\frac{\partial}{\partial \theta ^1}\xi D_{\theta ^2}\xi +\bar{\theta}_1\bar{\theta}_2\frac{\partial}{\partial \tau}\frac{\partial}{\partial \tau}\xi ^2.
\end{equation}

In summary, the most important properties of the SYK model are Gaussian randomness, which leads to a maximally entangled theory. The supersymmetric SYK model adds additional degree of freedom with symmetry, which also brings the same restrictions on the gravitational theory and reparametrization. And in this section, since the bosonic part of gravity should preserve as the original NAdS/SYK holography, we have built the fermionic extension from the supersymmetric relation between the SYK models. Moreover, the relationships between these extensions correspond to the N=1,2,4 supersymmetries.

The first-order correlation term is written in ansatz
\begin{equation}
\left< \bar{O}\left( t_{P}^{1} \right) O\left( t_{P}^{2} \right) \right> =\left< O\left( t_{P}^{1} \right) \bar{O}\left( t_{P}^{2} \right) \right> =\left| t_{P}^{1}-t_{P}^{2}-\theta _{\alpha}^{1}\bar{\theta}_{\alpha}^{2}-\bar{\theta}_{\alpha}^{1}\theta _{\alpha}^{2}-\theta _{\beta}^{1}\bar{\theta}_{\beta}^{2}-\bar{\theta}_{\beta}^{1}\theta _{\beta}^{2} \right|^{-2\varDelta}.
\end{equation}
For both N=2 and N=4 theories, the super-reparametrization differs significantly from that of the N=1 theory. These differences are crucial for understanding the distinct properties of each theory.

\section{Thermofield double state}

The thermofield double in the Majorana SYK model is defined as the entangled state. This state relates the original system to its time-reversed counterpart at a finite temperature
\begin{equation}
\left| TFD_{\beta} \right> =\frac{1}{\sqrt{Z_{\beta}}}\sum_n{\exp \left( -\beta E_n/2 \right) \left| n \right> _1}\otimes \left| \hat{n} \right> _2.
\end{equation}
where $\beta =1/T$. Since the N=1 SYK supercharge has the same structure as Majorana fermions, except that the parity has changed from odd to even. We can propose a projection from N=0 to N=1 by changing q to -q. The Hilbert space remains in its original form to directly represent the supercharge. It also indirectly relates to the system's Hamiltonian.
\begin{equation}
\left| TFD_{\beta} \right> =\frac{1}{\sqrt{Z_{\beta}}}\sum_n{\exp \left( -\beta Q_{n}^{2}/2 \right) \left| n \right> _1}\otimes \left| \hat{n} \right> _2.
\end{equation}
The interaction term is proportional to Majorana fermions. These fermions act as supercharges in Eq.(5) and Eq.(7).
\begin{equation}
\varPsi _i=\left( I+\theta Q \right) \psi _i=\left( I+\theta Q \right) \left( C_{i}^{\dagger}+C_i \right) ,
\end{equation}
which exhibits anticommutative relations
\begin{equation}
H_{int}=i\mu \partial _{\theta}\sum_i{\varPsi _{i}^{L}\varPsi _{i}^{R}}=i\mu \sum_i{\left( QC_{i}^{L\dagger}C_{i}^{R}-QC_{i}^{R\dagger}C_{i}^{L} \right)}.
\end{equation}
Then, we can explicitly define the eigenstates of the superspace operator
\begin{equation}
H\left| \varPsi _{n}^{r} \right> =Q^2\left| \varPsi _{n}^{r} \right> =E_n\left| \varPsi _{n}^{r} \right> ,
$$
$$
H^{\dagger} \left| \varPsi _{m}^{l} \right> =Q^{2\dagger}\left| \varPsi _{n}^{r} \right> =E_{m}^{*} \left| \varPsi _{m}^{l} \right> ,
\end{equation}
with
\begin{equation}
\sum_n \left|\right. \varPsi _{n}^{l} \left.\right> \left< \varPsi _{n}^{r} \right|=I,
$$
$$
\left. \left. \right< \varPsi _{n}^{l} \right.\left| \varPsi _{m}^{r} \right> =\delta _{nm}.
\end{equation}
The ground states of the interaction part are generated from the N=0 MQ model
\begin{equation}
\left| I \right> =\prod_i{\left( \left| 1 \right> _{L,i}\left| 0 \right> _{R,i}+i\left| 0 \right> _{L,i}\left| 1 \right> _{R,i} \right)},
\end{equation}
and
\begin{equation}
H_{int}^{\dagger}\left| \varPsi _{0}^{l} \right> =-\mu N Q_0  \left| \varPsi _{0}^{l} \right> ,
$$
$$
H_{int}\left| \varPsi _{0}^{r} \right> =-\mu N Q_0 \left| \varPsi _{0}^{r} \right> .
\end{equation}
The TFD $\beta$ =0 state shall be defined with an anti-unitary transformation, as described in\cite{37,48}
\begin{equation}
\left| TFD_0\left( 0 \right) \right> =\frac{1}{2^{N/2}}\sum_q{\sum_{n_q}{\left| n_q \right> _1\otimes}}\left( e^{\frac{-i\eta \pi \varGamma}{4}}e^{-iq\left( \frac{\pi}{2} \right)} \right) P\left| n_q \right> _2,
$$
$$
\varGamma =\left( -1 \right) ^{-q+\frac{N}{2}},
$$
$$
P^{-1}C_iP=\eta P_{i}^{\dagger},
$$
$$
\left<\right. \hat{n}_{-q^{'}} \left| \right. C_{i}^{R}\left| \hat{m}_{-q} \right> =-e^{-i\pi \left( -q+\frac{N}{2} \right)}\left< m_q \right|C_{i}^{L}\left|\right. n_{q^{'}} \left.\right> ,
$$
$$
\left<\right. \hat{n}_{-q^{'}}  \left| \right. C_{i}^{R\dagger}\left| \hat{m}_{-q} \right> =-e^{-i\pi \left( -q^{'}+\frac{N}{2} \right)}\left< m_q \right|C_{i}^{L\dagger}\left|\right. n_{q^{'}} \left.\right> .
\end{equation}
We now define the states and parity $P$ in terms of supercharges. These supercharges in the equation act like the Hamiltonian in the N=0 model. They give a constant without changing the states in Fock space.
Since the energy diagram of the N=1 SYK model can be projected onto N=0 theory\cite{49,50,51,52}, which could be considered as a special form of the N=0 SYK model. It is also referred to as 'fake superspace'. The ground states of the N=1 SYK model are isomorphic to the trivial homology of supercharges but not to cohomology. Furthermore, we know that the ground state of the N=1 SYK model is not a BPS state. However, we will see that the BPS ground states of the N=2 and N=4 SYK models have an infinite number of projections. These projections are onto supercharges or onto the N=0 SYK model.
\begin{equation}
\begin{split}
\left< TFD_{\beta =0} \right|H_{int}\left| TFD_{\beta =0} \right> &=\frac{1}{2^N}\sum_{qq^{'}}{\sum_{mn}{\left< \hat{n}_{-q^{'}} \right|_2\otimes \left< n_{q^{'}} \right|_1\left( i\mu \sum_i{\left( Q C_{i}^{L\dagger }C_{i}^{R} \right.} \right.}}
\\
&\left. \left. -Q C_{i}^{R\dagger} C_{i}^{L} \right) \right) \left| m_q \right> _1\otimes \left| \hat{m}_{-q} \right> _2
\\
&=\frac{1}{2^N}\sum_{qq^{'}}{\sum_{nm}{\sum_i{\left( i\mu \right.}}}e^{i\pi \left( -q+\frac{N}{2} \right)}\left< n_{q^{'}} \right|Q C_{i}^{L\dagger}\left| m_{q^{''}}^{'} \right> \left< \hat{n}_{-q^{'}} \right|C_{i}^{R}\left| \hat{m}_{-q} \right>
\\
&\left. -i\mu e^{i\pi \left( -q^{'}+\frac{N}{2} \right)}e^{2\alpha}\left< \hat{n}_{-q^{'}} \right|Q C_{i}^{R\dagger}\left| n_{q^{'''}}^{'} \right> \left< n_{q^{'}} \right| C_{i}^{R}\left| m_q \right> \right) ,
\end{split}
\end{equation}
gives
\begin{equation}
\begin{split}
\left< H_{int} \right> &=\frac{1}{2^N}\sum_{qq^{'}}{\sum_{nm}{\sum_i{\left[ -i\mu \right.}}}\left< n_{q^{'}} \right|Q C_{i}^{L\dagger}\left| m_q \right> \left< m_q \right| C_{i}^{L}\left| n_{q^{'}} \right>
\\
&+i\mu \left< m_q \right|Q C_{i}^{L\dagger}\left| n_{q^{'}} \right> \left< n_{q^{'}} \right| C_{i}^{L}\left| m_q \right>
\\
&=\sum_{qq^{'}}{\sum_{nm}{\sum_i{i\mu}}}/2^N\left< n_{q^{'}} \right|Q C_{i}^{L\dagger}\left| m_q \right> \left< m_q \right| C_{i}^{L}\left| n_{q^{'}} \right>
\\
&=-\frac{\mu}{2^N}\sum_{n,q}{\sum_i{\left< n_q \right|_1Q}}C_{i}^{L\dagger}QC_{i}^{L}\left| n_q \right> _1=Const.
\end{split}
\end{equation}

We can utilize the same process in the N=2,4 SYK model as described in Appendix A. This process also yields similar results.

Additionally, we can calculate the Witten index. This will help us confirm the presence of fermionic ground states.
\begin{equation}
\mathcal{I} \left( r \right) =Tr\left[ \left( -1 \right) ^Fe^{2\pi irQ_R} \right] .
\end{equation}
In the N=2 model, the Witten index is redefined to reflect specific properties or conditions relevant to the model
\begin{equation}
\mathcal{I} \left( r \right) =e^{i\pi N\left( \frac{r}{q}-\frac{1}{2} \right)}\left( 1-e^{\frac{2\pi ir}{q}} \right) ^N=\left( 2\sin \left( \frac{\pi r}{q} \right) \right) ^N.
\end{equation}
The Witten index is very similar for the N=4 SYK model. This model also incorporates anti-commutation within the Hamiltonian. The N=4 ground states enforce one of the anti-commutators of the supercharges to be zero, which provides a specific result
\begin{equation}
\mathcal{I} \left( r \right) =e^{i\pi N\left( \frac{2rq-r^2}{q^2}-\frac{1}{2} \right)}\left( 1-e^{\frac{2\pi i\left( 2rq-r^2 \right)}{q^2}} \right) ^N=\left( 2\sin \frac{\pi \left( 2rq-r^2 \right)}{q^2} \right) ^N.
\end{equation}
This quantity also vanishes when $r=0$. In this context, R-charge is defined as $Q_R=\frac{1}{q} Q$

\section{Transmission amplitude in Lorentz and out-of-time-ordered correlator}
\subsection{Real time correlation and transmission amplitude}
In both the N=1 and N=2 Supersymmetric SYK models, the Lorentz-Wightman correlation function can be derived from the properties of superfields
\begin{equation}
\mathcal{G} _{AB}^{>}\left( t_1,t_2 \right) =-i\mathcal{G} _{AB}^{>}\left( it_{1}^{-},it_{2}^{+} \right) =-i\underset{\epsilon \rightarrow -0}{\lim}\mathcal{G} _{AB}^{>}\left( it_1+\epsilon ,it_2-\epsilon \right) ,
$$
$$
\mathcal{G} _{AB}^{<}\left( t_1,t_2 \right) =-i\mathcal{G} _{AB}^{<}\left( it_{1}^{+},it_{2}^{-} \right) =-i\underset{\epsilon \rightarrow +0}{\lim}\mathcal{G} _{AB}^{<}\left( it_1-\epsilon ,it_2+\epsilon \right) ,
$$
$$
\mathcal{G} _{AB}^{R}\left( t_1,t_2 \right) =\vartheta \left( t_1-t_2 \right) \left( \mathcal{G} _{AB}^{>}\left( t_1,t_2 \right) -\mathcal{G} _{AB}^{<}\left( t_1,t_2 \right) \right) ,
$$
$$
\mathcal{G} _{AB}^{A}\left( t_1,t_2 \right) =\vartheta \left( t_2-t_1 \right) \left( \mathcal{G} _{AB}^{>}\left( t_1,t_2 \right) -\mathcal{G} _{AB}^{<}\left( t_1,t_2 \right) \right) .
\end{equation}
The Schwinger-Dyson equation in Lorentzian time version is defined as
\begin{equation}
iD_{t_1}G_{AB}^{>}\left( t_1,t_2 \right) =\mu\epsilon _{AC}\partial _{\theta}G_{CB}^{>}\left( t_1,t_2 \right) +\int{dtd\theta \left( \varSigma _{AC}^{R}\left( t_1,t \right) \mathcal{G} _{CB}^{>}\left( t,t_2 \right) +\varSigma _{AC}^{>}\left( t_1,t \right) \mathcal{G} _{CB}^{A}\left( t,t_2 \right) \right)}\,,
$$
$$
iD_{t_1}G_{AB}^{R}\left( t_1,t_2 \right) -\mu \epsilon_{AC}\partial _{\theta}G_{CB}^{R}\left( t_1,t_2 \right) -\int{dtd\theta \left( \varSigma _{AC}^{R}\left( t_1,t \right) \mathcal{G} _{CB}^{R}\left( t,t_2 \right) +\varSigma _{AC}^{A}\left( t_1,t \right) \mathcal{G} _{CB}^{A}\left( t,t_2 \right) \right)}
$$
$$
=\delta _{AB}\left( \bar{\theta}_1-\bar{\theta}_2 \right) \delta \left( t_1-t_2 \right) \,,
\end{equation}
The corresponding Lorentzian equation of motion
\begin{equation}
\varSigma _{AB}^{>}\left( t_1,t_2 \right) =J\mathcal{G} _{AB}^{>}\left( t_1,t_2 \right) \mathcal{G} _{AB}^{>}\left( t_1,t_2 \right)
$$
$$
\varSigma _{AB}^{R}\left( t_1,t_2 \right) =\vartheta \left( t_1-t_2 \right) \left( \varSigma _{AB}^{>}\left( t_1,t_2 \right) -\varSigma _{AB}^{<}\left( t_2,t_1 \right) \right) ,
\end{equation}
and written in components
\begin{equation}
\begin{split}
\varSigma _{\psi \psi ,AB}^{>}\left( t \right) &=-\left( q-1 \right) \left( -1 \right) ^{\left( q-1 \right) /2}JG_{\psi \psi ,AB}^{>q-2}\left( t \right) G_{bb,AB}^{>}\left( t \right) +\frac{\left( q-1 \right)}{2}\left( q-2 \right) \left( -1 \right) ^{\left( q-1 \right) /2}JG_{\psi \psi ,AB}^{>q-3}\left( t \right),
\\
&G_{\psi b,AB}^{>A}\left( t \right) G_{b\psi ,AB}^{>S}\left( t \right) +\frac{\left( q-1 \right)}{2}\left( q-2 \right) \left( -1 \right) ^{\left( q-1 \right) /2}JG_{\psi \psi ,AB}^{>q-2}\left( t \right) G_{\psi b,AB}^{>S}\left( t \right) G_{b\psi ,AB}^{>A}\left( t \right) ,
\\
\varSigma _{bb,AB}^{>}\left( t \right) &=-JG_{\psi \psi ,AB}^{>q}\left( t \right) ,
\\
\varSigma _{b\psi ,AB}^{>S}\left( t \right) &=\frac{\left( q-1 \right)}{2}JG_{\psi b,AB}^{>A}\left( t \right) G_{\psi \psi ,AB}^{>q-2}\left( t \right) ,
\\
\varSigma _{b\psi ,AB}^{>A}\left( t \right) &=\left( -1 \right) ^{\left( q-1 \right) /2}\frac{\left( q-1 \right)}{2}JG_{\psi b,AB}^{>S}\left( t \right) G_{\psi \psi ,AB}^{>q-2}\left( t \right) ,
\\
\varSigma _{\psi b,AB}^{>S}\left( \tau \right) &=\frac{\left( q-1 \right)}{2}JG_{b\psi ,AB}^{>A}\left( \tau \right) G_{\psi \psi ,AB}^{>}\left( \tau \right) ,
\\
\varSigma _{\psi b,AB}^{>A}\left( t \right) &=-\left( -1 \right) ^{\left( q-1 \right) /2}\frac{\left( q-1 \right)}{2}JG_{b\psi ,AB}^{>S}\left( \tau \right) G_{\psi \psi ,AB}^{>}\left( \tau \right) .
\end{split}
\end{equation}
We can rewrite the retarded component using both bosonic and fermionic partitions.
\begin{equation}
G_{\psi \psi ,AB}^{>}\left( \omega \right) =\frac{G_{\psi \psi ,AB}^{R}\left( \omega \right) -\left( G_{\psi \psi ,AB}^{R}\left( \omega \right) \right) ^*}{1+\exp \left( -\beta \omega \right)},
$$
$$
G_{bb,AB}^{>}\left( \omega \right) =\frac{G_{bb,AB}^{R}\left( \omega \right) -\left( G_{bb,AB}^{R}\left( \omega \right) \right) ^*}{-1+\exp \left( -\beta \omega \right)},
$$
$$
G_{b\psi ,AB}^{>S}\left( \omega \right) =\frac{G_{b\psi ,AB}^{RS}\left( \omega \right) -\left( G_{b\psi ,AB}^{RS}\left( \omega \right) \right) ^*}{-1+\exp \left( -\beta \omega \right)},
$$
$$
G_{\psi b,AB}^{>A}\left( \omega \right) =\frac{G_{\psi b,AB}^{RA}\left( \omega \right) -\left( G_{\psi b,AB}^{RA}\left( \omega \right) \right) ^*}{1+\exp \left( -\beta \omega \right)},
$$
$$
G_{b\psi ,AB}^{>A}\left( \omega \right) =\frac{G_{b\psi ,AB}^{RA}\left( \omega \right) -\left( G_{b\psi ,AB}^{RA}\left( \omega \right) \right) ^*}{1+\exp \left( -\beta \omega \right)},
$$
$$
G_{\psi b,AB}^{>S}\left( \omega \right) =\frac{G_{\psi b,AB}^{RS}\left( \omega \right) -\left( G_{\psi b,AB}^{RS}\left( \omega \right) \right) ^*}{-1+\exp \left( -\beta \omega \right)}.
\end{equation}
In this context, the bosonic and fermionic factors are incorporated into the contour integrations, rendering the traditional Matsubara methods obsolete. Since the frequency is discrete, we can now consider the original bosonic and fermionic factors, which can be analyzed in real time
\begin{equation}
D_{Lorentz}=\left( \begin{matrix}
	w_n-\varSigma _{LL,\psi \psi}&		-\varSigma _{LR,\psi \psi}&		-\varSigma _{LL,\psi b}^{A}&		-i\mu -\varSigma _{LR,\psi b}^{A}\\
	-\varSigma _{RL,\psi \psi}&		w_n-\varSigma _{RR,\psi \psi}&		i\mu -\varSigma _{RL,\psi b}^{A}&		-\varSigma _{RR,\psi b}^{A}\\
	-\varSigma _{LL,b\psi}^{A}&		i\mu -\varSigma _{LR,b\psi}^{A}&		-1-0&		-0\\
	-i\mu -\varSigma _{RL,b\psi}^{A}&		-\varSigma _{RR,b\psi}^{A}&		-0&		-1-0\\
\end{matrix} \right) ,
$$
$$
D_{Lorentz\,\,inverse}=\left( \begin{matrix}
	w_n-0&		-0&		-\varSigma _{LL,\psi b}^{S}&		i\mu -\varSigma _{LR,\psi b}^{S}\\
	-0&		w_n-0&		-i\mu -\varSigma _{RL,\psi b}^{S}&		-\varSigma _{RR,\psi b}^{S}\\
	-\varSigma _{LL,b\psi}^{S}&		-i\mu -\varSigma _{LR,b\psi}^{S}&		-1-\varSigma _{LL,bb}&		-\varSigma _{LR,bb}\\
	i\mu -\varSigma _{RL,b\psi}^{S}&		-\varSigma _{RR,b\psi}^{S}&		-\varSigma _{RL,bb}&		-1-\varSigma _{RR,bb}\\
\end{matrix} \right) ,
$$
$$
G_{\psi \psi ,AB}\left( \tau \right) =\frac{Det\left( A_{\psi \psi ,AB}\left( D_{Lorentz} \right) \right)}{Det\left( D_{Lorentz} \right)},
$$
$$
G_{bb,AB}\left( \tau \right) =-\frac{Det\left( A_{bb,AB}\left( D_{Lorentz\,\,inverse} \right) \right)}{Det\left( D_{Lorentz\,\,inverse} \right)},
$$
$$
G_{\psi b,AB}^{A}\left( \tau \right) =\frac{Det\left( A_{\psi b,AB}^{A}\left( D_{Lorentz} \right) \right)}{Det\left( D_{Lorentz} \right)},
$$
$$
G_{b\psi ,AB}^{S}\left( \tau \right) =-\frac{Det\left( A_{b\psi ,AB}^{S}\left( D_{Lorentz\,\,inverse} \right) \right)}{Det\left( D_{Lorentz\,\,inverse} \right)},
$$
$$
G_{\psi b,AB}^{S}\left( \tau \right) =\frac{Det\left( A_{\psi b,AB}^{S}\left( D_{Lorentz\,\,inverse} \right) \right)}{Det\left( D_{Lorentz\,\,inverse} \right)},
$$
$$
G_{b\psi ,AB}^{A}\left( \tau \right) =-\frac{Det\left( A_{b\psi ,AB}^{A}\left( D_{Lorentz} \right) \right)}{Det\left( D_{Lorentz} \right)}.
\end{equation}

\begin{figure}[!t]
\begin{minipage}{0.48\linewidth}
\centerline{\includegraphics[width=8cm]{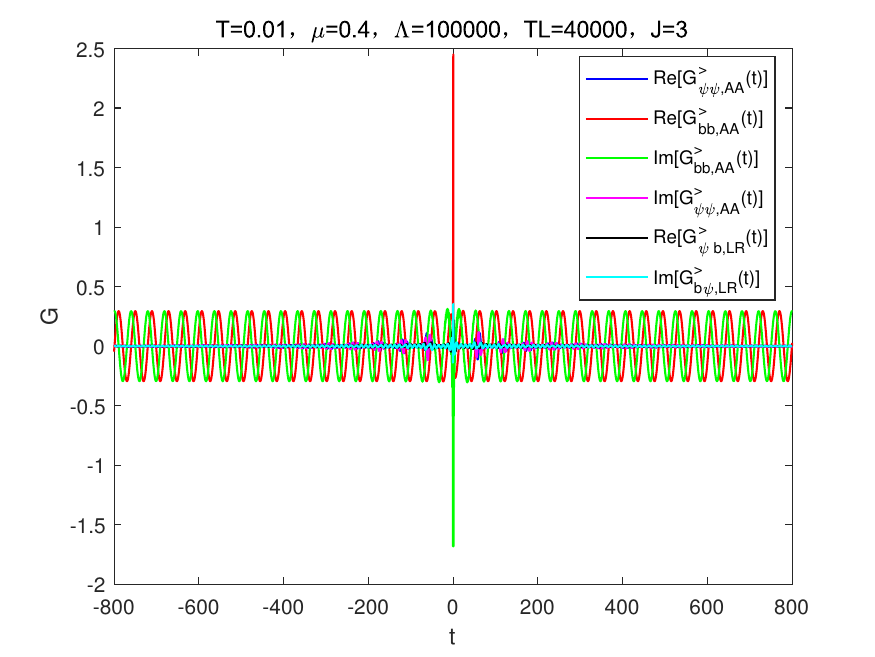}}
\centerline{(a)}
\end{minipage}
\hfill
\begin{minipage}{0.48\linewidth}
\centerline{\includegraphics[width=8cm]{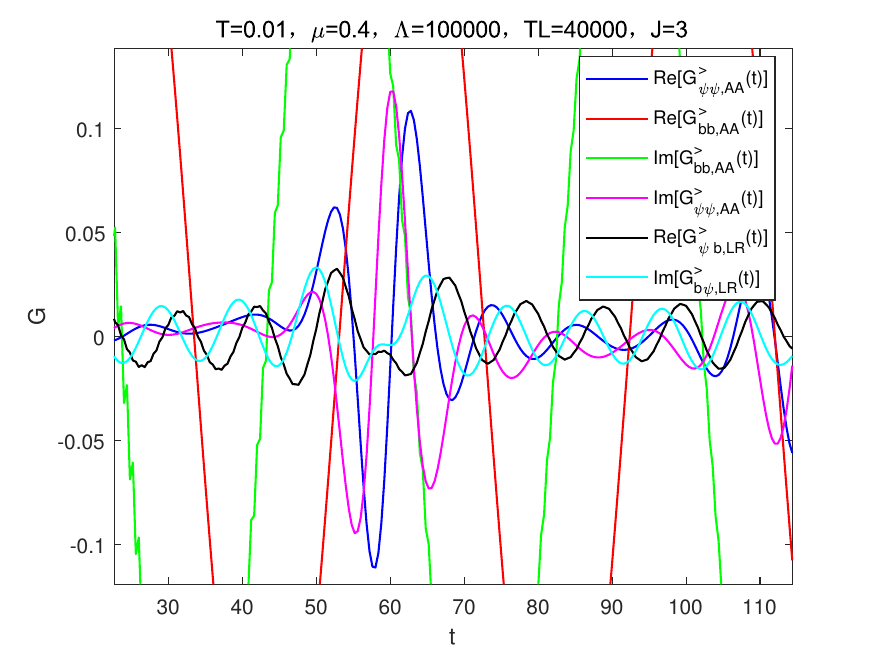}}
\centerline{(b)}
\end{minipage}
\vfill
\begin{minipage}{0.48\linewidth}
\centerline{\includegraphics[width=8cm]{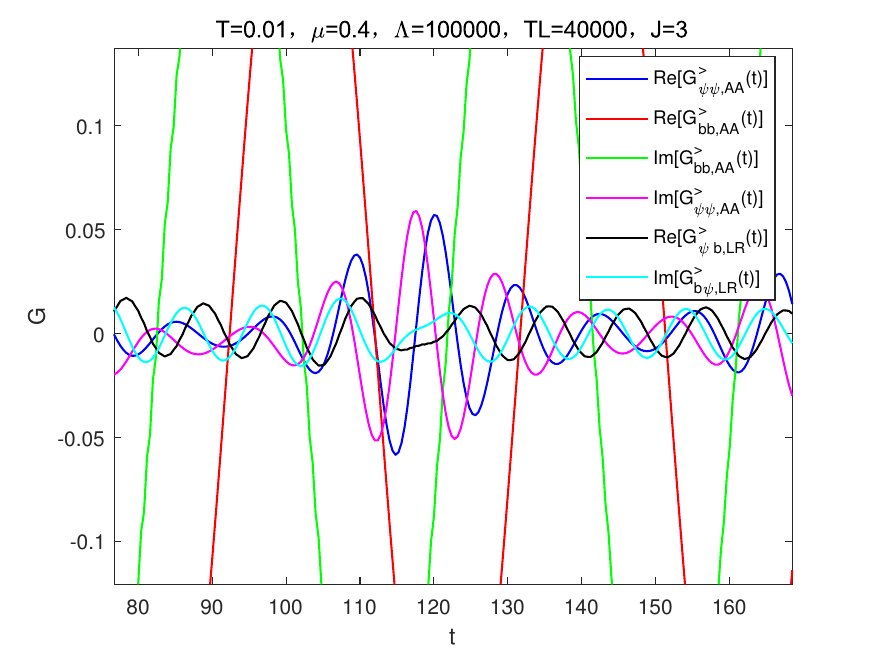}}
\centerline{(c)}
\end{minipage}
\hfill
\begin{minipage}{0.48\linewidth}
\centerline{\includegraphics[width=8cm]{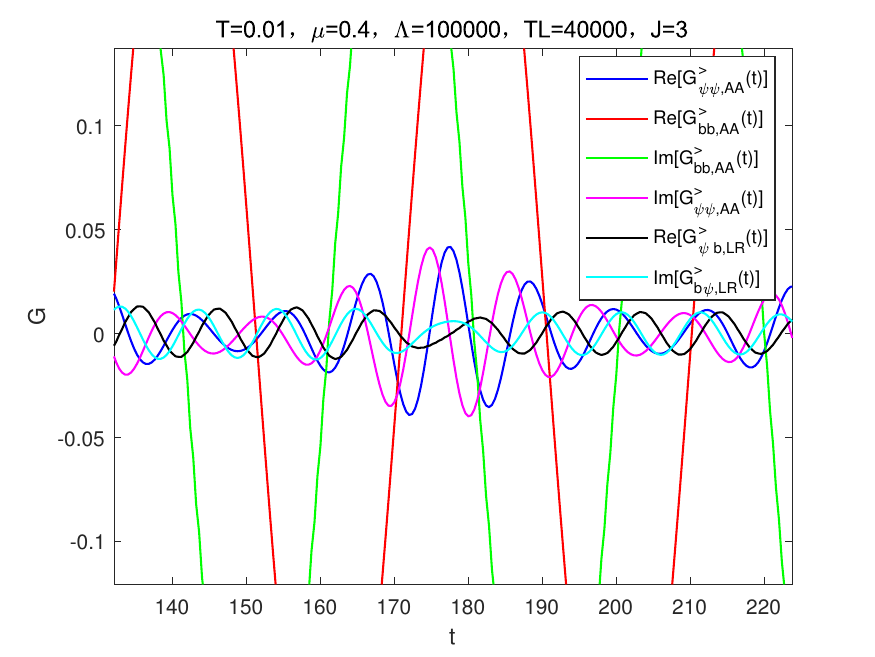}}
\centerline{(d)}
\end{minipage}
\caption{\label{Figure:figure 5} Package solution of Real-time Green's function fixed $J=3,q=5,\mu=0.4,T=0.01$ (a) 1st package  (b) 2nd package (c) 3rd package (d) 4th package}
\end{figure}
Here, we naturally ignore the excitation at initial time because the numerical approach around point 0 could lead to a divergence in the bosonic Retarded correlator. This implies that we must select the Lorentz time such that $t\in \left( -t_{\max},-1 \right) \cup \left( 1,t_{\max} \right)$. Since the frequency-dependent Green's functions encompass all the information within this system, omitting the zero point will not affect the system's properties, similar to how the Matsubara method was applied in previous works.
We have the wave package solution in numerical limit $\lambda_{max}=100000$ and long-time cutoff at $t_{max}=40000$ which has been plotted in Figure 5. This package fermionic solution gradually returns to 0, allowing us to employ the numerical long-time cutoff technique. One notable feature is that the bosonic component does not return to 0 in the long-time limit. However, it approaches a harmonic solution and becomes analytical, which we consider reliable when the number of oscillations is sufficiently large.

We have plotted the packages of the solution in Figure 5. When we introduce an excitation at the initial Lorentz time, this signal information will periodically recovers. Additionally, the excitation decays after recovery, exhibiting behavior similar to the dispersion of packages. The first revival excitation in Figure 5(a) occurs around Lorentz time $t_{1}=58$, and the numerical Green's function results between peaks are $\left| \max \left( \mathrm{Re}\left( G_{\psi \psi} \right) \right) -\min \left( \mathrm{Re}\left( G_{\psi \psi} \right) \right) \right|=0.219$ and $\left| \max \left( \mathrm{Im}\left( G_{\psi \psi} \right) \right) -\min \left( \mathrm{Im}\left( G_{\psi \psi} \right) \right) \right|=0.212$. The second revival excitation occurs around Lorentz time $t_{2}=119$ in Figure 5(b), and we have $\left| \max \left( \mathrm{Re}\left( G_{\psi \psi} \right) \right) -\min \left( \mathrm{Re}\left( G_{\psi \psi} \right) \right) \right|=0.115$ and $\left| \max \left( \mathrm{Im}\left( G_{\psi \psi} \right) \right) -\min \left( \mathrm{Im}\left( G_{\psi \psi} \right) \right) \right|=0.110$. The third revival excitation occurs around Lorentz time $t_{3}=178$, and $\left| \max \left( \mathrm{Re}\left( G_{\psi \psi} \right) \right) -\min \left( \mathrm{Re}\left( G_{\psi \psi} \right) \right) \right|=0.080$, $\left| \max \left( \mathrm{Im}\left( G_{\psi \psi} \right) \right) -\min \left( \mathrm{Im}\left( G_{\psi \psi} \right) \right) \right|=0.081$. There are also several othe revival excitation $t_{4}=234,t_{5}=292,t_{6}=349$ from Figure 5(a), and the peaks of the Green's functions decay with time.

For N=4 SYK model, the real-time processes do not work effectively. We introduce a long-time cutoff following the periodic condition from $0\sim \beta $. However, the bosonic ansatz $G_{\phi \phi}\sim 0.5\left| t \right|$ exhibits long-time divergence, rendering this saddle method unreliable directly. Consequently, we will further approximate this problem using the fully diagonal method in the next subsection.

\subsection{Out-of-time-ordered correlator}
In this subsection, we investigate the causal and chaotic behaviors using the exact diagonal method and out-of-time-ordered correlator, which allows us to examine these characteristics in a finite system. Here, we primarily consider the correlation between different sectors with a non-zero Hamiltonian.

First, as in the previous section, we replace the fermions in the supercharges with creation and annihilation operators
$$
\psi _i=C_{i}^{\dagger}+C_i.
$$
These operators are represented by $2\times2$ Pauli matrices. For the case of SYK model with $N$ fermions on single side, we can represent the theory using a $4^{N}\times4^{N}$ Fock matrix.

Here we define the regularized out-of-time-ordered correlator. In order to compare the results with previous work \cite{53,54}, we can shift the multi-side OTOCs by 1-F(t) and invert the phase factors
\begin{equation}
F\left( t \right) =1-\sum_{ij} \frac{Z\left( \beta \right) Tr\left( \rho W\left( t \right) \rho V\left( 0 \right) \rho W\left( t \right) \rho V\left( 0 \right) \right)}{Tr\left( \rho ^2W\rho ^2W \right) Tr\left( \rho ^2V\rho ^2V \right)},
\end{equation}
where the partition function $Z\left( \beta \right) = Tr\left( \exp \left( -\beta H \right) \right)$ and the density of the finite temperature propagator $\rho = \exp \left( -\beta H/4 \right)$. Here the operators $W$ and $V$ comprise the fermions, and $W\left( t \right)$ represents the time-evolution of $W$.

In N=1 theory, we can then define the fermionic operator.
\begin{equation}
W\left( t \right) =\exp \left( iH^{\dagger}t \right) \left( C_{i}^{L\dagger}+C_{i}^{L} \right) \exp \left( -iHt \right) ,
$$
$$
V\left( 0 \right) =C_{j}^{R\dagger}+C_{j}^{R}.
\end{equation}
and it can be treated as a special form of N=0 theory.

Analyzing the off-diagonal term in the Hamiltonian, we therefore define the effective N=2 operators.
\begin{equation}
W\left( t \right) =\exp \left( iH^{\dagger}t \right) \left( C_{i}^{L\dagger}+C_{i}^{L} \right) \exp \left( -iHt \right) ,
$$
$$
V\left( 0 \right) =\bar{b}_{j}^{R}=\left\{ Q^R,\bar{C}_{j}^{R\dagger}+\bar{C}_{j}^{R} \right\} =\left\{ Q^R,\bar{c}_{j}^{R\dagger}+\bar{c}_{j}^{R} \right\} .
\end{equation}
Here, we can simplify the chiral representation with the components $C_{i}^{L}$ and $\bar{C}_{i}^{R\dagger}$, or alternatively $C_{i}^{L\dagger}$ and $\bar{C}_{i}^{R}$.

For N=4 theory, the non-zero terms in the Hamiltonian are
\begin{equation}
W\left( t \right) =\exp \left( iH^{\dagger}t \right) \left( \phi _{i}^{L} \right) \exp \left( -iHt \right) =\exp \left( iH^{\dagger}t \right) \left( c^{\dagger}+c \right) _{i}^{\alpha L}\left( c^{\dagger}+c \right) _{i}^{\beta L}\exp \left( -iHt \right) ,
$$
$$
V\left( 0 \right) =\bar{F}^{Rj}=\bar{b}_{j}^{\alpha R}\bar{b}_{j}^{\beta R}=\left\{ Q,\bar{c}^{\dagger}+\bar{c} \right\} _{j}^{\alpha L}\left\{ Q,\bar{c}^{\dagger}+\bar{c} \right\} _{j}^{\beta R}.
\end{equation}
Here we have plotted the early-time behaviors of the OTOCs in N=2 and N=4 theories with 3 identical fermions in single-side supercharges. As shown in Figure 6, we calculate the average of 60 implementations for OTOCs. Our results show that the early-time OTOCs with larger coupling $J$ and inverse temperature $\beta$ decay much slower, as indicated by the slopes. It is also suggested that a smaller Lyapunov exponent should imply larger coupling and lower temperature.

\begin{figure}[!t]
\begin{minipage}{0.48\linewidth}
\centerline{\includegraphics[width=8cm]{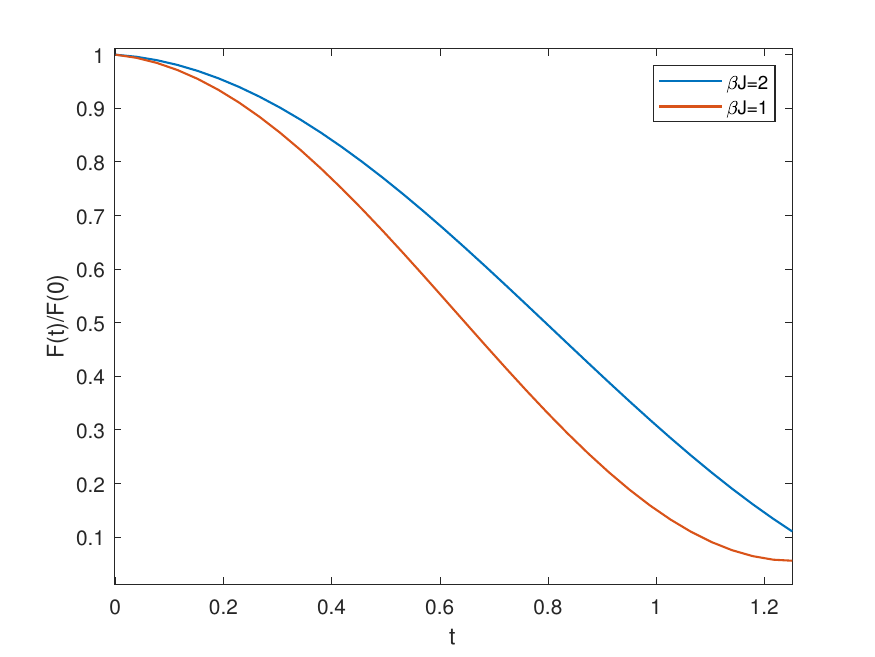}}
\centerline{(a)}
\end{minipage}
\hfill
\begin{minipage}{0.48\linewidth}
\centerline{\includegraphics[width=8cm]{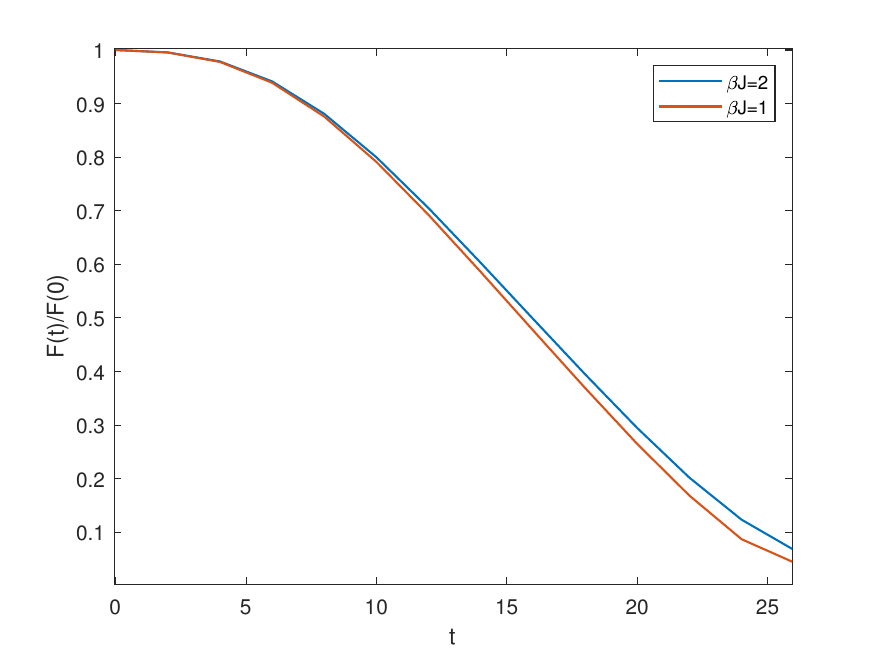}}
\centerline{(b)}
\end{minipage}
\caption{\label{Figure:figure 6} Averaged early-time OTOCs with 60 implement $\mu=0.15$ (a) multi-side OTOC between N=2 algebra $\psi_{L}$ and $\bar{b}_{R}$, (b) multi-side OTOC between N=4 algebra $\phi_{L}$ and $F_{R}$,}
\end{figure}

\section{Conclusion}

In this work, we have investigated the interaction terms between two identical supersymmetric SYK models.

First, we consider a special form of the N=4 SYK model that involves a supermultiplet. This model is generated from two identical N=2 SYK models. It is constrained by particle conservation, as represented by Grassmann integration, which introduces additional auxiliary bosons. Next, we introduce first-order interaction terms in the N=1,2,4 SYK models using superfields, ensuring that the boundary supersymmetry and solvability are maintained. In the thermal limit, we evaluate Green's function outside of superspace. Subsequently, we analyze both the free energy and the energy aspects. The N=1 and N=2 theories exhibit a Hawking-Page-like phase transition, similar to what is observed in the N=0 SYK model. However, unlike the N=0,1,2 SYK models, the N=4 SYK model is primarily bosonic. Its phase structure does not involve a wormhole-black hole transition.

The N=1 and N=2 effective actions in the low-energy limit are reparameterized as superconformal and transition back to the super Schwarzian form. They revert to the usual SYK when the supersymmetric part is ignored. The N=4 effective theory is also derived from two identical N=2 theories and exhibits superconformality under certain conditions. This work also discusses the holographic picture of supersymmetric SYK. We examine supersymmetric JT gravity, a natural consequence of supersymmetrized NAdS. A notable feature is the significant contribution of component $\phi$ in the IR limit. This contribution causes supersymmetric black holes in N=4 to undergo an evaporation process.

We also elaborated on these coupled N=1,2,4 SYK models concerning their entanglement properties with thermal field double. By definition, these supersymmetric TFD states can be derived from the N=0 case. We have verified that the interactive models are maximally entangled, a property that remains intact even after introducing supersymmetry to the SYK model. Additionally, we investigate the causality associated with possible wormholes. In N=2 theory, we study the real-time revival dynamics. We calculate the exact diagonal elements for both N=2 and N=4 theories and examine the out-of-time-ordered correlator.

\section*{Acknowledgements}
This work is supported by NSFC China(Grants No.12275166, No.11805117 and No.11875184)

\appendix
\section*{Appendix A: Calculation of N=2,4 thermalfield double state}
\addcontentsline{toc}{section}{Appendix A: Calculation of N=2,4 thermalfield double state}
\setcounter{equation}{0}
\renewcommand\theequation{A.\arabic{equation}}
N=2 TFD states are similar to N=1 and N=0 states. Due to the introduction of two independent supercharges with only one Hamiltonian constraint, the ground states are no longer parallel vacuum states, but these states are determined by the chiral anticommutation of the supercharges. We can still define maximally entangled states and the TFD states.
\begin{equation}
\left| TFD_{\beta} \right> =\frac{1}{\sqrt{Z_{\beta}}}\sum_n{\exp \left( -\beta \left\{ Q,\bar{Q} \right\} _n/2 \right) \left| n \right> _1}\otimes \left| \hat{n} \right> _2,
\end{equation}
the states exhibiting a dual nature are relevant to chiral supercharges with maximal entanglement
\begin{equation}
\left| n \right> _1=\left( \left| Q \right> _1\left| \bar{Q} \right> _1 \right) _n=\left( \left| \psi +i\bar{\psi} \right> _1\left| \psi -i\bar{\psi} \right> _1 \right) _n,
$$

$$
\left| \hat{n} \right> _2=\left( e^{\frac{-i\eta \pi \varGamma}{4}}e^{i\left( q+\bar{q} \right) \left( \phi-\frac{\pi}{2} \right)} \right) P\left( \left| \psi +i\bar{\psi} \right> _2\left| \psi -i\bar{\psi} \right> _2 \right),
$$
$$
\varGamma =\left( -1 \right) ^{-q-\bar{q}+\frac{N}{2}},
\end{equation}
The equations satisfy the anti-commutation relations, allowing us to construct the corresponding Hilbert space for a specific type of entanglement.
\begin{equation}
\left\{ \psi ^i+i\bar{\psi}_i,\psi ^j+i\bar{\psi}_j \right\} =i\left( \delta _{j}^{i}+\delta _{i}^{j} \right) ,
$$
$$
\left\{ \psi ^i-i\bar{\psi}_i,\psi ^j-i\bar{\psi}_j \right\} =-i\left( \delta _{j}^{i}+\delta _{i}^{j} \right)
$$
$$
\left\{ \psi ^i-i\bar{\psi}_i,\psi ^j+i\bar{\psi}_j \right\} =i\left( \delta _{j}^{i}-\delta _{i}^{j} \right)=0,
$$
$$
\left\{ \psi ^i+i\bar{\psi}_i,\psi ^j-i\bar{\psi}_j \right\} =-i\left( \delta _{j}^{i}-\delta _{i}^{j} \right)=0.
\end{equation}
The algebra can be expressed in terms of creation and annihilation operators
\begin{equation}
\psi ^i=c^{i\dagger}+c^i
$$
$$
\bar{\psi}_i=\bar{c}_{i\dagger}+\bar{c}_i
$$
$$
C^{\dagger}=c ^{i\dagger}+i\bar{c}_i,
$$
$$
C=c ^i+i\bar{c}_{i}^{\dagger},
$$
$$
\bar{C}=c ^{i\dagger}-i\bar{c}_i,
$$
$$
\bar{C}^{\dagger}=c ^i-i\bar{c}_{i}^{\dagger},
\end{equation}
we can further simplify the algebra using the N=2 anti-commutation relations
\begin{equation}
\psi ^i=c^{i\dagger}
$$
$$
\bar{\psi}_i=c_i
$$
$$
C^{\dagger}=\left( 1+i \right) c^{i\dagger},
$$
$$
C=0,
$$
$$
\bar{C}^{\dagger}=0,
$$
$$
\bar{C}=\left( 1-i \right) c^i,
\end{equation}
This can be generated from the symmetry breaking of N=1. This represents a simple form of supersymmetric algebra. We can also utilize the chiral conjugate by removing $C^{\dagger}$ and $\bar{C}$.

\begin{equation}
\left< \hat{n}_{-q^{'}} \right|C_{i}^{R}\left| n_{q^{'''}}^{'} \right> =-e^{i\pi \left( -q-\bar{q}+\frac{N}{2} \right)}e^{-i\phi}\left< m_{q^{'}} \right|C_{i}^{L}\left| n_{q^{'}} \right> ,
$$
$$
\left< \hat{n}_{-q^{'}} \right|\bar{C}_{i}^{R}\left| n_{q^{'''}}^{'} \right> =-e^{i\pi \left( -q-\bar{q}+\frac{N}{2} \right)}e^{-i\phi}\left< m_{q^{'}} \right|\bar{C}_{i}^{L}\left| n_{q^{'}} \right> ,
$$
$$
P^{-1}C_iP=\eta P_{i}^{\dagger}.
\end{equation}
recall the following
\begin{equation}
\varPsi ^i=\psi ^i+\theta \left\{ \bar{Q},\psi ^i \right\} ,
$$
$$
\bar{\varPsi}_i=\bar{\psi}_i+\bar{\theta}\left\{ Q,\bar{\psi}_i \right\} .
\end{equation}
We can rewrite the interaction terms involving complex chiral fields
\begin{equation}
H_{int}=i\mu \partial _{\theta}\left( \varPsi ^{Li}\bar{\varPsi}_{i}^{R}- \varPsi ^{Ri}\bar{\varPsi}_{i}^{L} \right),
$$
and
$$
\bar{H}_{int}=i\mu \partial _{\bar{\theta}}\left( \bar{\varPsi}_{i}^{L}\varPsi ^{Ri}-\bar{\varPsi}_{i}^{R}\varPsi ^{Li} \right) ,
\end{equation}
In N=2 superspace, the fermion algebra involves the anti-commutation relations between the creation and annihilation operators $D$ and $\bar{D}$
\begin{equation}
\begin{split}
H_{int}&=i\mu e^{-i\phi}\partial _{\theta}\left( \psi ^i+\theta \left\{ \bar{Q},\psi ^i \right\} \right) ^L\left( \bar{\psi}_i+\bar{\theta}\left\{ Q,\bar{\psi}_i \right\} \right) ^R-i\mu e^{i\phi}\partial _{\theta}\left( \psi ^i+\theta \left\{ \bar{Q},\psi ^i \right\} \right) ^R\left( \bar{\psi}_i+\bar{\theta}\left\{ Q,\bar{\psi}_i \right\} \right) ^L,
\\
&=i\mu e^{-i\phi}\left\{ \bar{Q},\psi ^i \right\} ^L\bar{\psi}_{i}^{R}-i\mu e^{i\phi}\left\{ \bar{Q},\psi ^i \right\} ^R\bar{\psi}_{i}^{L}
\\
&=i\mu e^{-i\phi}\left\{ \bar{Q},\left( c^{i\dagger}+c^i \right) \right\} ^L\left( \bar{c}_{i\dagger}+\bar{c}_i \right) ^R-i\mu e^{i\phi}\left\{ \bar{Q},\left( c^{i\dagger}+c^i \right) \right\} ^R\left( \bar{c}_{i\dagger}+\bar{c}_i \right) ^L
\\
&=i\mu e^{-i\phi}\left\{ \bar{Q},\frac{\left( C^{i\dagger}+\bar{C}^{i\dagger}+C^i+\bar{C}^i \right)}{2} \right\} ^L\left( \frac{C^{i\dagger}-\bar{C}^{i\dagger}+C^i-\bar{C}^i}{2i} \right) ^R
\\
&-i\mu e^{i\phi}\left\{ \bar{Q},\frac{\left( C^{i\dagger}+\bar{C}^{i\dagger}+C^i+\bar{C}^i \right)}{2} \right\} ^R\left( \frac{C^{i\dagger}-\bar{C}^{i\dagger}+C^i-\bar{C}^i}{2i} \right) ^L
\\
\bar{H}_{int}&=i\mu e^{-i\phi}\partial _{\bar{\theta}}\left( \bar{\psi}_i+\bar{\theta}\left\{ Q,\bar{\psi}_i \right\} \right) ^L\left( \psi ^i+\theta \left\{ \bar{Q},\psi ^i \right\} \right) ^R-i\mu e^{i\phi}\partial _{\bar{\theta}}\left( \bar{\psi}_i+\bar{\theta}\left\{ Q,\bar{\psi}_i \right\} \right) ^R\left( \psi ^i+\theta \left\{ \bar{Q},\psi ^i \right\} \right) ^L.
\\
&=i\mu e^{-i\phi}\left\{ Q,\bar{\psi}_i \right\} ^L\psi ^{iR}-i\mu \left\{ Q,\bar{\psi}_i \right\} ^R\psi ^{iL}
\\
&=i\mu e^{-i\phi}\left\{ Q,\left( \bar{c}_{i\dagger}+\bar{c}_i \right) \right\} ^L\left( c^{i\dagger}+c^i \right) ^R-i\mu e^{i\phi}\left\{ Q,\left( \bar{c}_{i\dagger}+\bar{c}_i \right) \right\} ^R\left( c^{i\dagger}+c^i \right) ^L
\\
&=i\mu e^{-i\phi}\left\{ Q,\frac{C^{i\dagger}-\bar{C}^{i\dagger}+C^i-\bar{C}^i}{2i} \right\} ^L\left( \frac{\left( C^{i\dagger}+\bar{C}^{i\dagger}+C^i+\bar{C}^i \right)}{2} \right) ^R
\\
&-i\mu e^{i\phi}\left\{ Q,\frac{C^{i\dagger}-\bar{C}^{i\dagger}+C^i-\bar{C}^i}{2i} \right\} ^R\left( \frac{\left( C^{i\dagger}+\bar{C}^{i\dagger}+C^i+\bar{C}^i \right)}{2} \right) ^L
\end{split}
\end{equation}
We can still simplify them into an algebraic form
\begin{equation}
\begin{split}
H_{int}&=i\mu e^{-i\phi}\left\{ \bar{Q},\frac{C^{i\dagger}}{2} \right\} ^L\left( \frac{C^i}{2i} \right) ^R-i\mu e^{-i\phi}\left\{ \bar{Q},\frac{\bar{C}^{i\dagger}}{2} \right\} ^L\left( \frac{\bar{C}^i}{2i} \right) ^R
\\
&-i\mu e^{i\phi}\left\{ \bar{Q},\frac{C^{i\dagger}}{2} \right\} ^R\left( \frac{C^i}{2i} \right) ^L+i\mu e^{i\phi}\left\{ \bar{Q},\frac{\bar{C}^{i\dagger}}{2} \right\} ^R\left( \frac{\bar{C}^i}{2i} \right) ^L
\\
\bar{H}_{int}&=i\mu e^{-i\phi}\left\{ Q,\frac{C^{i\dagger}}{2i} \right\} ^L\left( \frac{C^i}{2} \right) ^R-i\mu e^{-i\phi}\left\{ Q,\frac{\bar{C}^{i\dagger}}{2i} \right\} ^L\left( \frac{\bar{C}^i}{2} \right) ^R
\\
&-i\mu e^{i\phi}\left\{ Q,\frac{C^{i\dagger}}{2i} \right\} ^R\left( \frac{C^i}{2} \right) ^L+i\mu e^{i\phi}\left\{ Q,\frac{\bar{C}^{i\dagger}}{2i} \right\} ^R\left( \frac{\bar{C}^i}{2} \right) ^L
\end{split}
\end{equation}
N=2 Hamiltonian eigenstates are created by applying the appropriate creation and annihilation operators
\begin{equation}
H\left| \varPsi _{m}^{r},\bar{\varPsi}_{n}^{r} \right> =\left\{ Q,\bar{Q} \right\} \left| \varPsi _{m}^{r},\bar{\varPsi}_{n}^{r} \right> =E_{mn}\left| \varPsi _{m}^{r},\bar{\varPsi}_{n}^{r} \right> ,
$$
$$
H^{\dagger}\left| \varPsi _{m}^{l},\bar{\varPsi}_{n}^{l} \right> =\left\{ Q,\bar{Q} \right\} ^{\dagger}\left| \varPsi _{m}^{r},\bar{\varPsi}_{n}^{r} \right> =E_{mn}^{*}\left| \varPsi _{m}^{l},\bar{\varPsi}_{n}^{l} \right> ,
\end{equation}
with
\begin{equation}
\sum_{n,k}{\left| \varPsi _{n}^{l},\bar{\varPsi}_{k}^{l} \right>}\left< \varPsi _{n}^{r},\bar{\varPsi}_{k}^{r} \right|=I,
$$
$$
\left< \varPsi _{n}^{l},\bar{\varPsi}_{k}^{l} \right|\left. \varPsi _{m}^{r},\bar{\varPsi}_{l}^{r} \right> =\delta _{nm}\delta _{kl}.
\end{equation}
And we can further propose the ground states
\begin{equation}
\begin{split}
\left| \varPsi _0,\bar{\varPsi}_0 \right> &=\prod_j^N{\left( \frac{1}{\sqrt{2}}\left| 1 \right> _{L,j}\left| 0 \right> _{R,j}+\frac{i}{\sqrt{2}}\left| 0 \right> _{L,j}\left| 1 \right> _{R,j} \right)}
\\
&=\prod_j^N{\sum_{abcd}{\left( \frac{1}{\sqrt{2}}\left| a\pm 1,a \right> _{L,j}\left| b,b \right> _{R,j}+\frac{i}{\sqrt{2}}\left| c,c \right> _{L,j}\left| d\pm 1,d \right> _{R,j} \right)}},
\end{split}
\end{equation}
with
\begin{equation}
H_{int}^{\dagger}\left| \varPsi _{0}^{l},\bar{\varPsi}_{0}^{l} \right> =-\mu N\left( 1+\theta \bar{Q}_0 \right) \left( 1+\bar{\theta}Q_0 \right) \left| \varPsi _{0}^{l},\bar{\varPsi}_{0}^{l} \right> ,
$$
$$
H_{int}\left| \varPsi _{0}^{r},\bar{\varPsi}_{0}^{r} \right> =-\mu N\left( 1+\theta \bar{Q}_0 \right) \left( 1+\bar{\theta}Q_0 \right) \left| \varPsi _{0}^{r},\bar{\varPsi}_{0}^{r} \right> .
\end{equation}
The $Q_0$ is a constant which is relevant to the ground state supercharges. We use $a,b,c,d$ to represent the excited states with $a,b,c,d$ energy levels in each supercharge $Q_L,\bar{Q_L}$ and $Q_R,\bar{Q_R}$. In the N=2 theory in vacuum, we denote $Q=\bar{Q}$. The excitation of $Q$ is higher than $\bar{Q}$ by one, these are the first excited states and vice versa.

We have imposed the entangled states in the N=2 TFD
\begin{equation}
\left| m_q \right> _1\otimes \left| \bar{m}_{-q} \right> _2=\prod_j^N{\sum_q{\sum_{kl}{\left( \left| k_q\pm m_q,k_q \right> _{1,j}\overline{\left| l_{-q}\pm m_{-q},l_{-q} \right> }_{2,j} \right)}}}
\end{equation}
Indicate that
\begin{equation}
\begin{split}
&\left< TFD_{\beta =0} \right|H_{int}\left| TFD_{\beta =0} \right> =\frac{1}{2^N}\sum_{qq^{'}}{\sum_{mn}{\left< \hat{n}_{-q^{'}} \right|_2\otimes \left< n_{q^{'}} \right|_1\left( i\mu \sum_i{\left( e^{-i\phi}\left\{ \bar{Q},\frac{C^{i\dagger}}{2} \right\} ^L\left( \frac{C^i}{2i} \right) ^R \right.} \right.}}
\\
&\left. \left. -i\mu e^{-i\phi}\left\{ \bar{Q},\frac{\bar{C}^{i\dagger}}{2} \right\} ^L\left( \frac{\bar{C}^i}{2i} \right) ^R-i\mu e^{i\phi}\left\{ \bar{Q},\frac{C^{i\dagger}}{2} \right\} ^R\left( \frac{C^i}{2i} \right) ^L+i\mu e^{i\phi}\left\{ \bar{Q},\frac{\bar{C}^{i\dagger}}{2} \right\} ^R\left( \frac{\bar{C}^i}{2i} \right) ^L \right) \right) \left| m_q \right> _1\otimes \left| \hat{m}_{-q} \right> _2
\\
&=\frac{1}{2^N}\sum_{qq^{'}}{\sum_{nm}{\sum_i{\left( i\mu \right.}}}e^{i\pi \left( -q-\bar{q}+\frac{N}{2} \right)}\left< n_{q^{'}} \right|\left\{ \bar{Q},\frac{C^{i\dagger}}{2} \right\} ^L\left| m_{q^{''}}^{'} \right> \left< \hat{n}_{-q^{'}} \right|\left( \frac{C^i}{2i} \right) ^R\left| \hat{m}_{-q} \right>
\\
&-\left< n_{q^{'}} \right|\left\{ \bar{Q},\frac{\bar{C}^{i\dagger}}{2} \right\} ^L\left| m_{q^{''}}^{'} \right> \left< \hat{n}_{-q^{'}} \right|\left( \frac{\bar{C}^i}{2i} \right) ^R\left| \hat{m}_{-q} \right> -i\mu e^{i\pi \left( -q^{'}-\bar{q}^{'}+\frac{N}{2} \right)}e^{i\phi}\left< \hat{n}_{-q^{'}} \right|\left\{ \bar{Q},\frac{C^{i\dagger}}{2} \right\} ^R\left| n_{q^{'''}}^{'} \right>
\\
&\left. \left< n_{q^{'}} \right|\left( \frac{C^i}{2i} \right) ^L\left| m_q \right>+i\mu e^{i\pi \left( -q^{'}-\bar{q}^{'}+\frac{N}{2} \right)}e^{i\phi}\left< \hat{n}_{-q^{'}} \right|\left\{ \bar{Q},\frac{\bar{C}^{i\dagger}}{2} \right\} ^R\left| n_{q^{'''}}^{'} \right> \left< n_{q^{'}} \right|\left( \frac{\bar{C}^i}{2i} \right) ^L\left| m_q \right> \right)
\end{split}
\end{equation}

and
\begin{equation}
\begin{split}
\left< H_{int} \right> &=\frac{1}{2^N}\sum_{qq^{'}}{\sum_{nm}{\sum_i{\left\{ -i\mu \right.}}}e^{-i\phi}\left< n_{q^{'}} \right|\left\{ \bar{Q},\frac{C^{i\dagger}}{2} \right\} ^L\left| m_q \right> \left< m_q \right|\left( \frac{C^i}{2i} \right) ^L\left| n_{q^{'}} \right> +i\mu e^{-i\phi}\left< n_{q^{'}} \right|\left\{ \bar{Q},\frac{\bar{C}^{i\dagger}}{2} \right\} ^L\left| m_q \right>
\\
&\left< m_q \right|\left( \frac{\bar{C}^i}{2i} \right) ^L\left| n_{q^{'}} \right> +i\mu e^{i\phi}\left< m_q \right|\left\{ \bar{Q},\frac{C^{i\dagger}}{2} \right\} ^L\left| n_{q^{'}} \right> \left< n_{q^{'}} \right|\left( \frac{C^i}{2i} \right) ^L\left| m_q \right>
\\
&-i\mu e^{i\phi}\left< m_q \right|\left\{ \bar{Q},\frac{\bar{C}^{i\dagger}}{2} \right\} ^L\left| n_{q^{'}} \right> \left< n_{q^{'}} \right|\left( \frac{\bar{C}^i}{2i} \right) ^L\left| m_q \right>
\\
&=\sum_{qq^{'}}{\sum_{nm}{\sum_i{i\mu}}}/2^N\left< n_{q^{'}} \right|\left\{ \bar{Q},\frac{C^{i\dagger}}{2} \right\} ^L\left| m_q \right> \left< m_q \right|\left( \frac{C^i}{2i} \right) ^L\left| n_{q^{'}} \right> -\left< n_{q^{'}} \right|\left\{ \bar{Q},\frac{\bar{C}^{i\dagger}}{2} \right\} ^L\left| m_q \right> \left< m_q \right|\left( \frac{\bar{C}^i}{2i} \right) ^L\left| n_{q^{'}} \right>
\\
&=-\frac{\mu}{2^N}\sum_{n,q}{\sum_i{\left< n_q \right|_1}}\left\{ \bar{Q},\frac{C^{i\dagger}}{2} \right\} ^L\left( \frac{C^i}{2i} \right) ^L-\left\{ \bar{Q},\frac{\bar{C}^{i\dagger}}{2} \right\} ^L\left( \frac{\bar{C}^i}{2i} \right) ^L\left| n_q \right> _1=Const.
\end{split}
\end{equation}

The Hamiltonian of N=4 SYK model is
\begin{equation}
\begin{split}
H&=\epsilon _{\alpha \beta}\psi ^{\alpha}\psi ^{\beta}\left\{ \bar{Q}_{\alpha},\psi ^{\alpha} \right\} ^2 \epsilon _{\alpha \beta}\psi ^{\alpha}\psi ^{\beta}+\epsilon _{\alpha \beta}\psi ^{\alpha}\psi ^{\beta}\left\{ \bar{Q}_{\alpha},\psi ^{\alpha} \right\} \psi ^{\alpha}\left\{ \bar{Q}_{\alpha},\psi ^{\alpha} \right\} \psi ^{\beta}\epsilon _{\alpha \beta}+h.c
\\
&=\left\{ \bar{Q}_{\alpha},Q^{\alpha} \right\} \left\{ \bar{Q}_{\beta},Q^{\beta} \right\} +T\left\{ \bar{Q}_{\alpha},Q^{\alpha} \right\} \left\{ \bar{Q}_{\beta},Q^{\beta} \right\} +h.c
\\
&=\left\{ \bar{Q}_{\alpha},Q^{\alpha} \right\} \left\{ \bar{Q}_{\beta},Q^{\beta} \right\} +h.c
\end{split}
\end{equation}

Therefore, the N=4 supersymmetric SYK thermal field double is defined as
\begin{equation}
\left| TFD_{\beta} \right> =\frac{1}{\sqrt{Z_{\beta}}}\sum_n{\exp \left( -\beta \left\{ \bar{Q}_{\alpha},Q^{\alpha} \right\} \left\{ \bar{Q}_{\beta},Q^{\beta} \right\} _n/2 \right) \left| n \right> _1}\otimes \left| \hat{n} \right> _2,
\end{equation}
The Hilbert space also contains the eigenvectors of the supercharges
\begin{equation}
\left| n \right> _1=\left( \left| Q \right> _{\alpha 1}\left| \bar{Q} \right> _{\alpha 1}\left| Q \right> _{\beta 1}\left| \bar{Q} \right> _{\beta 1} \right) _n\sim \left( \left| \psi +i\bar{\psi} \right> _{\alpha 1}\left| \psi -i\bar{\psi} \right> _{\alpha 1}\left| \psi +i\bar{\psi} \right> _{\beta 1}\left| \psi -i\bar{\psi} \right> _{\beta 1} \right) _n,
$$
$$
\left| \hat{n} \right> _2=\left( e^{\frac{-i\eta \pi \varGamma}{4}}e^{i\left( q_{\alpha}+\bar{q}_{\alpha}+q_{\beta}+\bar{q}_{\beta} \right) \left( \phi -\frac{\pi}{2} \right)} \right) P\left( \left| \psi +i\bar{\psi} \right> _{\alpha 2}\left| \psi -i\bar{\psi} \right> _{\alpha 2}\left| \psi +i\bar{\psi} \right> _{\beta 2}\left| \psi -i\bar{\psi} \right> _{\beta 2} \right) ,
$$
$$
\varGamma =\left( -1 \right) ^{-q_{\alpha}-\bar{q}_{\alpha}-q_{\beta}-\bar{q}_{\beta}+\frac{N}{2}},
\end{equation}
Here, we have defined sets of new eigenvectors, which are also eigenvectors of the Hamiltonian. It is straightforward to verify that these new eigenvectors satisfy the properties required for being eigenvectors of the Hamiltonian
\begin{equation}
\left\{ \psi _{\alpha}^{i}+i\bar{\psi}_{\alpha i},\psi _{\beta}^{j}+i\bar{\psi}_{\beta j} \right\} =i\left( \delta _{j}^{i}+\delta _{i}^{j} \right) \delta _{\alpha \beta},
$$
$$
\left\{ \psi _{\alpha}^{i}-i\bar{\psi}_{\alpha i},\psi _{\beta}^{j}-i\bar{\psi}_{\beta j} \right\} =-i\left( \delta _{j}^{i}+\delta _{i}^{j} \right) \delta _{\alpha \beta}
$$
$$
\left\{ \psi _{\alpha}^{i}-i\bar{\psi}_{\alpha i},\psi _{\beta}^{j}+i\bar{\psi}_{\beta j} \right\} =i\left( \delta _{j}^{i}-\delta _{i}^{j} \right) \delta _{\alpha \beta}=0,
$$
$$
\left\{ \psi _{\alpha}^{i}+i\bar{\psi}_{\alpha i},\psi _{\beta}^{j}-i\bar{\psi}_{\beta j} \right\} =-i\left( \delta _{j}^{i}-\delta _{i}^{j} \right) \delta _{\alpha \beta}=0.
\end{equation}
We can also rewrite these operators within the Hilbert space
\begin{equation}
\psi _{\alpha}^{i}=c_{\alpha}^{i\dagger}+c_{\alpha}^{i}
$$
$$
\bar{\psi}_{\alpha i}=\bar{c}_{\alpha i}^{\dagger}+\bar{c}_{\alpha i}
$$
$$
{C_{\alpha}^{i}}^{\dagger}=c_{\alpha}^{i\dagger}+i\bar{c}_{\alpha i},
$$
$$
C_{\alpha i}=c_{\alpha}^{i}+i\bar{c}_{\alpha i}^{\dagger},
$$
$$
{\bar{C}_{\alpha}}^{i\dagger}=c_{\alpha}^{i\dagger}-i\bar{c}_{\alpha i},
$$
$$
\bar{C}_{\alpha}^{i}=c_{\alpha}^{i}-i\bar{c}_{\alpha i}^{\dagger},
\end{equation}
with
\begin{equation}
\left< \hat{n}_{-q^{'}} \right|C_{\alpha i}^{R}\left| n_{q^{'''}}^{'} \right> =-e^{i\pi \left( -q_{\alpha}-\bar{q}_{\alpha}-q_{\beta}-\bar{q}_{\beta}+\frac{N}{2} \right)}e^{-i\phi}\left< m_{q^{'}} \right|C_{\alpha i}^{L}\left| n_{q^{'}} \right> ,
$$
$$
\left< \hat{n}_{-q^{'}} \right|\bar{C}_{\alpha i}^{R}\left| n_{q^{'''}}^{'} \right> =-e^{i\pi \left( -q_{\alpha}-\bar{q}_{\alpha}-q_{\beta}-\bar{q}_{\beta}+\frac{N}{2} \right)}e^{-i\phi}\left< m_{q^{'}} \right|\bar{C}_{\alpha i}^{L}\left| n_{q^{'}} \right> ,
$$
$$
P^{-1}C_iP=\eta P_{i}^{\dagger}.
\end{equation}
We can further introduce the interaction Hamiltonian into superspace
\begin{equation}
\varPhi =\psi ^{\alpha}\psi ^{\beta}+\theta _{\alpha}\psi ^{\beta}\left\{ \bar{Q}_{\alpha},\psi ^{\alpha} \right\} +\theta _{\beta}\psi ^{\alpha}\left\{ \bar{Q}_{\beta},\psi ^{\beta} \right\} +\theta _{\alpha}\theta _{\beta}\left\{ \bar{Q}_{\alpha},\psi ^{\alpha} \right\} \left\{ \bar{Q}_{\beta},\psi ^{\beta} \right\}
$$
$$
\bar{\varPhi}=\bar{\psi}_{\alpha}\bar{\psi}_{\beta}-\bar{\theta}_{\alpha}\bar{\psi}_{\beta}\left\{ Q^{\alpha},\bar{\psi}_{\alpha} \right\} -\bar{\theta}_{\beta}\bar{\psi}_{\alpha}\left\{ Q^{\beta},\bar{\psi}_{\beta} \right\} +\bar{\theta}_{\alpha}\bar{\theta}_{\beta}\left\{ Q^{\alpha},\bar{\psi}_{\alpha} \right\} \left\{ Q^{\beta},\bar{\psi}_{\beta} \right\}
\end{equation}

The interaction Hamiltonian is defined as follow
\begin{equation}
H_{int}=\partial _{\theta}^{2}\left( \varPhi ^{Li}\bar{\varPhi}_{i}^{R}+\varPhi ^{Ri}\bar{\varPhi}_{i}^{L} \right)
$$
$$
\bar{H}_{int}=\partial _{\bar{\theta}}^{2}\left( \bar{\varPhi}_{i}^{L}\varPhi ^{Ri}+\bar{\varPhi}_{i}^{R}\varPhi ^{Li} \right) ,
\end{equation}
and
\begin{equation}
\begin{split}
H_{int}&=e^{-i\phi}\partial _{\theta}^{2}\left( \psi ^{\alpha}\psi ^{\beta}+\theta _{\alpha}\psi ^{\beta}\left\{ \bar{Q}_{\alpha},\psi ^{\alpha} \right\} +\theta _{\beta}\psi ^{\alpha}\left\{ \bar{Q}_{\beta},\psi ^{\beta} \right\} +\theta _{\alpha}\theta _{\beta}\left\{ \bar{Q}_{\alpha},\psi ^{\alpha} \right\} \left\{ \bar{Q}_{\beta},\psi ^{\beta} \right\} \right) ^L
\\
&\left( \bar{\psi}_{\alpha}\bar{\psi}_{\beta}-\bar{\theta}_{\alpha}\bar{\psi}_{\beta}\left\{ Q^{\alpha},\bar{\psi}_{\alpha} \right\} -\bar{\theta}_{\beta}\bar{\psi}_{\alpha}\left\{ Q^{\beta},\bar{\psi}_{\beta} \right\} +\bar{\theta}_{\alpha}\bar{\theta}_{\beta}\left\{ Q^{\alpha},\bar{\psi}_{\alpha} \right\} \left\{ Q^{\beta},\bar{\psi}_{\beta} \right\} \right) ^R
\\
&+e^{i\phi}\partial _{\theta}^{2}\left( \psi ^{\alpha}\psi ^{\beta}+\theta _{\alpha}\psi ^{\beta}\left\{ \bar{Q}_{\alpha},\psi ^{\alpha} \right\} +\theta _{\beta}\psi ^{\alpha}\left\{ \bar{Q}_{\beta},\psi ^{\beta} \right\} +\theta _{\alpha}\theta _{\beta}\left\{ \bar{Q}_{\alpha},\psi ^{\alpha} \right\} \left\{ \bar{Q}_{\beta},\psi ^{\beta} \right\} \right) ^R
\\
&\left( \bar{\psi}_{\alpha}\bar{\psi}_{\beta}-\bar{\theta}_{\alpha}\bar{\psi}_{\beta}\left\{ Q^{\alpha},\bar{\psi}_{\alpha} \right\} -\bar{\theta}_{\beta}\bar{\psi}_{\alpha}\left\{ Q^{\beta},\bar{\psi}_{\beta} \right\} +\bar{\theta}_{\alpha}\bar{\theta}_{\beta}\left\{ Q^{\alpha},\bar{\psi}_{\alpha} \right\} \left\{ Q^{\beta},\bar{\psi}_{\beta} \right\} \right) ^L,
\\
&=i\mu e^{-i\phi}\left( \left\{ \bar{Q}_{\alpha},\psi ^{\alpha} \right\} \left\{ \bar{Q}_{\beta},\psi ^{\beta} \right\} \right) ^L\left( \bar{\psi}_{\alpha}\bar{\psi}_{\beta} \right) ^R+i\mu e^{i\phi}\left( \left\{ \bar{Q}_{\alpha},\psi ^{\alpha} \right\} \left\{ \bar{Q}_{\beta},\psi ^{\beta} \right\} \right) ^R\left( \bar{\psi}_{\alpha}\bar{\psi}_{\beta} \right) ^L
\\
&=i\mu e^{-i\phi}\left( \left\{ \bar{Q}_{\alpha},\left( c_{\alpha}^{i\dagger}+c_{\alpha}^{i} \right) \right\} \left\{ \bar{Q}_{\beta},\left( c_{\beta}^{i\dagger}+c_{\beta}^{i} \right) \right\} \right) ^L\left( \left( \bar{c}_{\alpha i}^{\dagger}+\bar{c}_{\alpha i} \right) \left( \bar{c}_{\beta i}^{\dagger}+\bar{c}_{\beta i} \right) \right) ^R
\\
&+i\mu e^{i\phi}\left( \left\{ \bar{Q}_{\alpha},\left( c_{\alpha}^{i\dagger}+c_{\alpha}^{i} \right) \right\} \left\{ \bar{Q}_{\beta},\left( c_{\beta}^{i\dagger}+c_{\beta}^{i} \right) \right\} \right) ^R\left( \left( \bar{c}_{\alpha i}^{\dagger}+\bar{c}_{\alpha i} \right) \left( \bar{c}_{\beta i}^{\dagger}+\bar{c}_{\beta i} \right) \right) ^L
\\
&=i\mu e^{-i\phi}\left( \left\{ \bar{Q}_{\alpha},\left( \frac{C_{\alpha}^{i\dagger}+\bar{C}_{\alpha}^{i\dagger}+C_{\alpha}^{i}+\bar{C}_{\alpha}^{i}}{2} \right) \right\} \left\{ \bar{Q}_{\beta},\left( \frac{C_{\beta}^{i\dagger}+\bar{C}_{\beta}^{i\dagger}+C_{\beta}^{i}+\bar{C}_{\beta}^{i}}{2} \right) \right\} \right) ^L
\\
&\left( \left( \frac{C_{\alpha}^{i\dagger}-\bar{C}_{\alpha}^{i\dagger}+C_{\alpha}^{i}-\bar{C}_{\alpha}^{i}}{2i} \right) \left( \frac{C_{\beta}^{i\dagger}-\bar{C}_{\beta}^{i\dagger}+C_{\beta}^{i}-\bar{C}_{\beta}^{i}}{2i} \right) \right) ^R
\\
&+i\mu e^{i\phi}\left( \left\{ \bar{Q}_{\alpha},\left( \frac{C_{\alpha}^{i\dagger}+\bar{C}_{\alpha}^{i\dagger}+C_{\alpha}^{i}+\bar{C}_{\alpha}^{i}}{2} \right) \right\} \left\{ \bar{Q}_{\beta},\left( \frac{C_{\beta}^{i\dagger}+\bar{C}_{\beta}^{i\dagger}+C_{\beta}^{i}+\bar{C}_{\beta}^{i}}{2} \right) \right\} \right) ^R
\\
&\left( \left( \frac{C_{\alpha}^{i\dagger}-\bar{C}_{\alpha}^{i\dagger}+C_{\alpha}^{i}-\bar{C}_{\alpha}^{i}}{2i} \right) \left( \frac{C_{\beta}^{i\dagger}-\bar{C}_{\beta}^{i\dagger}+C_{\beta}^{i}-\bar{C}_{\beta}^{i}}{2i} \right) \right) ^L
\end{split}
\end{equation}

And
\begin{equation}
\begin{split}
\bar{H}_{int}&=e^{-i\phi}\partial _{\theta}^{2}\left( \bar{\psi}_{\alpha}\bar{\psi}_{\beta}-\bar{\theta}_{\alpha}\bar{\psi}_{\beta}\left\{ Q^{\alpha},\bar{\psi}_{\alpha} \right\} -\bar{\theta}_{\beta}\bar{\psi}_{\alpha}\left\{ Q^{\beta},\bar{\psi}_{\beta} \right\} +\bar{\theta}_{\alpha}\bar{\theta}_{\beta}\left\{ Q^{\alpha},\bar{\psi}_{\alpha} \right\} \left\{ Q^{\beta},\bar{\psi}_{\beta} \right\} \right) ^L
\\
&\left( \psi ^{\alpha}\psi ^{\beta}+\theta _{\alpha}\psi ^{\beta}\left\{ \bar{Q}_{\alpha},\psi ^{\alpha} \right\} +\theta _{\beta}\psi ^{\alpha}\left\{ \bar{Q}_{\beta},\psi ^{\beta} \right\} +\theta _{\alpha}\theta _{\beta}\left\{ \bar{Q}_{\alpha},\psi ^{\alpha} \right\} \left\{ \bar{Q}_{\beta},\psi ^{\beta} \right\} \right) ^R
\\
&+e^{i\phi}\partial _{\theta}^{2}\left( \bar{\psi}_{\alpha}\bar{\psi}_{\beta}-\bar{\theta}_{\alpha}\bar{\psi}_{\beta}\left\{ Q^{\alpha},\bar{\psi}_{\alpha} \right\} -\bar{\theta}_{\beta}\bar{\psi}_{\alpha}\left\{ Q^{\beta},\bar{\psi}_{\beta} \right\} +\bar{\theta}_{\alpha}\bar{\theta}_{\beta}\left\{ Q^{\alpha},\bar{\psi}_{\alpha} \right\} \left\{ Q^{\beta},\bar{\psi}_{\beta} \right\} \right) ^R
\\
&\left( \psi ^{\alpha}\psi ^{\beta}+\theta _{\alpha}\psi ^{\beta}\left\{ \bar{Q}_{\alpha},\psi ^{\alpha} \right\} +\theta _{\beta}\psi ^{\alpha}\left\{ \bar{Q}_{\beta},\psi ^{\beta} \right\} +\theta _{\alpha}\theta _{\beta}\left\{ \bar{Q}_{\alpha},\psi ^{\alpha} \right\} \left\{ \bar{Q}_{\beta},\psi ^{\beta} \right\} \right) ^L,
\\
&=i\mu e^{-i\phi}\left( \left\{ Q^{\alpha},\bar{\psi}_{\alpha} \right\} _{\alpha}\left\{ Q^{\beta},\bar{\psi}_{\beta} \right\} \right) ^L\left( \psi ^{\alpha}\psi ^{\beta} \right) ^R+i\mu e^{i\phi}\left( \left\{ Q^{\alpha},\bar{\psi}_{\alpha} \right\} \left\{ Q^{\beta},\bar{\psi}_{\beta} \right\} \right) ^R\left( \psi ^{\alpha}\psi ^{\beta} \right) ^L
\\
&=i\mu e^{-i\phi}\left( \left\{ \bar{Q}_{\alpha},\left( \bar{c}_{\alpha i}^{\dagger}+\bar{c}_{\alpha i} \right) \right\} \left\{ \bar{Q}_{\beta},\left( \bar{c}_{\beta i}^{\dagger}+\bar{c}_{\beta i} \right) \right\} \right) ^L\left( \left( c_{\alpha}^{i\dagger}+c_{\alpha}^{i} \right) \left( c_{\beta}^{i\dagger}+c_{\beta}^{i} \right) \right) ^R
\\
&+i\mu e^{i\phi}\left( \left\{ \bar{Q}_{\alpha},\left( \bar{c}_{\alpha i}^{\dagger}+\bar{c}_{\alpha i} \right) \right\} \left\{ \bar{Q}_{\beta},\left( \bar{c}_{\beta i}^{\dagger}+\bar{c}_{\beta i} \right) \right\} \right) ^R\left( \left( c_{\alpha}^{i\dagger}+c_{\alpha}^{i} \right) \left( c_{\beta}^{i\dagger}+c_{\beta}^{i} \right) \right) ^L
\\
&=i\mu e^{-i\phi}\left( \left\{ \bar{Q}_{\alpha},\left( \frac{C_{\alpha}^{i\dagger}-\bar{C}_{\alpha}^{i\dagger}+C_{\alpha}^{i}-\bar{C}_{\alpha}^{i}}{2i} \right) \right\} \left\{ \bar{Q}_{\beta},\left( \frac{C_{\beta}^{i\dagger}-\bar{C}_{\beta}^{i\dagger}+C_{\beta}^{i}-\bar{C}_{\beta}^{i}}{2i} \right) \right\} \right) ^L
\\
&\left( \left( \frac{C_{\alpha}^{i\dagger}+\bar{C}_{\alpha}^{i\dagger}+C_{\alpha}^{i}+\bar{C}_{\alpha}^{i}}{2} \right) \left( \frac{C_{\beta}^{i\dagger}+\bar{C}_{\beta}^{i\dagger}+C_{\beta}^{i}+\bar{C}_{\beta}^{i}}{2} \right) \right) ^R
\\
&+i\mu e^{i\phi}\left( \left\{ \bar{Q}_{\alpha},\left( \frac{C_{\alpha}^{i\dagger}-\bar{C}_{\alpha}^{i\dagger}+C_{\alpha}^{i}-\bar{C}_{\alpha}^{i}}{2i} \right) \right\} \left\{ \bar{Q}_{\beta},\left( \frac{C_{\beta}^{i\dagger}-\bar{C}_{\beta}^{i\dagger}+C_{\beta}^{i}-\bar{C}_{\beta}^{i}}{2i} \right) \right\} \right) ^R
\\
&\left( \left( \frac{C_{\alpha}^{i\dagger}+\bar{C}_{\alpha}^{i\dagger}+C_{\alpha}^{i}+\bar{C}_{\alpha}^{i}}{2} \right) \left( \frac{C_{\beta}^{i\dagger}+\bar{C}_{\beta}^{i\dagger}+C_{\beta}^{i}+\bar{C}_{\beta}^{i}}{2} \right) \right) ^L
\end{split}
\end{equation}
In this section, we aim to investigate whether the interaction Hamiltonian affects the supersymmetry thermal field double states. We should concentrate on the diagonal terms within the Hilbert space
\begin{equation}
\begin{split}
H_{int}&=i\mu e^{-i\phi}\left( \left\{ \bar{Q}_{\alpha},\frac{C_{\alpha}^{i\dagger}}{2} \right\} \left\{ \bar{Q}_{\beta},\frac{C_{\beta}^{i\dagger}}{2} \right\} \right) ^L\left( \frac{C_{\alpha}^{i}}{2i}\frac{C_{\beta}^{i}}{2i} \right) ^R+i\mu e^{-i\phi}\left( \left\{ \bar{Q}_{\alpha},\frac{\bar{C}_{\alpha}^{i\dagger}}{2} \right\} \left\{ \bar{Q}_{\beta},\frac{\bar{C}_{\beta}^{i\dagger}}{2} \right\} \right) ^L\left( \frac{\bar{C}_{\alpha}^{i}}{2i}\frac{\bar{C}_{\beta}^{i}}{2i} \right) ^R
\\
&-i\mu e^{-i\phi}\left( \left\{ \bar{Q}_{\alpha},\frac{C_{\alpha}^{i\dagger}}{2} \right\} \left\{ \bar{Q}_{\beta},\frac{\bar{C}_{\beta}^{i\dagger}}{2} \right\} \right) ^L\left( \frac{C_{\alpha}^{i}}{2i}\frac{\bar{C}_{\beta}^{i}}{2i} \right) ^R-i\mu e^{-i\phi}\left( \left\{ \bar{Q}_{\alpha},\frac{\bar{C}_{\alpha}^{i\dagger}}{2} \right\} \left\{ \bar{Q}_{\beta},\frac{C_{\beta}^{i\dagger}}{2} \right\} \right) ^L\left( \frac{\bar{C}_{\alpha}^{i}}{2i}\frac{C_{\beta}^{i}}{2i} \right) ^R
\\
&+i\mu e^{-i\phi}\left( \left\{ \bar{Q}_{\alpha},\frac{C_{\alpha}^{i\dagger}}{2} \right\} \left\{ \bar{Q}_{\beta},\frac{C_{\beta}^{i\dagger}}{2} \right\} \right) ^R\left( \frac{C_{\alpha}^{i}}{2i}\frac{C_{\beta}^{i}}{2i} \right) ^L+i\mu e^{-i\phi}\left( \left\{ \bar{Q}_{\alpha},\frac{\bar{C}_{\alpha}^{i\dagger}}{2} \right\} \left\{ \bar{Q}_{\beta},\frac{\bar{C}_{\beta}^{i\dagger}}{2} \right\} \right) ^R\left( \frac{\bar{C}_{\alpha}^{i}}{2i}\frac{\bar{C}_{\beta}^{i}}{2i} \right) ^L
\\
&-i\mu e^{-i\phi}\left( \left\{ \bar{Q}_{\alpha},\frac{C_{\alpha}^{i\dagger}}{2} \right\} \left\{ \bar{Q}_{\beta},\frac{\bar{C}_{\beta}^{i\dagger}}{2} \right\} \right) ^R\left( \frac{C_{\alpha}^{i}}{2i}\frac{\bar{C}_{\beta}^{i}}{2i} \right) ^L-i\mu e^{-i\phi}\left( \left\{ \bar{Q}_{\alpha},\frac{\bar{C}_{\alpha}^{i\dagger}}{2} \right\} \left\{ \bar{Q}_{\beta},\frac{C_{\beta}^{i\dagger}}{2} \right\} \right) ^R\left( \frac{\bar{C}_{\alpha}^{i}}{2i}\frac{C_{\beta}^{i}}{2i} \right) ^L
\end{split}
\end{equation}

\begin{equation}
\begin{split}
\bar{H}_{int}&=i\mu e^{-i\phi}\left( \left\{ \bar{Q}_{\alpha},\frac{C_{\alpha}^{i\dagger}}{2i} \right\} \left\{ \bar{Q}_{\beta},\frac{C_{\beta}^{i\dagger}}{2i} \right\} \right) ^L\left( \frac{C_{\alpha}^{i}}{2}\frac{C_{\beta}^{i}}{2} \right) ^R+i\mu e^{-i\phi}\left( \left\{ \bar{Q}_{\alpha},\frac{\bar{C}_{\alpha}^{i\dagger}}{2i} \right\} \left\{ \bar{Q}_{\beta},\frac{\bar{C}_{\beta}^{i\dagger}}{2i} \right\} \right) ^L\left( \frac{\bar{C}_{\alpha}^{i}}{2}\frac{\bar{C}_{\beta}^{i}}{2} \right) ^R
\\
&-i\mu e^{-i\phi}\left( \left\{ \bar{Q}_{\alpha},\frac{C_{\alpha}^{i\dagger}}{2i} \right\} \left\{ \bar{Q}_{\beta},\frac{\bar{C}_{\beta}^{i\dagger}}{2i} \right\} \right) ^L\left( \frac{C_{\alpha}^{i}}{2}\frac{\bar{C}_{\beta}^{i}}{2} \right) ^R-i\mu e^{-i\phi}\left( \left\{ \bar{Q}_{\alpha},\frac{\bar{C}_{\alpha}^{i\dagger}}{2i} \right\} \left\{ \bar{Q}_{\beta},\frac{C_{\beta}^{i\dagger}}{2i} \right\} \right) ^L\left( \frac{\bar{C}_{\alpha}^{i}}{2}\frac{C_{\beta}^{i}}{2} \right) ^R
\\
&+i\mu e^{-i\phi}\left( \left\{ \bar{Q}_{\alpha},\frac{C_{\alpha}^{i\dagger}}{2i} \right\} \left\{ \bar{Q}_{\beta},\frac{C_{\beta}^{i\dagger}}{2i} \right\} \right) ^R\left( \frac{C_{\alpha}^{i}}{2}\frac{C_{\beta}^{i}}{2} \right) ^L+i\mu e^{-i\phi}\left( \left\{ \bar{Q}_{\alpha},\frac{\bar{C}_{\alpha}^{i\dagger}}{2i} \right\} \left\{ \bar{Q}_{\beta},\frac{\bar{C}_{\beta}^{i\dagger}}{2i} \right\} \right) ^R\left( \frac{\bar{C}_{\alpha}^{i}}{2}\frac{\bar{C}_{\beta}^{i}}{2} \right) ^L
\\
&-i\mu e^{-i\phi}\left( \left\{ \bar{Q}_{\alpha},\frac{C_{\alpha}^{i\dagger}}{2i} \right\} \left\{ \bar{Q}_{\beta}\frac{\bar{C}_{\beta}^{i\dagger}}{2i} \right\} \right) ^R\left( \frac{C_{\alpha}^{i}}{2}\frac{\bar{C}_{\beta}^{i}}{2} \right) ^L-i\mu e^{-i\phi}\left( \left\{ \bar{Q}_{\alpha},\frac{\bar{C}_{\alpha}^{i\dagger}}{2i} \right\} \left\{ \bar{Q}_{\beta},\frac{C_{\beta}^{i\dagger}}{2i} \right\} \right) ^R\left( \frac{\bar{C}_{\alpha}^{i}}{2}\frac{C_{\beta}^{i}}{2} \right) ^L
\end{split}
\end{equation}
Then the turning terms give
\begin{equation}
\begin{split}
&\left< TFD_{\beta =0} \right|H_{int}\left| TFD_{\beta =0} \right> =\frac{1}{2^N}\sum_{qq^{'}}{\sum_{mn}{\left< \hat{n}_{-q^{'}} \right|_2\otimes \left< n_{q^{'}} \right|_1\left( i\mu \sum_i{\left( e^{-i\phi}\left( \left\{ \bar{Q}_{\alpha},\frac{C_{\alpha}^{i\dagger}}{2} \right\} \left\{ \bar{Q}_{\beta},\frac{C_{\beta}^{i\dagger}}{2} \right\} \right) ^L\left( \frac{C_{\alpha}^{i}}{2i}\frac{C_{\beta}^{i}}{2i} \right) ^R \right.} \right.}}
\\
&+i\mu e^{-i\phi}\left( \left\{ \bar{Q}_{\alpha},\frac{\bar{C}_{\alpha}^{i\dagger}}{2} \right\} \left\{ \bar{Q}_{\beta},\frac{\bar{C}_{\beta}^{i\dagger}}{2} \right\} \right) ^L\left( \frac{\bar{C}_{\alpha}^{i}}{2i}\frac{\bar{C}_{\beta}^{i}}{2i} \right) ^R-i\mu e^{-i\phi}\left( \left\{ \bar{Q}_{\alpha},\frac{C_{\alpha}^{i\dagger}}{2} \right\} \left\{ \bar{Q}_{\beta},\frac{\bar{C}_{\beta}^{i\dagger}}{2} \right\} \right) ^L\left( \frac{C_{\alpha}^{i}}{2i}\frac{\bar{C}_{\beta}^{i}}{2i} \right) ^R
\\
&-i\mu e^{-i\phi}\left( \left\{ \bar{Q}_{\alpha},\frac{\bar{C}_{\alpha}^{i\dagger}}{2} \right\} \left\{ \bar{Q}_{\beta},\frac{C_{\beta}^{i\dagger}}{2} \right\} \right) ^L\left( \frac{\bar{C}_{\alpha}^{i}}{2i}\frac{C_{\beta}^{i}}{2i} \right) ^R+i\mu e^{i\phi}\left( \left\{ \bar{Q}_{\alpha}\frac{C_{\alpha}^{i\dagger}}{2} \right\} \left\{ \bar{Q}_{\beta},\frac{C_{\beta}^{i\dagger}}{2} \right\} \right) ^R\left( \frac{C_{\alpha}^{i}}{2i}\frac{C_{\beta}^{i}}{2i} \right) ^L
\\
&+i\mu e^{i\phi}\left( \left\{ \bar{Q}_{\alpha},\frac{\bar{C}_{\alpha}^{i\dagger}}{2} \right\} \left\{ \bar{Q}_{\beta},\frac{\bar{C}_{\beta}^{i\dagger}}{2} \right\} \right) ^R\left( \frac{\bar{C}_{\alpha}^{i}}{2i}\frac{\bar{C}_{\beta}^{i}}{2i} \right) ^L-i\mu e^{i\phi}\left( \left\{ \bar{Q}_{\alpha},\frac{C_{\alpha}^{i\dagger}}{2} \right\} \left\{ \bar{Q}_{\beta},\frac{\bar{C}_{\beta}^{i\dagger}}{2} \right\} \right) ^R\left( \frac{C_{\alpha}^{i}}{2i}\frac{\bar{C}_{\beta}^{i}}{2i} \right) ^L
\\
&\left. \left. -i\mu e^{i\phi}\left( \left\{ \bar{Q}_{\alpha},\frac{\bar{C}_{\alpha}^{i\dagger}}{2} \right\} \left\{ \bar{Q}_{\beta},\frac{C_{\beta}^{i\dagger}}{2} \right\} \right) ^R\left( \frac{\bar{C}_{\alpha}^{i}}{2i}\frac{C_{\beta}^{i}}{2i} \right) ^L \right) \right) \left| m_q \right> _1\otimes \left| \hat{m}_{-q} \right> _2
\\
&=\frac{1}{2^N}\sum_{qq^{'}}{\sum_{nm}{\sum_i{\left( i\mu \right.}}}e^{i\pi \left( -q_{\alpha}-\bar{q}_{\alpha}-q_{\beta}-\bar{q}_{\beta}+\frac{N}{2} \right)}\left< n_{q^{'}} \right|\left( \left\{ \bar{Q}_{\alpha},\frac{C_{\alpha}^{i\dagger}}{2} \right\} \left\{ \bar{Q}_{\beta},\frac{C_{\beta}^{i\dagger}}{2} \right\} \right) ^L\left| m_{q^{''}}^{'} \right> \left< \hat{n}_{-q^{'}} \right|\left( \frac{C_{\alpha}^{i}}{2i}\frac{C_{\beta}^{i}}{2i} \right) ^R\left| \hat{m}_{-q} \right>
\\
&+\left< n_{q^{'}} \right|\left( \left\{ \bar{Q}_{\alpha},\frac{\bar{C}_{\alpha}^{i\dagger}}{2} \right\} \left\{ \bar{Q}_{\beta},\frac{\bar{C}_{\beta}^{i\dagger}}{2} \right\} \right) ^L\left| m_{q^{''}}^{'} \right> \left< \hat{n}_{-q^{'}} \right|\left( \frac{\bar{C}_{\alpha}^{i}}{2i}\frac{\bar{C}_{\beta}^{i}}{2i} \right) ^R\left| \hat{m}_{-q} \right>
\\
&-\left< n_{q^{'}} \right|\left( \left\{ \bar{Q}_{\alpha},\frac{C_{\alpha}^{i\dagger}}{2} \right\} \left\{ \bar{Q}_{\beta},\frac{\bar{C}_{\beta}^{i\dagger}}{2} \right\} \right) ^L\left| m_{q^{''}}^{'} \right> \left< \hat{n}_{-q^{'}} \right|\left( \frac{C_{\alpha}^{i}}{2i}\frac{\bar{C}_{\beta}^{i}}{2i} \right) ^R\left| \hat{m}_{-q} \right>
\\
&-\left< n_{q^{'}} \right|\left( \left\{ \bar{Q}_{\alpha},\frac{\bar{C}_{\alpha}^{i\dagger}}{2} \right\} \left\{ \bar{Q}_{\beta},\frac{C_{\beta}^{i\dagger}}{2} \right\} \right) ^L\left| m_{q^{''}}^{'} \right> \left< \hat{n}_{-q^{'}} \right|\left( \frac{\bar{C}_{\alpha}^{i}}{2i}\frac{C_{\beta}^{i}}{2i} \right) ^R\left| \hat{m}_{-q} \right>
\\
&+i\mu e^{i\pi \left( -q_{\alpha}-\bar{q}_{\alpha}-q_{\beta}-\bar{q}_{\beta}+\frac{N}{2} \right)}e^{i\phi}\left< n_{q^{'}} \right|\left( \left\{ \bar{Q}_{\alpha},\frac{C_{\alpha}^{i\dagger}}{2} \right\} \left\{ \bar{Q}_{\beta},\frac{C_{\beta}^{i\dagger}}{2} \right\} \right) ^R\left| m_{q^{''}}^{'} \right> \left< \hat{n}_{-q^{'}} \right|\left( \frac{C_{\alpha}^{i}}{2i}\frac{C_{\beta}^{i}}{2i} \right) ^L\left| \hat{m}_{-q} \right>
\\
&+\left< n_{q^{'}} \right|\left( \left\{ \bar{Q}_{\alpha},\frac{\bar{C}_{\alpha}^{i\dagger}}{2} \right\} \left\{ \bar{Q}_{\beta},\frac{\bar{C}_{\beta}^{i\dagger}}{2} \right\} \right) ^R\left| m_{q^{''}}^{'} \right> \left< \hat{n}_{-q^{'}} \right|\left( \frac{\bar{C}_{\alpha}^{i}}{2i}\frac{\bar{C}_{\beta}^{i}}{2i} \right) ^L\left| \hat{m}_{-q} \right>
\\
&-\left< n_{q^{'}} \right|\left( \left\{ \bar{Q}_{\alpha},\frac{C_{\alpha}^{i\dagger}}{2} \right\} \left\{ \bar{Q}_{\beta},\frac{\bar{C}_{\beta}^{i\dagger}}{2} \right\} \right) ^R\left| m_{q^{''}}^{'} \right> \left< \hat{n}_{-q^{'}} \right|\left( \frac{C_{\alpha}^{i}}{2i}\frac{\bar{C}_{\beta}^{i}}{2i} \right) ^L\left| \hat{m}_{-q} \right>
\\
&-\left< n_{q^{'}} \right|\left( \left\{ \bar{Q}_{\alpha},\frac{\bar{C}_{\alpha}^{i\dagger}}{2} \right\} \left\{ \bar{Q}_{\beta},\frac{C_{\beta}^{i\dagger}}{2} \right\} \right) ^R\left| m_{q^{''}}^{'} \right> \left< \hat{n}_{-q^{'}} \right|\left( \frac{\bar{C}_{\alpha}^{i}}{2i}\frac{C_{\beta}^{i}}{2i} \right) ^L\left| \hat{m}_{-q} \right>
\end{split}
\end{equation}
\begin{equation}
\begin{split}
&\left< H_{int} \right> =\frac{1}{2^N}\sum_{qq^{'}}{\sum_{nm}{\sum_i{\left[ -i\mu \right.}}}e^{-i\phi}\left< n_{q^{'}} \right|\left( \left\{ \bar{Q}_{\alpha},\frac{C_{\alpha}^{i\dagger}}{2} \right\} \left\{ \bar{Q}_{\beta},\frac{C_{\beta}^{i\dagger}}{2} \right\} \right) ^L\left| m_q \right> \left< m_q \right|\left( \frac{C_{\alpha}^{i}}{2i}\frac{C_{\beta}^{i}}{2i} \right) ^R\left| n_{q^{'}} \right>
\\
&-i\mu e^{-i\phi}\left< n_{q^{'}} \right|\left( \left\{ \bar{Q}_{\alpha},\frac{\bar{C}_{\alpha}^{i\dagger}}{2} \right\} \left\{ \bar{Q}_{\beta},\frac{\bar{C}_{\beta}^{i\dagger}}{2} \right\} \right) ^L\left| m_q \right> \left< m_q \right|\left( \frac{\bar{C}_{\alpha}^{i}}{2i}\frac{\bar{C}_{\beta}^{i}}{2i} \right) ^R\left| n_{q^{'}} \right>
\\
&+i\mu e^{-i\phi}\left< n_{q^{'}} \right|\left( \left\{ \bar{Q}_{\alpha},\frac{C_{\alpha}^{i\dagger}}{2} \right\} \left\{ \bar{Q}_{\beta},\frac{\bar{C}_{\beta}^{i\dagger}}{2} \right\} \right) ^L\left| m_q \right> \left< m_q \right|\left( \frac{C_{\alpha}^{i}}{2i}\frac{\bar{C}_{\beta}^{i}}{2i} \right) ^R\left| n_{q^{'}} \right>
\\
&+i\mu e^{-i\phi}\left< n_{q^{'}} \right|\left( \left\{ \bar{Q}_{\alpha},\frac{\bar{C}_{\alpha}^{i\dagger}}{2} \right\} \left\{ \bar{Q}_{\beta},\frac{C_{\beta}^{i\dagger}}{2} \right\} \right) ^L\left| m_q \right> \left< m_q \right|\left( \frac{\bar{C}_{\alpha}^{i}}{2i}\frac{C_{\beta}^{i}}{2i} \right) ^R\left| n_{q^{'}} \right>
\\
&-i\mu e^{i\phi}\left< m_q \right|\left( \left\{ \bar{Q}_{\alpha},\frac{C_{\alpha}^{i\dagger}}{2} \right\} \left\{ \bar{Q}_{\beta},\frac{C_{\beta}^{i\dagger}}{2} \right\} \right) ^R\left| n_{q^{'}} \right> \left< n_{q^{'}} \right|\left( \frac{C_{\alpha}^{i}}{2i}\frac{C_{\beta}^{i}}{2i} \right) ^L\left| m_q \right>
\\
&-i\mu e^{i\phi}\left< m_q \right|\left( \left\{ \bar{Q}_{\alpha},\frac{\bar{C}_{\alpha}^{i\dagger}}{2} \right\} \left\{ \bar{Q}_{\beta},\frac{\bar{C}_{\beta}^{i\dagger}}{2} \right\} \right) ^R\left| n_{q^{'}} \right> \left< n_{q^{'}} \right|\left( \frac{\bar{C}_{\alpha}^{i}}{2i}\frac{\bar{C}_{\beta}^{i}}{2i} \right) ^L\left| m_q \right>
\\
&+i\mu e^{-i\phi}\left< m_q \right|\left( \left\{ \bar{Q}_{\alpha},\frac{C_{\alpha}^{i\dagger}}{2} \right\} \left\{ \bar{Q}_{\beta},\frac{\bar{C}_{\beta}^{i\dagger}}{2} \right\} \right) ^R\left| n_{q^{'}} \right> \left< n_{q^{'}} \right|\left( \frac{C_{\alpha}^{i}}{2i}\frac{\bar{C}_{\beta}^{i}}{2i} \right) ^L\left| m_q \right>
\\
&+i\mu e^{-i\phi}\left< m_q \right|\left( \left\{ \bar{Q}_{\alpha},\frac{\bar{C}_{\alpha}^{i\dagger}}{2} \right\} \left\{ \bar{Q}_{\beta},\frac{C_{\beta}^{i\dagger}}{2} \right\} \right) ^R\left| n_{q^{'}} \right> \left< n_{q^{'}} \right|\left( \frac{\bar{C}_{\alpha}^{i}}{2i}\frac{C_{\beta}^{i}}{2i} \right) ^L\left| m_q \right>
\\
&=\sum_{qq^{'}}{\sum_{nm}{\sum_i{i\mu}}}/2^N\left< n_{q^{'}} \right|\left( \left\{ \bar{Q}_{\alpha},\frac{C_{\alpha}^{i\dagger}}{2} \right\} \left\{ \bar{Q}_{\beta},\frac{C_{\beta}^{i\dagger}}{2} \right\} \right) ^L\left| m_q \right> \left< m_q \right|\left( \frac{C_{\alpha}^{i}}{2i}\frac{C_{\beta}^{i}}{2i} \right) ^L\left| n_{q^{'}} \right>
\\
&+\left< n_{q^{'}} \right|\left( \left\{ \bar{Q}_{\alpha},\frac{\bar{C}_{\alpha}^{i\dagger}}{2} \right\} \left\{ \bar{Q}_{\beta},\frac{\bar{C}_{\beta}^{i\dagger}}{2} \right\} \right) ^L\left| m_q \right> \left< m_q \right|\left( \frac{\bar{C}_{\alpha}^{i}}{2i}\frac{\bar{C}_{\beta}^{i}}{2i} \right) ^L\left| n_{q^{'}} \right>
\\
&-\left< n_{q^{'}} \right|\left( \left\{ \bar{Q}_{\alpha},\frac{C_{\alpha}^{i\dagger}}{2} \right\} \left\{ \bar{Q}_{\beta},\frac{\bar{C}_{\beta}^{i\dagger}}{2} \right\} \right) ^L\left| m_q \right> \left< m_q \right|\left( \frac{C_{\alpha}^{i}}{2i}\frac{\bar{C}_{\beta}^{i}}{2i} \right) ^L\left| n_{q^{'}} \right>
\\
&-\left< n_{q^{'}} \right|\left( \left\{ \bar{Q}_{\alpha},\frac{\bar{C}_{\alpha}^{i\dagger}}{2} \right\} \left\{ \bar{Q}_{\beta},\frac{C_{\beta}^{i\dagger}}{2} \right\} \right) ^L\left| m_q \right> \left< m_q \right|\left( \frac{\bar{C}_{\alpha}^{i}}{2i}\frac{C_{\beta}^{i}}{2i} \right) ^L\left| n_{q^{'}} \right>
\\
&=-\frac{\mu}{2^N}\sum_{n,q}{\sum_i{\left< n_q \right|_1}}\left( \left\{ \bar{Q}_{\alpha},\frac{C_{\alpha}^{i\dagger}}{2} \right\} \left\{ \bar{Q}_{\beta},\frac{C_{\beta}^{i\dagger}}{2} \right\} \right) ^L\left( \frac{C_{\alpha}^{i}}{2i}\frac{C_{\beta}^{i}}{2i} \right) ^L
\\
&+\left( \left\{ \bar{Q}_{\alpha},\frac{\bar{C}_{\alpha}^{i\dagger}}{2} \right\} \left\{ \bar{Q}_{\beta},\frac{\bar{C}_{\beta}^{i\dagger}}{2} \right\} \right) ^L\left( \frac{\bar{C}_{\alpha}^{i}}{2i}\frac{\bar{C}_{\beta}^{i}}{2i} \right) ^L\left| n_q \right> _1
\\
&-\left( \left\{ \bar{Q}_{\alpha},\frac{C_{\alpha}^{i\dagger}}{2} \right\} \left\{ \bar{Q}_{\beta},\frac{\bar{C}_{\beta}^{i\dagger}}{2} \right\} \right) ^L\left( \frac{C_{\alpha}^{i}}{2i}\frac{\bar{C}_{\beta}^{i}}{2i} \right) ^L\left| n_q \right> _1
\\
&-\left( \left\{ \bar{Q}_{\alpha},\frac{\bar{C}_{\alpha}^{i\dagger}}{2} \right\} \left\{ \bar{Q}_{\beta},\frac{C_{\beta}^{i\dagger}}{2} \right\} \right) ^L\left( \frac{\bar{C}_{\alpha}^{i}}{2i}\frac{C_{\beta}^{i}}{2i} \right) ^L\left| n_q \right> _1=Const.
\end{split}
\end{equation}
where
\begin{equation}
\sum_{n,q}{\sum_i{\left< n_q \right|}}\left\{ \bar{Q}_{\alpha},\frac{C_{\alpha}^{i\dagger}}{2} \right\} \left\{ \bar{Q}_{\beta},\frac{C_{\beta}^{i\dagger}}{2} \right\} \left( \frac{C_{\alpha}^{i}}{2i}\frac{C_{\beta}^{i}}{2i} \right) \left| n_q \right> =-\frac{1}{4}\sum_{n,q}{N\left( Q_{n_q}+Q_{n_q}+1 \right) ^2}\,.
\end{equation}
We can easily verify that the chiral-conjugate component satisfies $\left< \bar{H_{int}} \right>=Const$ by calculating its value directly from the definitions.

\section*{Appendix B: Supersymmetric traversable wormhole via boundary interaction} 
\addcontentsline{toc}{section}{Appendix B: Supersymmetric traversable wormhole via boundary interaction}
\setcounter{equation}{0}
\renewcommand\theequation{B.\arabic{equation}}
In this appendix, we mainly consider the wormhole traversability corresponding to the interaction of two SYK models. 

We consider the processes in previous work as double trace deformations in AdS spacetime \cite{46,55,56}. In other words, we can add the interaction term as a path integral into the maximally entangled thermal field on both sides. The same method can also be applied to supersymmetric systems, which contain interactions with fermionic-bosonic correlators. After incorporating these interaction terms, the two boundaries are directly coupled, and we can use various methods to verify these kinds of traversable wormholes.

In gravitational analysis, we can also define interactions with certain dimensions, known as a set of operators with specific dimensions. Additionally, we can use fermionic interactions in coupled SYK models, which are related to the gravitational action on NAdS boundaries as the reparametrization Schwarzian action. The interaction part can be written as

\begin{equation}
gV=g\sum_i{\int{du}\left( O_{L}^{i}\left( u \right) O_{R}^{i}\left( u \right) \right)}\sim i\mu \sum_j{\left( \psi _{L}^{j}\psi _{R}^{j}-\psi _{R}^{j}\psi _{L}^{j} \right)}.
\end{equation}

In order to preserve the interaction terms in a supersymmetric manner, we can simply expand it into a supersymmetric version.

\begin{equation}
gV=g\sum_i{\int{dud\theta}\left( O_{L}^{i}\left( u,\theta \right) O_{R}^{i}\left( u,\theta \right) \right)}\sim i\mu \partial _{\theta}\sum_j{\left( \varPsi _{L}^{j}\varPsi _{R}^{j}-\varPsi _{R}^{j}\varPsi _{L}^{j} \right)}.
\end{equation}

After considering the additional chiral restriction and R symmetry, we obtain the N=2 and N=4 formalisms

\begin{equation}
gV=g\sum_i{\int{dud\theta d\bar{\theta}}\left( O_{L}^{i}\left( u,\theta ,\bar{\theta} \right) \bar{O}_{R}^{i}\left( u,\theta ,\bar{\theta} \right) +\bar{O}_{L}^{i}\left( u,\theta ,\bar{\theta} \right) O_{R}^{i}\left( u,\theta ,\bar{\theta} \right) \right)}
$$
$$
\sim i\mu \partial _{\theta}\sum_j{\left( \varPsi _{L}^{j}\bar{\varPsi}_{R}^{j}-\varPsi _{R}^{j}\bar{\varPsi}_{L}^{j} \right)}+i\mu \partial _{\bar{\theta}}\sum_j{\left( \bar{\varPsi}_{L}^{j}\varPsi _{R}^{j}-\bar{\varPsi}_{R}^{j}\varPsi _{L}^{j} \right)},
$$

$$
gV=g\sum_i{\int{dud^2\theta d^2\bar{\theta}}\left( O_{L}^{i}\left( u,\theta ^1,\theta ^2,\bar{\theta}_1,\bar{\theta}_2 \right) \bar{O}_{R}^{i}\left( u,\theta ^1,\theta ^2,\bar{\theta}_1,\bar{\theta}_2 \right) +\bar{O}_{L}^{i}\left( u,\theta ^1,\theta ^2,\bar{\theta}_1,\bar{\theta}_2 \right) O_{R}^{i}\left( u,\theta ^1,\theta ^2,\bar{\theta}_1,\bar{\theta}_2 \right) \right)}
$$
$$
\sim i\mu \partial _{\theta}^{2}\sum_j{\left( \varPhi _{L}^{j}\bar{\varPhi}_{R}^{j}-\varPsi _{R}^{j}\bar{\varPhi}_{L}^{j} \right)}+i\mu \partial _{\bar{\theta}}^{2}\sum_j{\left( \bar{\varPhi}_{L}^{j}\varPhi _{R}^{j}-\bar{\varPhi}_{R}^{j}\varPhi _{L}^{j} \right)}.
\end{equation}

The previous work \cite{55} has investigated traversability with observables. We can use simple entanglement operators with reparametrization to measure traversability.
\begin{equation}
C\equiv \left< \left[ \exp \left( -igV \right) \phi _{L}^{i}\left( t_L \right) \exp \left( igV \right) ,\phi _{R}^{i}\left( t_R \right) \right] \right> .
\end{equation}

Here, we use the commutation relationship and interaction operators that exchange information between two non-supersymmetric sides. We can further measure the traversability between boundary operators.

In the supersymmetric case, we have used $\Phi$ to denote the gravitational operators on the boundaries, and we can also rewrite the supersymmetric version.
\begin{equation}
C\equiv \partial _{\theta}\left< \left[ \exp \left( -igV \right) \varPhi _{L}^{i}\left( t_L,\theta _L \right) \exp \left( igV \right) ,\varPhi _{R}^{i}\left( t_R,\theta _R \right) \right] \right> .
\end{equation}
In chiral measurement, we can define more explicitly the operators in N=2 and N=4 supersymmetric SYK systems.
\begin{equation}
\begin{split}
C&\equiv \partial _{\theta}\partial _{\bar{\theta}}\left< \left[ \exp \left( -igV \right) \varPhi _{L}^{i}\left( t_L,\theta _L,\bar{\theta}_L \right) \exp \left( igV \right) ,\bar{\varPhi}_{R}^{i}\left( t_R,\theta _R,\bar{\theta}_R \right) \right]\right.
\\
& \left.+\left[ \exp \left( -igV \right) \bar{\varPhi}_{L}^{i}\left( t_L,\theta _L,\bar{\theta}_L \right) \exp \left( igV \right) ,\varPhi _{R}^{i}\left( t_R,\theta _R,\bar{\theta}_R \right) \right] \right> 
\\
C&\equiv \partial _{\theta}^{2}\partial _{\bar{\theta}}^{2}\left< \left[ \exp \left( -igV \right) \varPhi _{L}^{i}\left( t_L,\theta _{L}^{1},\theta _{L}^{2},\bar{\theta}_{1L},\bar{\theta}_{2L} \right) \exp \left( igV \right) ,\bar{\varPhi}_{R}^{i}\left( t_R,\theta _{R}^{1},\theta _{R}^{2},\bar{\theta}_{1R},\bar{\theta}_{2R} \right) \right] \right. 
\\
&\left. +\left[ \exp \left( -igV \right) \bar{\varPhi}_{L}^{i}\left( t_L,\theta _{L}^{1},\theta _{L}^{2},\bar{\theta}_{1L},\bar{\theta}_{2L} \right) \exp \left( igV \right) ,\varPhi _{R}^{i}\left( t_R,\theta _{R}^{1},\theta _{R}^{2},\bar{\theta}_{1R},\bar{\theta}_{2R} \right) \right] \right> .
\end{split}
\end{equation}

When the interactions are small, we consider the entanglement in perturbation theory, and the path integral can be included up to the first order
\begin{equation}
C\sim \sum_i{ig}\left< \left[ \phi _L,O_{L}^{i} \right] \left[ \phi _R,O_{R}^{i} \right] \right> +o\left( g^2 \right) .
\end{equation}
In this case, the theory is investigated in weak coupling limit. We can simply expand it into the N=1 supersymmetric SYK model

\begin{equation}
C\sim \partial _{\theta}\sum_i{ig}\left< \left[ \varPhi _L,O_{L}^{i} \right] \left[ \varPhi _R,O_{R}^{i} \right] \right> +o\left( g^2 \right) .
\end{equation}
where the dilaton action and the interaction operators in superspace are also restricted by Grassmann variables. Similarly, the N=2 and N=4 form is given by
\begin{equation}
C\sim \partial _{\theta}\partial _{\bar{\theta}}\sum_i{ig}\left< \left[ \varPhi _L,\bar{O}_{L}^{i} \right] \left[ \bar{\varPhi}_R,O_{R}^{i} \right] +\left[ \bar{\varPhi}_L,O_{L}^{i} \right] \left[ \varPhi _R,\bar{O}_{R}^{i} \right] \right> +o\left( g^2 \right) ,
$$
$$
C\sim \partial _{\theta}^{2}\partial _{\bar{\theta}}^{2}\sum_i{ig}\left< \left[ \varPhi _L,\bar{O}_{L}^{i} \right] \left[ \bar{\varPhi}_R,O_{R}^{i} \right] +\left[ \bar{\varPhi}_L,O_{L}^{i} \right] \left[ \varPhi _R,\bar{O}_{R}^{i} \right] \right> +o\left( g^2 \right) .
\end{equation}

These interaction operators are positive to evolve the boundary time, and the messages can be considered traversable. The interaction energy is also produced negatively. 

However, this traversability measurement is not unique. For example, we can make the first-order correlation between Majorana fermions, and we can also consider the SYK interaction in the strong coupling and low temperature limit

\begin{equation}
C\equiv \frac{1}{2}\left< \left\{ \exp \left( -igV \right) \psi _{L}^{i}\left( t \right) \exp \left( igV \right) ,\psi _{R}^{i}\left( t \right) \right\} \right> \sim \mu \sum_i{\left< \left\{ \psi _{L}^{i}\left( t \right) ,\psi _{L}^{j}\left( 0 \right) \right\} \left\{ \psi _{R}^{i}\left( t \right) ,\psi _{R}^{j}\left( 0 \right) \right\} \right>}.
\end{equation}

The previous work has investigated traversability with observables, then evaluated the first-order expectation. After averaging over time translation, the anti-commutation would be quantified as
\begin{equation}
\left\{ \psi _{A}^{i}\left( t \right) ,\psi _{A}^{j}\left( 0 \right) \right\} \sim it\left\{ \left[ H,\psi _{A}^{i}\left( 0 \right) \right] ,\psi _{A}^{j}\left( 0 \right) \right\} =-itJ_{ijkl}\psi _{A}^{k}\psi _{A}^{l},
$$
$$
C=gt_Lt_R\sum_i{\left< J_{ijkl}\psi _{A}^{k}\psi _{A}^{l}J_{i'j'k'l'}\psi _{B}^{k'}\psi _{B}^{l'} \right>}\sim gJ^2t_Lt_R.
\end{equation}

Note that the fermionic operators commute with the Hamiltonian and supercharges. As in the supersymmetric version, we consider the correlation operators between fermions and bosons
\begin{equation}
\begin{split}
&C\equiv \frac{1}{2}\left< \left\{ \exp \left( -igV \right) b_{L}^{i}\left( t \right) \exp \left( igV \right) ,\psi _{R}^{i}\left( t \right) \right\} \right> \sim \mu \sum_j{\left< \left[ b_{L}^{i}\left( t \right) ,b_{L}^{j}\left( 0 \right) \right] \left\{ \psi _{R}^{i}\left( t \right) ,\psi _{R}^{j}\left( 0 \right) \right\} \right>}
\\
&\left\{ \psi _{A}^{i}\left( t \right) ,\psi _{A}^{j}\left( 0 \right) \right\} \sim it\left\{ \left[ H,\psi _{A}^{i}\left( 0 \right) \right] ,\psi _{A}^{j}\left( 0 \right) \right\} =tC_{ijk}b_{A}^{k},
\\
&\left[ b_{L}^{i}\left( t \right) ,b_{L}^{j}\left( 0 \right) \right] \sim it\left[ \left[ H,b_{A}^{i}\left( 0 \right) \right] ,b_{A}^{j}\left( 0 \right) \right] =it\left[ \left[ H,\left\{ Q,\psi _{A}^{j}\left( 0 \right) \right\} \right] ,\left\{ Q,\psi _{A}^{j}\left( 0 \right) \right\} \right] \sim tC_{ij'k'}\left[ \psi _{A}^{j'}\left( 0 \right) \psi _{A}^{k'}\left( 0 \right) ,b_{A}^{j}\left( 0 \right) \right] 
\\
&C=gt_Lt_R\sum_j{\left< C_{ijk}b_{A}^{k}C_{i'j'k'}\left[ \psi _{A}^{j'}\left( 0 \right) \psi _{A}^{k'}\left( 0 \right) ,b_{A}^{j}\left( 0 \right) \right] \right>}\sim gJ^2t_Lt_R.
\end{split}
\end{equation}

For the N=2 supersymmetric SYK model, we consider one of these components $b \bar{\psi}$ with chirals
\begin{equation}
\begin{split}
&C\equiv \frac{1}{2}\left< \left\{ \exp \left( -igV \right) b_{L}^{i}\left( t \right) \exp \left( igV \right) ,\bar{\psi}_{R}^{i}\left( t \right) \right\} \right> \sim \mu \sum_i{\left< \left[ b_{L}^{i}\left( t \right) ,b_{L}^{j}\left( 0 \right) \right] \left\{ \bar{\psi}_{R}^{i}\left( t \right) ,\bar{\psi}_{R}^{j}\left( 0 \right) \right\} \right>}
\\
&\left\{ \bar{\psi}_{A}^{i}\left( t \right) ,\bar{\psi}_{A}^{j}\left( 0 \right) \right\} \sim it\left\{ \left[ H,\bar{\psi}_{A}^{i}\left( t \right) \right] ,\bar{\psi}_{A}^{j}\left( 0 \right) \right\} =t\bar{C}_{ijk}\bar{b}_{A}^{k},
\\
&\left[ b_{L}^{i}\left( t \right) ,b_{L}^{j}\left( 0 \right) \right] \sim it\left[ \left[ H,b_{A}^{i}\left( 0 \right) \right] ,b_{A}^{j}\left( 0 \right) \right] =it\left[ \left[ H,\left\{ Q,\psi _{A}^{j}\left( 0 \right) \right\} \right] ,\left\{ Q,\psi _{A}^{j}\left( 0 \right) \right\} \right] \sim tC_{ij'k'}\left[ \psi _{A}^{j'}\left( 0 \right) \psi _{A}^{k'}\left( 0 \right) ,b_{A}^{j}\left( 0 \right) \right] 
\\
&C=gt_Lt_R\sum_i{\left< \bar{C}_{ijk}\bar{b}_{A}^{k}tC_{i'j'k'}\left[ \psi _{A}^{j'}\left( 0 \right) \psi _{A}^{k'}\left( 0 \right) ,b_{A}^{j}\left( 0 \right) \right] \right>}\sim gJ^2t_Lt_R.
\end{split}
\end{equation}

We can also utilize a similar process in the N=4 SYK correlation
\begin{equation}
\begin{split}
&C\equiv \frac{1}{2}\left< \left[ \exp \left( -igV \right) \phi _{L}^{i}\left( t \right) \exp \left( igV \right) ,\bar{F}_{R}^{i}\left( t \right) \right] \right> \sim \mu \sum_i{\left< \left[ \phi _{L}^{i}\left( t \right) ,\phi _{L}^{j}\left( 0 \right) \right] \left[ \bar{F}_{R}^{i}\left( t \right) ,\bar{F}_{R}^{j}\left( 0 \right) \right] \right>}
\\
&\left[ \bar{F}_{A}^{i}\left( t \right) ,\bar{F}_{A}^{j}\left( 0 \right) \right] \sim it\left[ \left[ H,\bar{F}_{A}^{i}\left( t \right) \right] ,\bar{F}_{A}^{j}\left( 0 \right) \right] =-3it\bar{C}_{ij'k'}\left[ \left[ \bar{\phi}_{A}^{j'}\bar{\phi}_{A}^{k'}\bar{F}_{A}^{i'}+\epsilon ^{\alpha \beta}\bar{\psi}_{\alpha A}^{j'}\bar{\psi}_{\beta A}^{k'}\bar{\phi}_{A}^{i'},\bar{F}_{A}^{i} \right] ,\bar{F}_{A}^{j} \right] ,
\\
&\left[ \phi _{A}^{i}\left( t \right) ,\phi _{A}^{j}\left( 0 \right) \right] \sim it\left[ \left[ H,\phi _{A}^{i}\left( t \right) \right] ,\phi _{A}^{j}\left( 0 \right) \right] =-3itC_{ij'k'}\left[ \phi _{A}^{j'}F_{A}^{k'}+\epsilon ^{\alpha \beta}\psi _{\alpha A}^{j'}\psi _{\beta A}^{k'},\phi _{A}^{j} \right] 
\\
&C=9gt_Lt_R\sum_i{\left< \bar{C}_{i'j'k'}\left[ \left[ \bar{\phi}_{A}^{j'}\bar{\phi}_{A}^{k'}\bar{F}_{A}^{i'}+\epsilon ^{\alpha \beta}\bar{\psi}_{\alpha A}^{j'}\bar{\psi}_{\beta A}^{k'}\bar{\phi}_{A}^{i'},\bar{F}_{A}^{i'} \right] ,\bar{F}_{A}^{j} \right] C_{i'j'k'}\left[ \phi _{A}^{j'}F_{A}^{k'}+\epsilon ^{\alpha \beta}\psi _{\alpha A}^{j'}\psi _{\beta A}^{k'},\phi _{A}^{j} \right] \right>}\sim gJ^2t_Lt_R.
\end{split}
\end{equation}

\begin{figure}[!t]
\begin{minipage}{0.28\linewidth}
\centerline{\includegraphics[width=6cm]{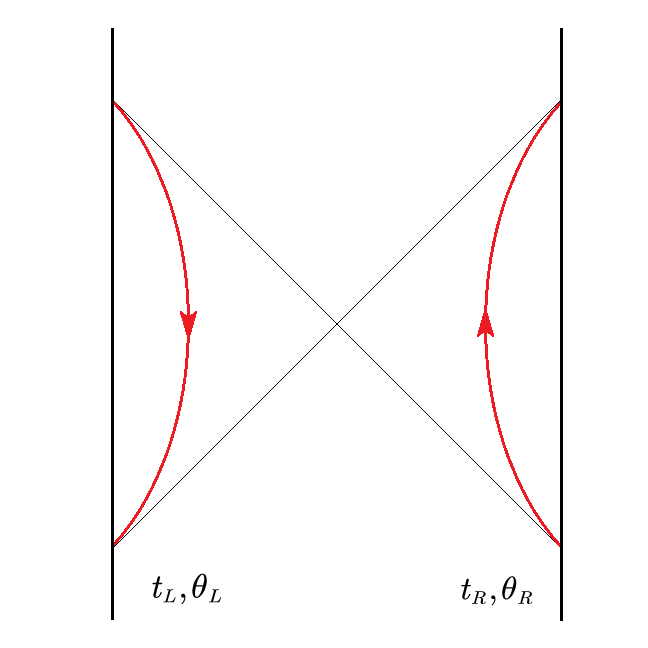}}
\centerline{(a)}
\end{minipage}
\hfill
\begin{minipage}{0.28\linewidth}
\centerline{\includegraphics[width=6cm]{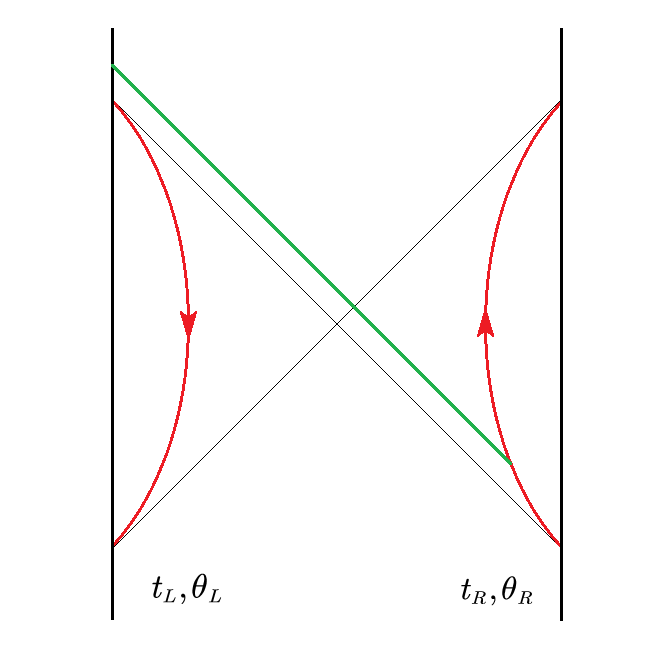}}
\centerline{(b)}
\end{minipage}
\hfill
\begin{minipage}{0.28\linewidth}
\centerline{\includegraphics[width=6cm]{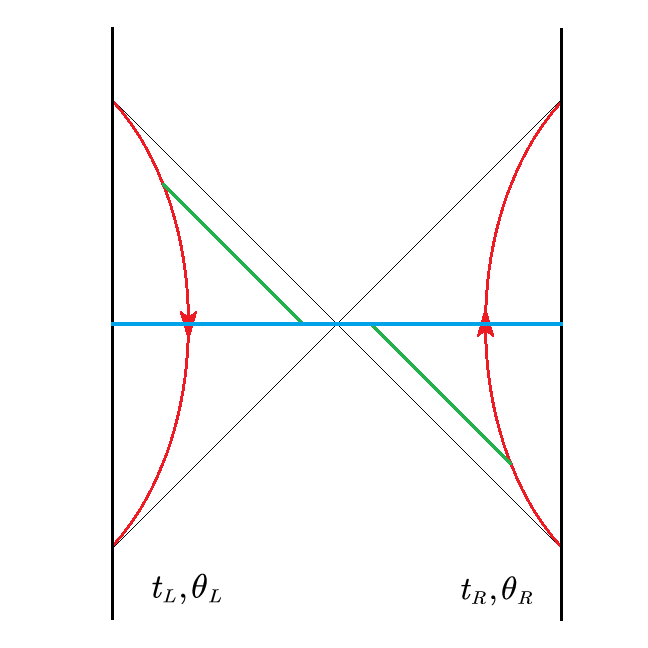}}
\centerline{(c)}
\end{minipage}

\caption{\label{fig:Figure 7}  Traversable wormhole via double trace deformation (a)The Lorentzian $NAdS_{2}$ geometry as maximally-entangled systems with Grassmann variables $\theta$ (b)Time-ordered messages sent from the right boundary do not reach the left. (c)The interaction term make the wormhole traversable}
\end{figure}

In Fig. 7(a), we have described the $NAdS_{2}$ Penrose diagram with two reparametrized boundaries. In Fig. 7(b), the message cannot arrive at the left side. However, after interacting with the boundary connection as shown in Fig. 7(c), the operators would be equivalent to those applied with a time evolution, and the messages would arrive at the left. We can also consider this effect as the shift onto the boundaries, where the infection of double trace deformation is now converted into the operator $exp(igV)$ for traversability. Here, we use the green line to express the signal, while the blue line denotes the deformations via boundary interactions.

When the interactions become larger and the theory returns to non-perturbation, the 'evolving operator' $exp(igV)$ would be replaced by the saddles, which corresponds to the results in Section II. We have proven the wormhole solution in this appendix and the black hole solution without the off-diagonal coupling of the supersymmetric SYK model.

{99}

\end{document}